\documentclass{ar-1col-S2O}

\usepackage{amssymb}
\usepackage[comma]{natbib}
\usepackage{url}
\usepackage{amsmath}
\usepackage{siunitx}

\jname{Annual Review of Astronomy and Astrophysics}
\jvol{AA}
\jyear{2023}
\doi{10.1146/((please add article doi))}

\def\nqso{531} %all
\def\nqsowithspectra{527} %quasars for which we have spectra
\def\nqsosix{275} %z>6
\def\nqsoseven{8} %z>7
\def\nqsomgii{113} %z>5.9 with MgII BH mass measurements

\def\lya     {\ensuremath{\text{Ly}\alpha}}

\def\cii     {\ensuremath{\text{[C\,\textsc{ii]}}}}
\def\ciilong    {\ensuremath{\text{[C\,\textsc{ii]}}_{158\,\mu\text{m}}}}
\def\civ     {\ensuremath{\text{C\,\textsc{iv}}}}
\def\mgii     {\ensuremath{\text{Mg\,\textsc{ii}}}}

\def\hi {\ensuremath{\text{H\,\textsc{i}}}}
\def\heii {\ensuremath{\text{He\,\textsc{ii}}}}
\def\xhi {\ensuremath{X_\text{H\,\textsc{i}}}}

\def\kms{{\rm\,km\,s^{-1}}}
\def\ergs{{\rm\,erg\,s^{-1}}}

\begin{document}

% Page header
\markboth{Fan et al.}{Quasars and IGM at Cosmic Dawn}

% Title
\title{Quasars and the Intergalactic Medium at Cosmic Dawn}

%Authors, affiliations address.
\author{Xiaohui Fan,$^1$  Eduardo Ba\~nados,$^2$ and Robert A. Simcoe$^3$
\affil{$^1$Steward Observatory, University of Arizona, 933 North Cherry Avenue, Tucson, AZ 85721, USA; email: xfan@arizona.edu}
\affil{$^2$Max-Planck-Institut f\"{u}r Astronomie, K\"{o}nigstuhl 17, D-69117 Heidelberg, Germany; email: banados@mpia.de}
\affil{$^3$MIT Kavli Institute for Astrophysics and Space Research, 77 Massachusetts Ave., Cambridge, MA 02139, USA; email: simcoe@space.mit.edu}}

\begin{abstract}
Quasars at cosmic dawn provide powerful probes of the formation and growth of the earliest supermassive black holes (SMBHs) in the universe, their connections to galaxy and structure formation, and the evolution of the intergalactic medium (IGM) at the epoch of  reionization (EoR). 
Hundreds of quasars have been discovered in the first billion years of cosmic history, with the quasar redshift frontier extended to $z\sim 7.6$. 
%, deep into the epoch of reionization. 
Observations of quasars at cosmic dawn show that:
 
\begin{itemize}
\vspace{0.2cm}
\begin{minipage}{0.8\linewidth}

\item The number density of luminous quasars declines exponentially at $z>5$, suggesting that the earliest quasars emerge at $z\sim 10$; the lack of strong evolution in their average spectral energy distribution indicates a rapid buildup of the AGN environment.
%their average spectral energy distribution shows insignificant evolution from lower redshift, indicating a rapid buildup of the AGN structure.
\item Billion-solar-mass BHs already exist at $z>7.5$; they must form and grow in less than 700\,Myr, by a combination of massive early BH  seeds with highly efficient and sustained accretion.  
\item The rapid quasar growth is accompanied by strong star formation and feedback activity in their host galaxies, which  
%, sites of early massive galaxy assembly and feedback. 
show diverse morphological and kinetic properties, with typical dynamical mass of lower than that implied by the local BH/galaxy scaling relations.  
\item HI absorption in quasar spectra probes the tail end of cosmic reionization at $z\sim 5.3 - 6$, and indicates the EoR midpoint at $6.9 < z < 7.6$, with large spatial fluctuations in IGM ionization.
%and we may be approaching the EoR midpoint  at $6.9 < z < 7.6$. Long dark gaps at $z<6$ and transmission spikes at $z>6$ both indicate there are large spatial fluctuations in IGM ionization.  
Observations of heavy element absorption lines suggest that the circumgalactic medium also experiences evolution in its ionization structure and metal enrichment during the EoR.
%, or alternatively that  it was not yet fully enriched with metals from star formation. 
\end{minipage}

\end{itemize}
 
\end{abstract}

%Keywords, etc.

\maketitle

%Table of Contents
\tableofcontents

\section{Introduction}
\label{section: intro} 

Quasars, and active galactic nuclei (AGN) in general, are powered by accretion onto the supermassive black holes (SMBHs) at the center of their host galaxies. Quasars are the most luminous non-transient sources in the Universe, and have been observed up to a redshift of $z\sim 7.6$ \citep{2021ApJ...907L...1W}. 
SMBH activities are a key ingredient of galaxy formation. Quasars provide crucial probes of galaxy evolution and cosmology across cosmic history at three critical spatial scales:
\begin{enumerate}
 \item At the scale of the AGN central engine ($<1$pc), quasar emission originates from well within the SMBH sphere of influence, at which the gravitational influence of the BH dominates. Quasars are fundamental in understanding the physics of BH accretion and growth and the physics of AGN activity. 
\item At the scale of quasar host galaxies ($\sim 1 - 10$ kpc), the evolution of quasars and galaxies are strongly coupled, as shown by the tight correlation between SMBH mass and the mass/velocity dispersion of their host galaxies seen at low redshift \citep[e.g, the M-$\sigma$ relation,][]{2013ARA&A..51..511K}. Quasars are key to understand the assembly, growth and quenching of massive galaxies. 
\item At the scale of galaxy clusters and superclusters ($>$Mpc),  quasars can be used to probe the growth of early large scale structure. They also provide sightlines to study  the properties of the intergalactic medium (IGM), including the history of cosmic reionization, the chemical enrichment of the IGM, the distribution of baryons and the evolution of ionization state of the IGM, and its connection to galaxy formation through the circumgalactic medium (CGM). 
\end{enumerate} 

The history of quasar studies is driven and shaped by innovations in systematic surveys and multiwavelength observations of quasars. These observations aim to reaching higher redshift, covering the full range of quasar luminosities, and spatially resolving quasar hosts and their central AGN structure. Studies of high-redshift quasars within the first few billion years of cosmic history started with the advent of large area digital or digitized sky surveys, highlighted by progress made using data from the Sloan Digital Sky Survey \citep[SDSS,][]{2000AJ....120.1579Y}, which resulted in the first detections of quasars at $z>5$ and $z>6$ \citep{2001AJ....122.2833F}. Observations of $z\sim 6$ quasars in the early 2000s provided fundamental insights in the three key scales of galaxy evolution mentioned above:
\begin{enumerate}
\item SMBHs with masses up to a few billion $M_{\odot}$ already existed in the Universe within one billion years after the Big Bang, requiring a combination of early massive BH seeds and rapid BH accretion. 
\item Early luminous quasars are sites of intensive galaxy-scale star formation and the assembly of early massive galaxies. 
\item Detection of strong IGM absorption in quasar spectra, especially the emergence of complete Gunn-Peterson absorption troughs \citep{1965ApJ...142.1633G}  shows a rapid transition of the ionization state of the IGM at $z\sim 5-6$, marking this epoch as the end of cosmic reionization. 
\end{enumerate} 

In this review, we will focus on the progress since then, including the quests for the earliest quasars, and the detailed studies of quasars during and right after the epoch of reionization (EoR). We will limit our discussions to:
%\begin{itemize}
(1) the redshift range of $z>5.3$; this is the lower redshift limit at which there are still detectable signatures of reionization activity. This redshift corresponds to 1.1 Gyr after the Big Bang, when the Universe was at 8\% of its current age. We consider $z>5.3$ as ``cosmic dawn'' in the context of this review.
(2) Luminous Type-1 quasars, for which secure spectroscopic observations and systematic surveys are currently possible. 
(3) IGM studies using quasar absorption spectra; we will not discuss other probes of EoR in detail. 
%\end{itemize} 
This review will focus mainly on observations, and we will discuss theories and simulations largely in the context of understanding and predicting observations.

At the time of this review, JWST has started routine observations, and initial JWST results have begun to appear in the literature. We will not include any early JWST results in this article. JWST observations of high-redshift quasars will undoubtedly provide many new insights that will result in discoveries and challenges that should be the subject of a future review. 

There have been a number of excellent Annual Review articles on related topics which we will not repeat: \cite{2020ARA&A..58...27I} reviewed topics related to the initial BH seeds and early growth;  \cite{2013ARA&A..51..511K} presented a very detailed discussion of BH/galaxy co-evolution; \cite{2013ARA&A..51..105C} reviewed  early (pre-ALMA) sub/mm observations of high-redshift galaxies;  including quasar hosts. \cite{2006ARA&A..44..415F} discussed early observational studies of reionization;   \cite{ 2016ARA&A..54..313M} provided a general discussion of the evolution of the IGM. 

This review is organized as follows:
In Section~\ref{sec:survey}, we will review the progress in searching for the highest redshift quasars and in establishing large sample of quasars at cosmic dawn,  as a result of the new generations of wide-field sky surveys and the developments in data mining and machine learning. In this Section, we will present a database of all published Type-1 quasars at $z>5.3$. 
In Section~\ref{sec:evolution}, we will discuss the evolution of quasars as a population at high redshift,  and present the measurements of quasar luminosity function, and the trend in the evolution of quasar intrinsic properties in early epochs. We will also highlight special populations of high-redshift quasars. 
In Section~\ref{sec:SMBH}, we will discuss the use of high-redshift quasars as probes to the history of SMBH growth in the early Universe, and review statistics of measurements of quasar BH masses and accretion characteristics. 
In Section~\ref{sec:host}, we will review the observations of quasar host galaxies from the rest-frame UV to far-IR, in the context of the co-evolution of early SMBH growth and galaxy formation, and the roles quasar played in early galaxy and structure formation.
In Section~\ref{sec:reionization}, we will review the the progress in using IGM absorption in quasar sight lines and properties of quasar proximity zones to probe the history of cosmic reionization and IGM chemical enrichment. 
%We will conclude with a discussion of the future of quasar research in probing the earliest BH formation, galaxy formation and reionization, in particular the facilities and capabilities that will be available in the next decade in Section \ref{sec:future}.
Throughout this review, we assume a spatially flat LCDM cosmological model with $\Omega_m = 0.3$ and $H_0 = 70$ km s$^{-1}$ Mpc$^{-1}$, consistent with the final Planck results. 
%When discussing photometric measurements, we will use AB magnitude for optical observations, and Vega-based magnitude for near-IR, following the convention in the literature. 
 \section{THE QUASAR REDSHIFT FRONTIER}
\label{sec:survey}

 Advances in the studies of high-redshift quasars 
 %-- the topic of this review --
 are first and foremost driven by advances in high-redshift quasar surveys and new discoveries.  The quasar redshift frontier continues to expand as a result of new sky surveys: the first quasar discoveries at $z>4$ in the 1980s were made possible by digital or digitized large sky surveys and the first implementations of color drop-out selection techniques \citep[e.g.,][]{1987Natur.325..131W}. %1989AJ.....98.1951S};
 After the SDSS discoveries of the first $z>5$ \citep{1999AJ....118....1F} and $z>6$ \citep{2001AJ....122.2833F} quasars in the early 2000s, 
% high-redshift quasar survey based on imaging and spectroscopic data from the Sloan Digital Sky Survey allowed discoveries of quasars at $z>5$ \citep{1999AJ....118....1F} and $z>6$ \citep{2001AJ....122.2833F} in early 2000s; 
wide-field near-infrared (NIR) sky surveys in the following decade, such as UKIDSS \citep{2007MNRAS.379.1599L} led to the detection of the first quasars at $z>7$ \citep[e.g.,][]{2011Natur.474..616M}, deep into the EoR. At the time of this review, the quasar redshift frontier stands at $z=7.64$ \citep{2021ApJ...907L...1W}. Meanwhile, increasingly large samples of high-redshift quasars are established by continued mining and systematic spectroscopic followup observations based on these new surveys. Currently, about 1000 quasars have been discovered at $z>5$, and more than 200 at $z>6$. 
%These samples are the basis of statistical studies of high-redshift quasar populations and their connections to galaxy formation and cosmic reionization discussed in later sections of this review. 
%In this section, we review the methodology of high-redshift quasar selection and present a database that include all published quasars at $z>5.3$ at the time of this review.

\subsection{Progress in High-Redshift Quasar Surveys}
\label{subsec:surveyprogress}

Surveys of the highest redshift quasars face three technical challenges. First, quasars at cosmic dawn are among the rarest objects in the Universe.  The final SDSS $z\sim 6$ quasar sample covers more than 11,000 deg$^2$, but contains only 52 quasars \citep{2016ApJ...833..222J}.  Their discoveries require large surveys that cover a significant fraction of the sky. The second challenge is often referred to as ``finding needles in a haystack''. Most high-redshift quasar surveys are based on Lyman break dropout selections using optical and NIR photometric survey data. However, other populations of celestial objects,  in particular cool galactic dwarfs with spectral types M, L and T (usually referred as MLTs) and compact early-type intermediate-redshift galaxies,  have similar optical and NIR colors. \cite{2019A&A...631A..85E} show that these contaminant populations outnumber $z\sim 7$ quasars by 2--4 orders of magnitude in deep photometric surveys. A number of photometric selection techniques have been developed, and the choices of how to apply these techniques require careful consideration of the balance between selection efficiency and completeness. Finally, spectroscopic identifications of candidate quasars require observations on large aperture telescopes. Until recently, such observations were only possible with single-object spectroscopy because the low spatial density of high-redshift quasars. The demand of telescope resources for discovery drives the need for both high selection efficiency and sometimes special observing strategies. For example, \cite{2017ApJ...839...27W} improved spectroscopic identification efficiency by using low-resolution ($R\sim 300$) long-slit NIR spectroscopy, which could capture the prominent Lyman break features in the quasar spectra and reject contaminants with shorter exposure than higher resolution spectra, which is more common for quasar spectroscopy followup work, would require. 

\begin{figure}[h]
\includegraphics[width=4in]{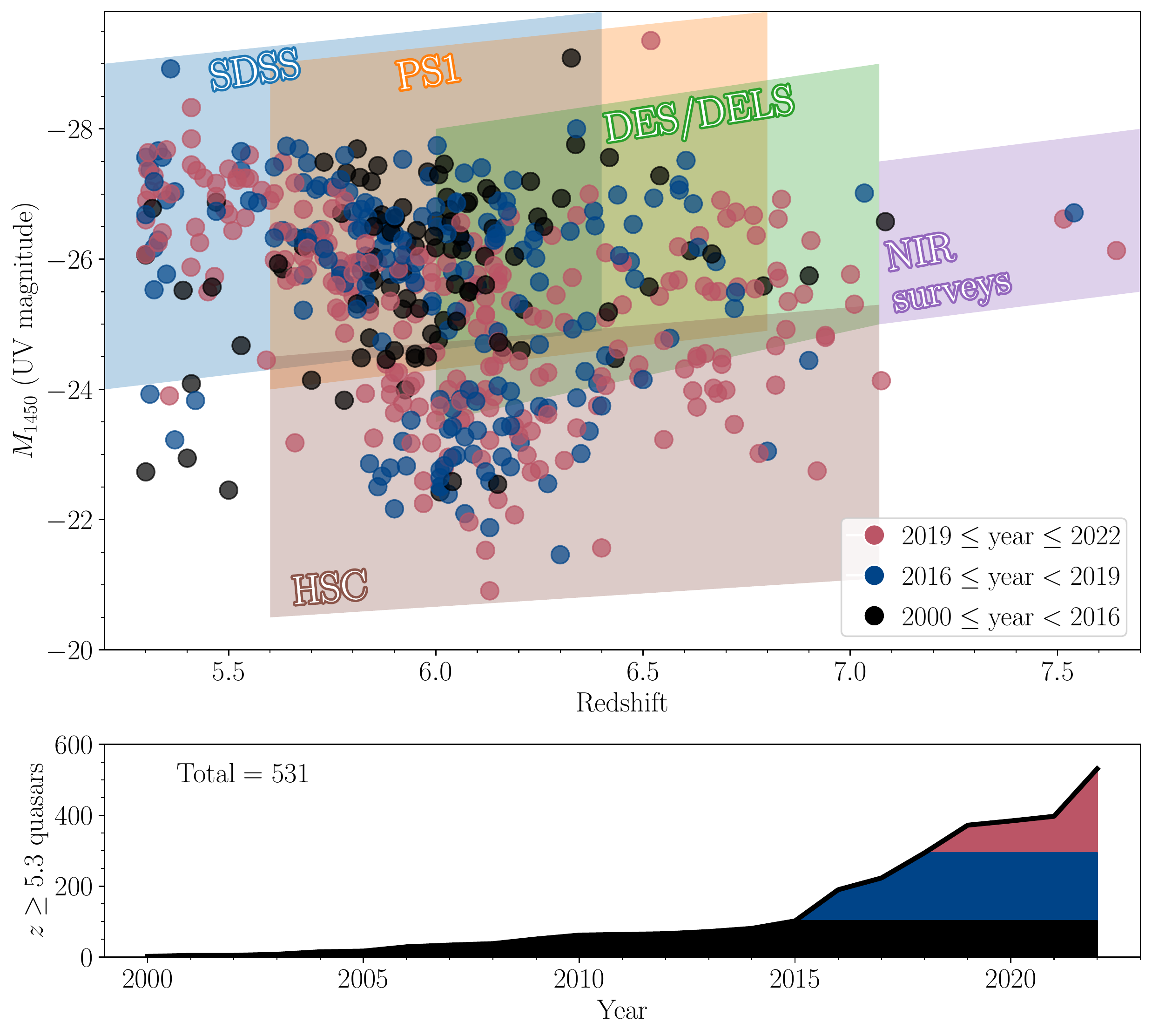}
%\vspace{-1cm}
\caption{Distribution of all known $z\geq5.3$ quasars in the absolute magnitude-redshift plane (top panel). The shaded areas are the parameter space probed by selected large quasar survey programs. The bottom panel shows the cumulative number of quasars known at $z\geq5.3$ as a function of year of publication. }
\label{fig:quasarsample}
\end{figure}

 Fig.~\ref{fig:quasarsample} presents the distribution of all published $z\geq5.3$ quasars, as of Dec 2022, on the absolute magnitude-redshift plane, highlighting the major survey programs from which most of these quasars are selected. The progress illustrated by the bottom panel is a result  of the availability of large scale optical and NIR sky surveys, improvements in selection and contaminant rejection, and efficient spectroscopic identification. 

%\subsection{Progress in High-Redshift Quasar Surveys}
%\label{subsec:surveyprogress} 

%The vast majority of high-redshift quasars are discovered by first using wide-field multi-color optical/IR photometric surveys to select quasar candidates, then carrying out spectroscopic observations for identifications and redshift measurements. 
%Regardless of the detailed selection methodology (Sec~\ref{subsec:selection}), q
Quasar candidates are separated from other point sources in photometric surveys because of their distinct spectral energy distributions (SEDs). The intrinsic spectra of quasars (Fig~\ref{fig:noevo}) in the rest-frame UV and optical are characterized by a blue power-law continuum and a number of strong broad emission lines. At $z>3$, the strong IGM neutral hydrogen (HI) absorption from Lyman series lines and Lyman continuum redshifts into the observed optical wavelength. Thus high-redshift quasars are ``dropout'' objects with a strong Lyman break with the dropout bands correspond to the observed wavelength of the Lyman break. %Quasar broad-band SEDs are a combination of strong Lyman break, or red color indices at shorter wavelengths, and blue continuum colors at longer wavelengths.
At longer wavelength, quasars have blue broad-band colors due to their power law continuum. 
The Lyman dropout selection method for quasars has been used since the discoveries of the first $z>4$ quasars \citep{1987Natur.325..131W}. A more recent development is the inclusion of mid-IR (MIR) photometric surveys in the candidate selection  \citep[e.g.][]{2015Natur.518..512W}. The long wavelength baseline from NIR to MIR, in particular using the WISE \citep{2010AJ....140.1868W} data,  allows more effective separation of high-redshift quasars and MLT dwarfs by colors. 

\begin{figure}[h]
\includegraphics[width=3in]{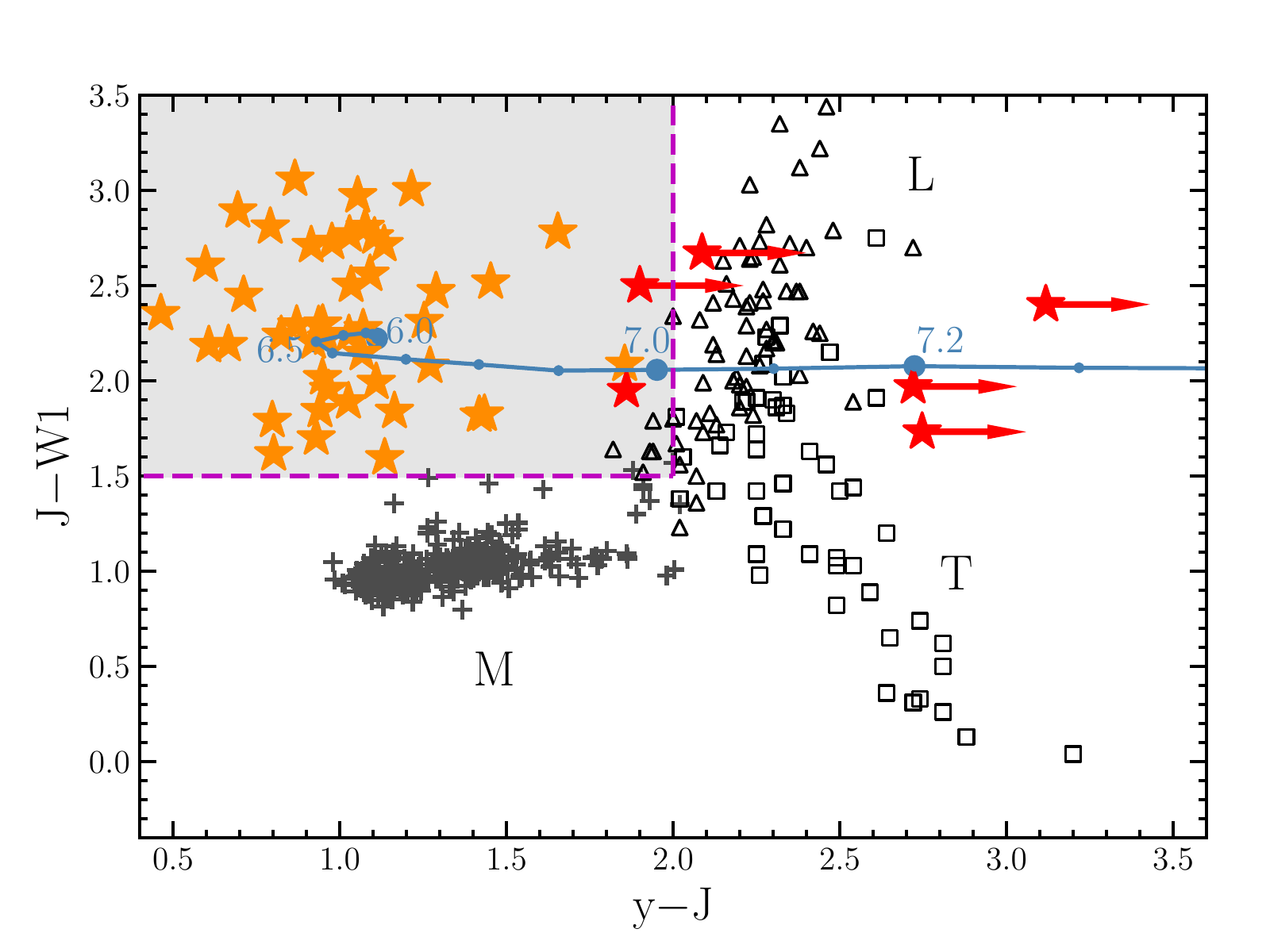}
\caption{Example of y$-$J vs.\ J$-$W1 color-color diagram used for high-redshift quasar selections. The cyan line represents the average color-redshift relation for quasars at $z=6-7.3$ based on simulated quasar spectra. For a typical quasar spectrum, see Fig~\ref{fig:noevo} Orange stars are known quasars at $6.5< z< 7.0$; magenta line shows the selection criteria used by \cite{2019ApJ...884...30W} for this redshift range. Red stars are known quasars at $z>7.0$; black symbols are known MLT contaminants.  Adapted from \cite{2019ApJ...884...30W}.}
\label{fig:colorcolor}
\end{figure}

Fig.~\ref{fig:colorcolor} illustrates how to use the combination of optical and IR colors to select $z>6.5$ quasars. At $z \sim 6.5$, the strong Ly$\alpha$ emission line is in the $y$ band. At higher redshift, this band starts to be dominated by IGM absorption and quasars become ``y-dropouts'' with increasing red $y-J$ colors. Meanwhile, the $J-W1$ color is determined by the quasar power law continuum and has a near constant value, which is redder than most of the sub/stellar objects with spectral types earlier than L. The main contaminants for this redshift range are L and T dwarfs.

\cite{2020ARA&A..58...27I} presented a summary table of the photometric surveys used in the discoveries of $z>6$ quasars in the past fifteen years. 
 Early surveys such as SDSS and PS1 \citep{2016arXiv161205560C} used 2--4 meter class telescopes with relatively short exposure times, thus they are optimized to discover rare, luminous quasars in large sky areas.  Selection of fainter quasars requires photometric observations on large aperture telescopes: the SHELLQs project \citep{2022ApJS..259...18M}  is based on the Hyper Suprime-Cam (HSC)  survey on the 8.2m Subaru Telescope \citep{2018PASJ...70S...8A}.  
As shown in Fig~\ref{fig:colorcolor}, at $z>7$, quasars begin to drop out in the $y$ band at the red limit of CCD detector sensitivities. The discovery of quasars at this redshift requires the combination of wide-field IR surveys that provide detection of the quasar rest-frame UV continuum, and deep optical surveys that sample the strong Lyman break. Following the first quasar discovery at $z>7$ using photometric data from the UKIDSS survey \citep{2011Natur.474..616M}, 
there have been eight quasars published at $z>7$ (see Table~\ref{tab:z7}). The current redshift frontier is represented by the three luminous quasars known at $z > 7.5$, selected using a combination of  NIR,  optical, and MIR (WISE) surveys: J1342+0928 at $z=7.54$ \citep{2018Natur.553..473B}; J1007+2115 at $z=7.52$ \citep[``Pōniuā'ena'',][]{2020ApJ...897L..14Y};  and J0313-1806 at $z=7.64$ \citep{2021ApJ...907L...1W}. 
%These objects provide powerful probes to the history of reionzation and SMBH growth at the earliest available epoch. 

\subsection{Optimization of Photometric Selection of High-Redshift Quasars}
\label{subsec:selection} 

The vast majority of high-redshift quasars have been discovered using color selection. 
%discovered using wide-field photometric surveys are color selected: based on their multi-band colors or spectral energy distribution, and  having point-like morphology. 
The  most commonly used method is to implement a series of ``color cuts'': select objects that meet a set of criteria in the flux/flux-error space (usually referred to as the ``color space'') as high-redshift quasar candidates. These cuts are often initially guided by a combination of photometry of existing high-redshift quasars and simulated colors based on synthetic quasar spectra, as well as observed and simulated data of the contaminant populations. 
These cuts are refined as surveys progress with larger training sets and better understanding of the contaminant populations. \cite{1999AJ....117.2528F} presented an early example of color space simulation of  different populations of objects with compact morphology (quasars, normal stars, white dwarfs and compact galaxies) in the SDSS photometric system. This simulation was used to formulate color selections for both the SDSS main spectroscopic survey \citep{2002AJ....123.2945R} and selections of $z\sim 6$ quasars \citep{2001AJ....122.2833F}. Works such as \cite{2006MNRAS.367..454H}, \cite{2013ApJ...768..105M}, \cite{2019A&A...631A..85E},  and \cite{2021MNRAS.508..737T} expanded these simulations by including more realistic quasar population models and models of MLT and intermediate-redshift galaxies, extending to higher redshifts, and including NIR and MIR photometric bands. 

High-redshift  surveys based on SDSS \citep[e.g.,][]{2016ApJ...833..222J}, PS1 \citep[e.g., ][]{2016ApJS..227...11B} as well as  works using the DESI Legacy Survey \citep[e.g.,][]{2019ApJ...884...30W} used color cuts as their primary selection method. 
%In those surveys, all objects that satisfy the selection criteria are classified as quasar candidates, even though some priority based on the exact location of the candidates in the color space is also used when conducting spectroscopic identifications. 
The color cut method is simple to implement, and 
%A key advantage of color cuts is that it is simple to apply. In addition, the candidate selection
has high selection completeness even for objects that have unusual spectral features such as broad absorption lines (BALs), weak emission lines or those with modest reddening, because the color cuts are usually fairly loose and cover a large portion of color space, and the strong Lyman break is not strongly affected by the intrinsic quasar SEDs. On the other hand, because all candidates that satisfy the cuts are selected without considering the relative density distribution of the targeted (quasar) population and the contaminant populations (primary MLTs and compact galaxies), or how the candidate SEDs match the quasar template, color cut selection tends to have a higher contamination rate. For example, \cite{2019ApJ...884...30W} find a 30\% spectroscopic success rate when searching for $z>6.5$ quasars. The contamination grows significantly worse for $z\sim 7.5$ (F. Wang, private communication). 

A  number of techniques have been applied to improve the efficiency of color selection. These improvements are especially important for quasars at the highest redshift ($z> 6.5$), where quasars are increasingly rare compared to the contaminant populations,  and at fainter fluxes for which increasing photometric errors results in more contaminants scatter into the selected area in color space. 
\cite{2012MNRAS.419..390M} first introduced a selection algorithm based on Bayesian model comparison (BMC). In BMC, the  posterior quasar probability  for a given object with survey photometry $\bf{d}$ is given by :
\begin{equation}
P_q = p (q | \mathbf{d} ) = \frac{W_q(\mathbf{d})}{W_q(\mathbf{d}) + W_b(\mathbf{d}) + W_g(\mathbf{d})},
\end{equation}
where q, b, and g represent quasar, brown dwarf (MLTs), and galaxy populations, respectively. The weights W for each population are calculated using a population model that describes its surface density distribution integrated over a Gaussian likelihood function based on model colors. BMC wss applied by  \cite{2011Natur.474..616M} to data from the UKIDSS survey to identify the first quasar at $z>7$. The SHELLQs survey \citep{2022ApJS..259...18M} used BMC to select faint quasars in the HSC survey with great success in achieving a high spectroscopic identification rate ($\sim 75\%$) even at the low-luminosity end at $z\sim 6$. \cite{2022A&A...660A..22W} presented a similar probabilistic approach that also includes radio data in the selection. 

\cite{2017MNRAS.468.4702R} applied a SED fitting method for high-redshift quasar selection. In this method, after the initial color cuts,  reduced $\chi^2$ values are calculated for each candidate, matching the observed colors of the candidate observed SEDs to a series of model SEDs, including different MLT spectral types as well as early type galaxies at various redshifts. Objects with high reduced $\chi^2$ for the contaminant populations, and low values for the quasar SED models are selected as quasar candidates. 
%Although the color-space density information of the contaminant populations is not directly used, the selection does prioritize objects with SEDs most similar to those of quasars compared to those of the contaminants. 

\cite{2021MNRAS.501.1663B} presented a detailed comparison study of different selection techniques in searching for $z>6.5$ quasars using VIKING survey data. They show that BMC is highly complete in recovering previously known quasars in the survey area, and at the same time effectively rejects confirmed contaminants. Using simulations, they calculate the survey selection function -- completeness as a function of quasar redshift and luminosity, and find that both BMC and the SED method represent a significant improvement in selection completeness and depth compared to simple color cuts. 

\cite{2021AJ....162...72W} used a random forest supervised machine learning method to select $z>5$ quasars from PS1 data with high efficiency. \cite{2022MNRAS.515.3224N} proposed an alternative probabilistic approach, in which the density of the contaminant populations in color space is modeled as Gaussian mixtures based on observed photometric data and calculated using the extreme deconvolution technique, instead of using simulated population models. 

High-redshift quasar selection can also be considered in the context of supervised machine learning. The training sets of existing objects are small,  and in most case are subject to the bias of previously selected samples or extrapolations from low-redshift populations. The choice of high-redshift quasar selection algorithm in a given survey is  a tradeoff between completeness and efficiency. Selections that rely on SED and contaminant population models run the risk of missing objects that are highly valuable but have unusual properties. An example is the discovery of J0100+2802, the most luminous unlensed quasar currently known at $z>5$ \citep{2015Natur.518..512W}. The object was assigned as low priority in the original SDSS survey due to its relatively red color and bright apparently magnitude, indicating high probability of it being a brown dwarf.  It is not surprising that record-breaking discovery is made often -- but not always \citep[e.g.,][]{2011Natur.474..616M} -- using the less restrictive color-cut method, and probabilistic methods are more effective once a large training set has been established. 

 If high-redshift quasar candidates could be part of the overall target selection in a large automated spectroscopic survey program, then the color selection could be relaxed to be more complete and less sensitive to templates or model assumptions, because high-redshift quasar candidates are rare and are thus only a small fraction of the total targets. The SDSS main quasar survey \citep{2002AJ....123.2945R} targeted quasars up to $z\sim 5.4$. The Dark Energy Spectroscopic Instrument \citep[DESI,][]{2022arXiv220510939A} is using spectrographs with 5000 fibers on the 4-m Mayall Telescope at Kitt Peak to conduct a 5-year spectroscopic survey of $\sim 40$ million galaxies, quasars and stars. 
 %Yang et al. (2022, submitted) described the first results from 
 %a high-redshift ($z \gtrsim 5$) quasar survey using DESI, and report the discovery of more than 200 new quasars at $5 < z < 6.5$ using the first year DESI data, with a success rate of $22\%$. 
 DESI is expected to more than double the number of quasars known at $z>5$ when completed. A similar survey is being planned for the 4MOST project \citep{2019most.confE..23M}.

%In addition to selection methods based on broad-band SEDs in the optical/near-IR wavelengths described above, quasars are also selected using a wide array of methods (and often in combination), 
Other established quasar selection methods include variability, astrometry \citep[i.e., lack of proper motion;][]{2009AJ....137.4400L}, detections in X-ray and radio wavelengths, MIR colors, as well as wide-field slitless spectroscopy. Except for the last method \citep[e.g.,][]{1989AJ.....98.1951S, 1999AJ....117...40S}, most selection methods are not sensitive to the quasar redshifts. Therefore, some color cuts are usually used to identify high-redshift quasar candidates. 
\cite{2006ApJ...652..157M} and \cite{2011ApJ...736...57Z} report the first discoveries of radio-loud $z>6$ quasars by matching radio surveys such as FIRST with optical photometric surveys. X-ray observations provide fundamental probes of AGN evolution across cosmic time.  New wide-field X-ray surveys such as e-ROSITA \citep{2021A&A...647A...5W} will enable selections of luminous X-ray quasars at the highest redshifts. We will discuss the X-ray and radio properties of high-redshift quasars in Sec \ref{subsec:SED}. 

\begin{figure}[ht]
\includegraphics[width=1.25\textwidth]{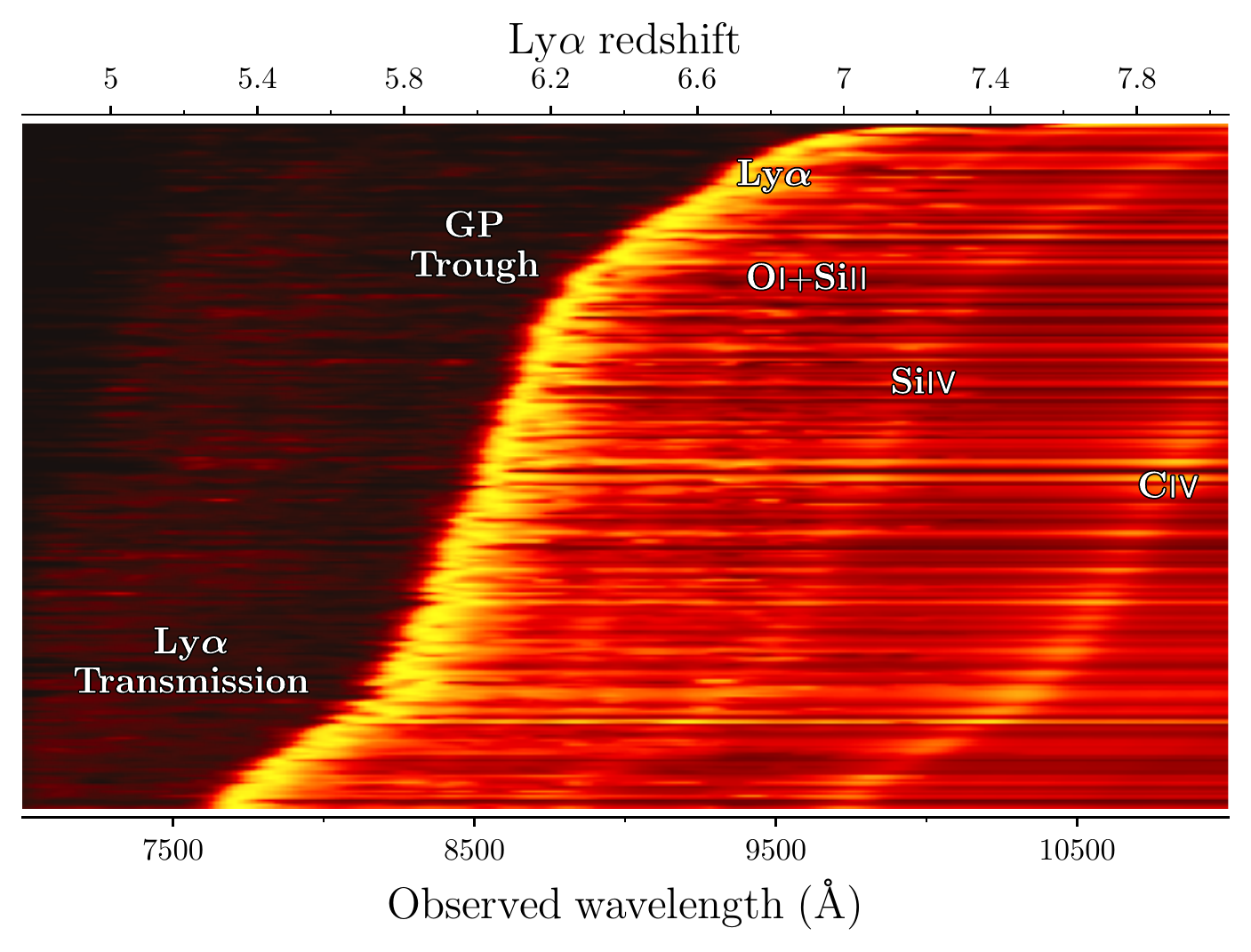}
\caption{Two-dimensional image representation of \nqsowithspectra{}/\nqso{} of all published spectra of $z\geq5.3$ quasars. Traces of major emission lines are labeled. On the blue side of Ly$\alpha$ line, there are clear signatures of the Ly$\alpha$ transmission at $z<6$, while at higher redshift, the spectra are dominated by long Gunn-Peterson absorption troughs.}
\label{fig:allspectra}
\end{figure}

\subsection{A Database of $z>5.3$ Quasars}
\label{subsec:database}

At the time of this review, there are \nqso{} $z\geq5.3$ quasars in the published literature. We present their spectra in Fig~\ref{fig:allspectra}.
The locations of the $z\geq5.3$ quasars on the redshift-luminosity plane is shown in Fig~\ref{fig:quasarsample}. 
We include a database with their basic properties in the {\bf Supplementary Material} associated with this review (follow the {\bf Supplemental Material} link from the Annual Reviews home page at http://www.annualreviews.org). 
%and include a database with their basic properties in the {\em supplementary material} associated with this review. For reference, selected entries for the five most distant quasars known to date are shown in Table~\ref{tab:database}.  \textbf{[discuss whether this is necessary.}]
%The locations of the $z>5.3$ quasars on the redshift-luminosity plane is shown in Fig~\ref{fig:quasarsample}. 
To be included in this database, we require the object to :
(1)  have a spectroscopically confirmed redshift $z\geq5.3$; we do not include objects with only photometric redshifts; 
(2) have at least one broad emission line (FWHM $> 2000$ km s$^{-1}$) in the rest-frame UV; we do not include narrow-line (Type-2) quasars or obscured AGN at high redshift. 
In addition to the quasar name, coordinates, redshift and references, the database also include measurements of the quasar's continuum luminosity at the rest-frame 1450\,\AA\ and 3000\,\AA,  properties of the \mgii\ emission line, and BH mass derived from \mgii\ measurements quoted in the literature, estimated using Equation \ref{eq:V09bhmass}. We only consider \mgii\ measurements for quasars at $z>5.9$, as \mgii\ lines are usually in regions highly affected by telluric absorption in the NIR at lower redshift.   

%(3) have UV continuum luminosity $M_{1450} < -22$. 

% \begin{table}
% \caption{Database of all $z>5.3$ quasars in the published literature. Only selected entries for the five most distant quasars currently known are shown here, the full database can be found in the \textit{supplementary material}. \label{tab:database}}
% \begin{tabular}{cccccc}
% \hline \hline
% Quasar Name & Redshift & $M_{1450}$ & $M_{\rm BH}$ & Disc.\ ref. & $M_{\rm BH}$\ ref. \\
%  &  & $\mathrm{mag}$ & $10^8\,M_\odot$ &  &  \\
% \hline
% J031343.84--180636.40 & 7.6423 & $-26.13$ & 16.1 & \cite{2021ApJ...907L...1W} & \cite{2021ApJ...923..262Y} \\
% J134208.11+092838.61 & 7.54 & $-26.71$ & 8.1 & \cite{2018Natur.553..473B} & \cite{2021ApJ...923..262Y} \\
% J100758.27+211529.21 & 7.5149 & $-26.62$ & 14.3 & \cite{2020ApJ...897L..14Y} & \cite{2021ApJ...923..262Y} \\
% J112001.48+064124.30 & 7.0848 & $-26.58$ & 13.5 & \cite{2011Natur.474..616M} & \cite{2021ApJ...923..262Y} \\
% J124353.93+010038.50 & 7.0749 & $-24.13$ & 3.6 & \cite{2019ApJ...872L...2M} & \cite{2019ApJ...872L...2M} \\
% \hline
% \end{tabular}
% \end{table}

\begin{table}
\caption{Quasars at $z>7$ \label{tab:z7}}
\begin{tabular}{cS[table-format=3.2]cccc}
\hline \hline
Quasar & $z$ & $M_{1450}$ & $M_{\rm BH}$ & Disc.\ ref. & $M_{\rm BH}$\ ref. \\
 &  & $\mathrm{mag}$ & $10^8\,M_\odot$ &  &  \\
\hline
J031343.84--180636.40 & 7.6423 & $-26.13$ & $16.1$ & \cite{2021ApJ...907L...1W} & \cite{2021ApJ...923..262Y} \\
J134208.11+092838.61 & 7.54 & $-26.71$ & $8.1$ & \cite{2018Natur.553..473B} & \cite{2021ApJ...923..262Y} \\
J100758.27+211529.21 & 7.5149 & $-26.62$ & $14.3$ & \cite{2020ApJ...897L..14Y} & \cite{2021ApJ...923..262Y} \\
J112001.48+064124.30 & 7.0848 & $-26.58$ & $13.5$ & \cite{2011Natur.474..616M} & \cite{2021ApJ...923..262Y} \\
J124353.93+010038.50 & 7.0749 & $-24.13$ & $3.6$ & \cite{2019ApJ...872L...2M} & \cite{2019ApJ...872L...2M} \\
J003836.10--152723.60 & 7.034 & $-27.01$ & $13.6$ & \cite{2018ApJ...869L...9W} & \cite{2021ApJ...923..262Y} \\
J235646.33+001747.30 & 7.01 & $-25.31$ & -- & \cite{2019ApJ...883..183M} & -- \\
J025216.64--050331.80 & 7.0006 & $-25.77$ & $12.8$ & \cite{2019AJ....157..236Y} & \cite{2021ApJ...923..262Y} \\
\hline
\end{tabular}
\end{table}

The high-redshift quasar database includes \nqsosix{} objects at $z>6$ and \nqsoseven{} at $z>7$.  
At the time of writing, there are \nqsomgii\ $z>5.9$ quasars with robust \mgii-based black hole mass estimates.
We list the properties of quasars currently known at $z>7$ in Table~\ref{tab:z7}.
This database expands the compilation in the Supplementary Material of \cite{2020ARA&A..58...27I}. The quasar luminosities lie between $M_{1450} = -20.9$ and $-29.4$; the median luminosity of these objects is $M_{1450}=-25.8$. The brightest object is J0439+1634, a gravitationally lensed quasar at $z=6.5$ \citep{2019ApJ...870L..11F}. 
Fig~\ref{fig:allspectra} shows a two-dimensional image representation of all spectra of the quasars in the database. In this image, each row is the one-dimensional spectrum of a quasar, ordered in ascending redshift.  The flux level of each column is normalized by its peak Ly$\alpha$ flux. The image shows how the quasar Ly$\alpha$ emission line move to near-IR wavelengths as the redshift increases from $z\sim 5$ to $>7$. On the blue side of the Ly$\alpha$ emission, the spectra show the extent of the highly ionized quasar proximity zone (Sec~\ref{subsec:proximity}), where the flux does not immediately drop to zero. Further blueward, the spectrum is dominated by strong Gunn-Peterson absorption. Complete Gunn-Peterson absorption troughs can be seen at $z>6$. At lower redshift, the presence of transmission spikes indicates that the IGM is, on average, highly ionized (Sec~\ref{sec:reionization}). On the red side of Ly$\alpha$ emission, broad emission lines such as OI+SiII$\lambda1306$, SiIV+OIV]$\lambda1402$ and CIV$\lambda1549$ are visible. 
 
\subsection{Future Quasar Surveys}
\label{ref:futuresurvey} 
  
 New surveys are on the horizon to further expand the quasar redshift frontier. The Legacy Survey of Space and Time (LSST) by the Vera C. Rubin Observatory will cover the southern sky at optical wavelengths to unprecedented depths. LSST will reach 5$\sigma$ co-added depths in the $z$ and $y$ bands at $25-26$ mag level in 5-10 years  \citep{2019ApJ...873..111I}, allowing selections of several thousand quasars and AGN at $z\sim 6-7.5$ using both color and variablity selection methods, while providing deep photometry in the dropout bands for selection of $z>7$ quasars. ESA's Euclid mission will provide deep near-IR photometry not possible with ground-based observations. It will cover 15,000 deg$^2$ of the sky to a 5$\sigma$ depth of 24 mag in Y, J and H bands  \citep{2022A&A...662A.112E}, enabling selection of quasar candidates up to $z\sim 10$.  
% short summary of the main conclusions and challenges in each sections.
\cite{2019A&A...631A..85E} predicted that Euclid + LSST will allow discoveries of $\sim 25$ quasars at $z>7.5$, including $\sim 8$ beyond $z>8$, although the exact yields strongly depend on the assumed evolution of quasar luminosity function. Further in the future is NASA's Nancy Grace Roman Space Telescope, which will cover a smaller area in its High Latitude Survey but reaching about two magnitude deeper. 

\cite{2019BAAS...51c.121F} show that by extrapolating the current measurement of the quasar luminosity function, there will be only one ``SDSS''-like quasar ($M_{1450} < -26$, powered by billion-M$_{\odot}$ SMBH) over the entire observable universe at $z>9$. The combination of LSST, Euclid and Roman will allow the discovery of the earliest luminous quasars in the Universe. However, in addition to the challenges of their selection \citep{2019A&A...631A..85E, 2022MNRAS.515.3224N},  their spectroscopic identification will require IR spectrographs more powerful than those with the current ground-based telescopes: 30m-class extremely large telescopes and JWST will be the primary tools for their confirmation and follow-up observations. 
\section{AN EVOLVING QUASAR POPULATION}
\label{sec:evolution}

In this section, we first review the evolution of the quasar luminosity function (QLF) at high redshift, which directly constrains the growth history of early SMBHs (Sec~\ref{subsec:QLF}). The number density of quasars is found to decline rapidly towards high redshift, %-- they are the rarest objects in the early universe.  
in sharp constrast to the  lack of strong evolution in the SEDs of quasars (Sec~\ref{subsec:SED}), from X-ray to radio, and in particular in the rest-frame UV to NIR, at which the quasar SED peaks, although a number of sub-types of quasars appear to be more common at $z>5-6$. 

\subsection {Evolution of Quasar Luminosity Function}
\label{subsec:QLF}

%We will focus our review on QLF based on rest-frame UV observations, and briefly discuss the bolometric QLF at the end of this subsection. 
Soon after the initial discovery of quasars, \cite{1968ApJ...151..393S} found that their density rises sharply with redshift up to $z\gtrsim 2$. Wide-field surveys such as 2dF \citep{2000MNRAS.317.1014B} and SDSS \citep{2006AJ....131.2766R} established that the density of luminous quasars peaks at $z\sim 2-3$. \cite{1982ApJ...253...28O} presented the first evidence that quasar density at $z>3.5$ appears to be declining.  \cite{1995AJ....110...68S} characterized this  decline as an exponential function with redshift:
\begin{equation}
\rho(z) \propto 10^{kz}, 
\label{eq:DenEvo}
\end{equation}
where $k$ is measured to be $\sim -0.5$ at $z\sim 3-5$. 
The shape of the QLF is usually described as a double, or broken, power law:
\begin{equation}
 \Phi(M_\mathrm{1450},z) = \frac{\Phi^{\ast}(z)}{10^{0.4(\alpha+1)(M_\mathrm{1450}-M^{\ast}_\mathrm{1450})}+10^{0.4(\beta+1)(M_\mathrm{1450}-M^{\ast}_\mathrm{1450})}}
\label{eq:DPLQLF}
\end{equation}
%and the spatial density decline rate $k$ is embedded within $\Phi^{\ast}(z)$,
%\begin{equation}
%\Phi^{\ast}(z) = \Phi^{\ast}(z=6) \times 10^{k(z-6)}
%\label{eq:Phi_star_z}
%\end{equation}
where $M_\mathrm{1450}$ is the absolute magnitude of the quasar continuum at rest-frame 1450 \AA, $\alpha$ and $\beta$ are the faint-end and bright-end slopes, respectively, $M^{\ast}_\mathrm{1450}$ is the characteristic absolute magnitude or break magnitude measured at 1450 \AA, and $\Phi^{\ast}$ is the normalization of the LF which has an exponential decline described in Eq.~\ref{eq:DenEvo}.

Early high-redshift surveys such as the SDSS were only sensitive to the most luminous quasars. 
Based on a sample of nine SDSS quasars at $z>5.7$, \cite{2004AJ....128..515F} found that at  $z\sim 6$, the co-moving spatial density of quasars with $M_{1450} < -26.7$ is $ 6 \times 10^{-10} \rm Mpc^{-3}$, consistent with extrapolation from %from \cite{1995AJ....110...68S}  and \cite{2001AJ....121...54F} at 
lower redshift trends. %Because of the limited luminosity range covered by this sample, even the bright-end slope is still poorly constrained with these earlier data. 
\cite{2009AJ....138..305J} extended the SDSS quasar survey to the ``stripe 82'' region, reaching two magnitudes deeper than the SDSS main survey. They found a bright end slope $\beta$ between --2.6 and --3.1. \cite{2010AJ....139..906W} combined the faint quasars discovered in the Canada-France High-z Quasar Survey (CFHQS) with the brighter SDSS quasars, and presented the first measurement of QLF at $z\sim 6$ using a sample of 40 quasars. They found a bright end slope $\beta = -2.81$ and a break magnitude $M^*_{1450} = -25.1$; the faint end slope was still poorly constrained at this redshift. 
\cite{2016ApJ...833..222J} presented the final results of the $z\sim 6$ quasar survey in the SDSS footprint. Using this sample of 52 quasar at $5.7 < z < 6.4$, they found a bright end slope of $\beta = -2.8 \pm 0.2$. In addition, they found a density evolution with the exponential evolution parameter $k=-0.72\pm 0.11$, a significantly steeper slope than the value at $z\sim 3-5$, suggesting an accelerated evolution of luminous quasars between $z\sim 5$ and 6.  

The SHELLQs project \citep{2016ApJ...828...26M} extended the $z\sim 6$ quasar sample to significantly lower luminosity. \cite{2018ApJ...869..150M} presented measurements of the QLF down to $M_{1450} \sim -22$, which is well fit with a double power law that has a faint-end slope $\alpha = -1.23^{+0.44}_{-0.34}$, a bright-end slope $\beta = -2.73^{+0.23}_{-0.31}$, and a break magnitude $M^*_{1450} = -24.90^{+0.75}_{-0.90}$. Their measured QLF showed  a strong break and a significant flattening at the faint end at $z\sim 6$. 
\cite{2022arXiv221204179S} used the combination of the bright PS1 quasar sample that includes 125 quasars at $z\sim 5.7 - 6.2$ with the fainter quasars from SHELLQs for a new QLF measurement at $z\sim 6$. They found a steeper  bright-end slope of $\beta = -3.84^{+0.63}_{-1.21}$, as well as a steeper faint-end slope of $\alpha = -1.70^{+0.29}_{-0.19}$, with a bright break magnitude of $M^*_{1450} = -26.38^{+0.79}_{-0.60}$. Their study yields a constant redshift evolution of $k \sim -0.7$ over a wide redshift range of $z=4-7$.   
   
%Their results show a strong decline in quasar density and steeping of bright end slope at high-redshift compared to earlier SDSS works. 

Measurement of the QLF becomes increasingly difficult at higher redshift due to the rapid decline in spatial density of quasars. \cite{2019ApJ...884...30W} conducted the first measurement of the QLF at redshift approaching seven based on a sample of 17 quasars at  $6.45 < z < 7.05$ \citep[see also][]{2013ApJ...779...24V}. They found a quasar spatial comoving density of $ \rho(M_{1450} < -26) = 0.39 \pm 0.11 \rm Gpc^{-3}$ at $z\sim 6.7$, and an exponential density evolution parameter $k= -0.78 \pm 0.18$. The density of luminous quasars declines by a factor of $\sim 6$ per unit redshift. The \textit{e}-folding time  of quasar density evolution and that of black hole accretion (Sec~\ref{sec:SMBH}) become comparable, underlying the strong constraints quasar evolution could place on the SMBH accretion mode. 

 \begin{figure}[ht]
% \centering
 \begin{minipage}{0.5\linewidth}
  %\centering 
%\includegraphics[width=3.5in]{SchindlerQLF.001.png}
  % \vspace{-10cm}
  % \hspace{-2cm}
% \includegraphics[width=2.8in, angle=-90]{fig/fig_QLF1.pdf}
\includegraphics[width=2.5in]{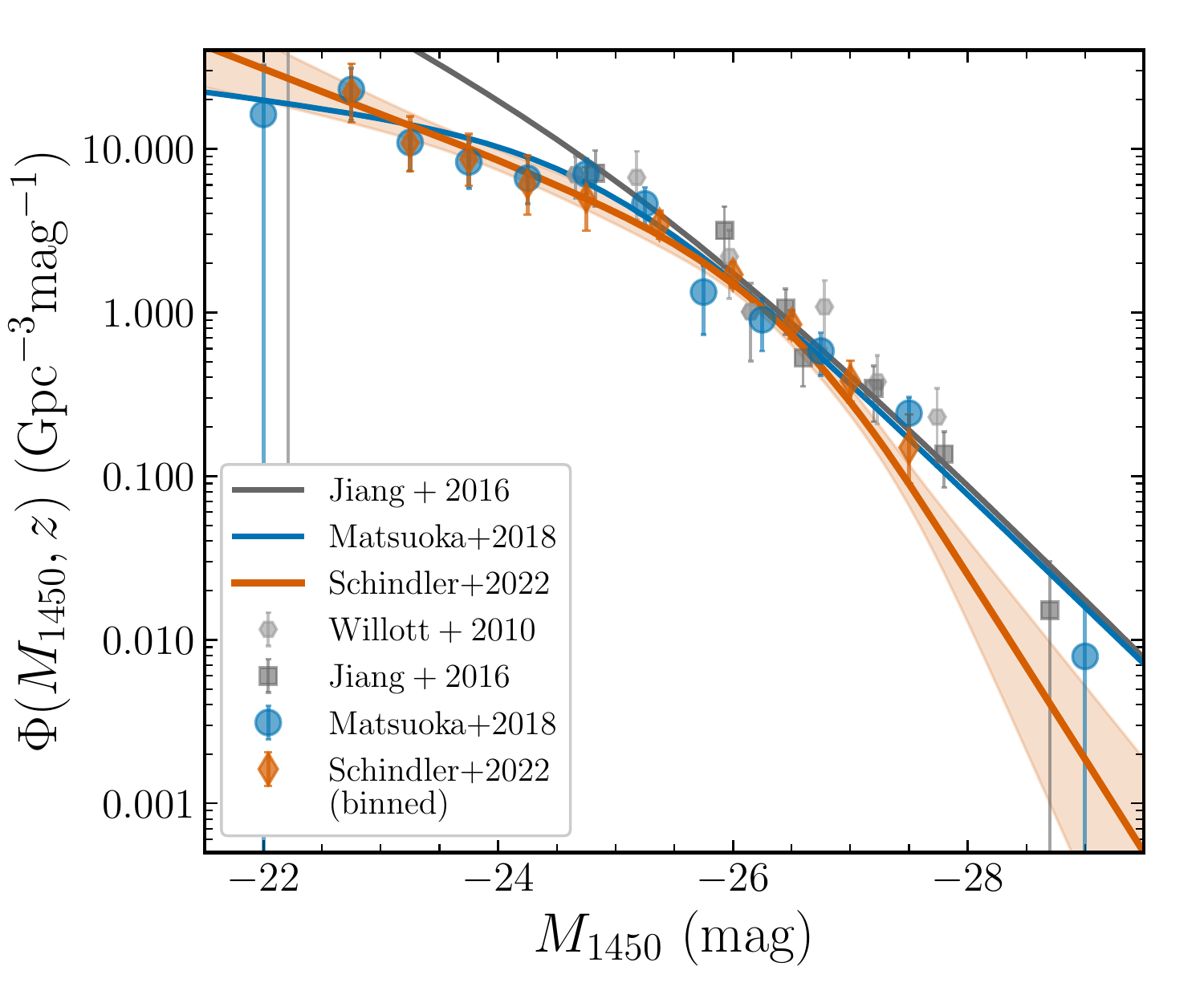}
 \end{minipage} 
\begin{minipage}{0.5\linewidth}
  %\centering 
  \hspace{-4cm}
  \includegraphics[width=3.3in]{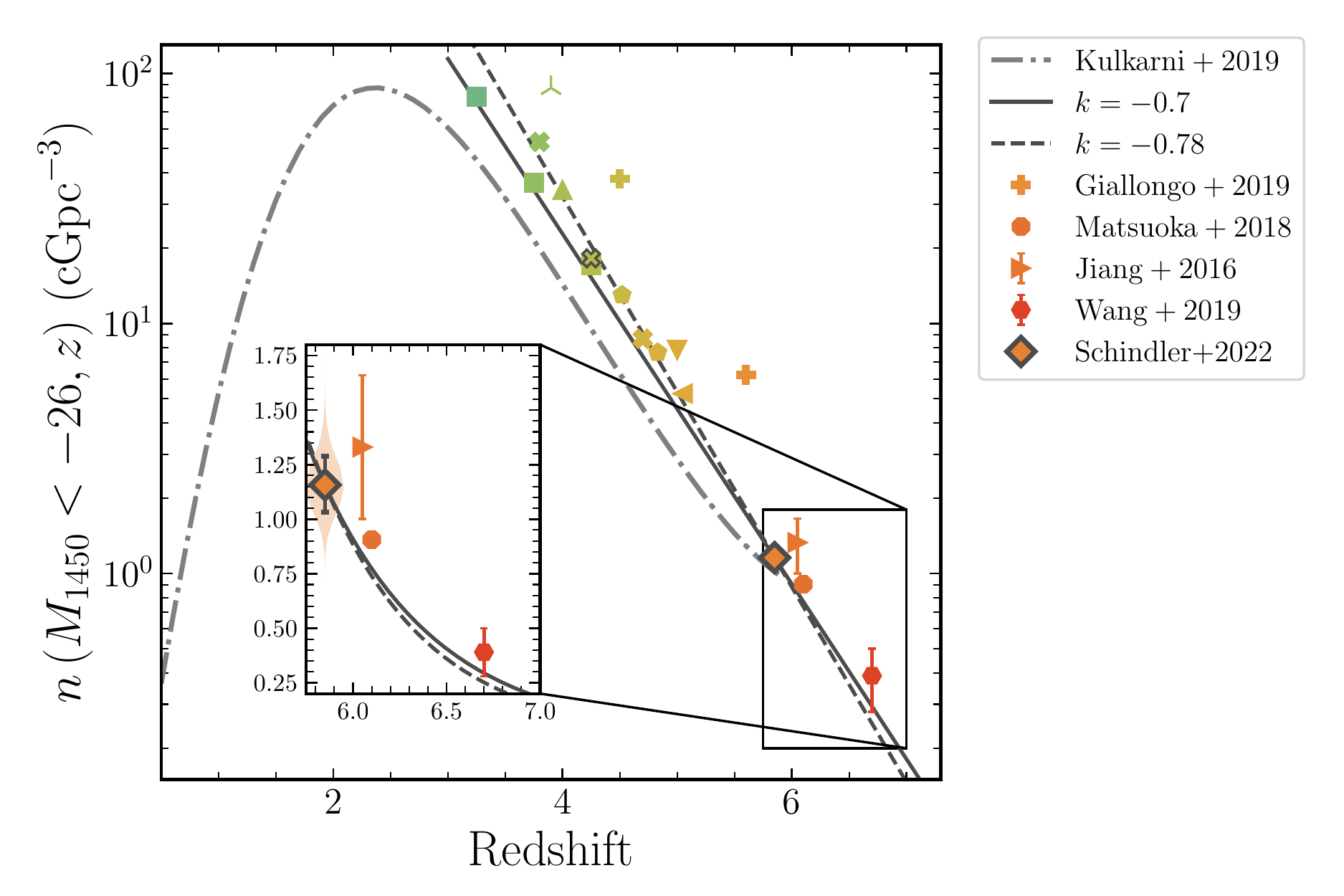}
 \end{minipage} 
%\end{figure} 
\vspace{1cm}
%\begin{figure}
%\includegraphics[width=3in]{fig6_Schindler_quasarLFz6.pdf} 
%\includegraphics[width=1in]{banados_figures/qlf_density_evolution.pdf}
%\includegraphics[width=6in]{QLFcombined.pdf} 
\caption{Quasar luminosity function measurements at high-redshift. (left): the QLF at $z\sim 6$ from SDSS, CFHTQS, SHELLQs, and PS1 surveys and best-fit double power law results. (right) The density of luminous ($M_{1450} < -26$) quasars as a function of redshift (for measurements at $z<5.3$, see references in \citealt{2022arXiv221204179S}). Figures adapted from \cite{2022arXiv221204179S}. 
}
\label{fig:QLF}
\end{figure}
%Schindler 2022 2022arXiv221204179S
%Banados 2022  2022arXiv221204452B

Fig~\ref{fig:QLF} (right panel) illustrates the overall density evolution of luminous ($M_{1450} < -26$) quasars with measurements from various surveys at $z=3 - 7$. 
While a strong exponential decline in the comoving spatial density of luminous quasars has been well established, there are still significant uncertainties in the evolution of the shape of the QLF (Fig~\ref{fig:QLF}, left panel). As discussed in \cite{2022arXiv221204179S}, the bright-end slope determination is strongly influenced by small number statistics. At the faint end, selection incompleteness could be a significant factor when a completeness correction of $\gtrsim 2$ is often needed.  
%The incompleteness correction (the selection function) is usually calculated using Monte Carlo simulations of quasar colors, and feeding them to the quasar selection criteria used in the survey, and it becomes significant close to the selection limit at which most quasars reside. In addition, the slopes and break magnitude have strong covariance.

There have been a number of ambitious efforts to combine QLF measurements across the electromagnetic spectrum and different cosmic epochs to obtain a complete picture of quasar/AGN evolution from reionization to the current epoch. \cite{2007ApJ...654..731H} presented the bolometric QLF up to $z\sim 6$. They explicitly modeled the X-ray column density distribution and SED shape variation of quasars to allow determinations of bolometric luminosities of high-redshift populations.  \cite{2019MNRAS.488.1035K} constructed a sample of more than 80,000 color-selected quasars and AGN  with a homogeneous treatment of survey selection effects, and derived the AGN UV luminosity function from $z=0$ to 7.5. Their measurement suggested a continued steeping of the faint end slope $\beta$ and a brightening of the break magnitude. \cite{2020MNRAS.495.3252S} updated the early Hopkins et al. work. They found bolometric QLF evolution consistent with those based on UV-selected samples alone, and concluded that the faint-end slope measurement is still uncertain at high-redshift. 

For the purpose of comparison with theoretical models or making predictions of future surveys, we suggest using the combined QLF measurements presented in \cite{2019MNRAS.488.1035K} or \cite{2020MNRAS.495.3252S} for $z\lesssim 6$, where the QLF is well measured over a wide range of luminosity. At $z\gtrsim 6$, the uncertainty is still large for both the QLF shape and its redshift evolution; \cite{2022arXiv221204179S} presented the most up-to-date measurements. 

A key limitation of current high-redshift quasar surveys is that the quasar selection methods assume a blue power-law continuum in the rest-frame UV. 
It is  possible that current surveys are missing significant number of red or reddened quasars. 
\cite{2020PASJ...72...84K} reported the discovery of two dust-reddened quasars at $z>5.6$ in the SHELLQs survey based on their mid-IR WISE detection. 
\cite{2022arXiv220600018E} discovered an obscured radio-loud AGN at $z=6.85$ in the 1.5 deg$^2$ COSMOS field with bolometric luminosity comparable to that of luminous SDSS $z\sim 6$ quasars.  \cite{2020MNRAS.495.2135N} studied the obscuration of high redshift quasars using the BLUETIDES simulation, and predicted that the dust-extincted UV LF is about 1.5 dex lower than the intrinsic LF, and the vast majority of $z\sim 7$ AGNs have been missed by  UV-based surveys due to dust extinction.   This would imply a much higher bolometric luminosity density of early quasars, and would have a profound impact on the growth of SMBHs and early galaxy evolution overall.  A reliable determination of the obscured fraction of quasars at $z\gtrsim 5$ needs a combination of wide-field deep X-ray survey and effective spectroscopic identification of faint, obscured sources with either JWST or NOEMA/ALMA.  

\subsection{(A Lack of) Evolution of the Quasar Spectral Energy Distributions}
\label{subsec:SED} 

\begin{figure}[ht]
\includegraphics[width=4in]{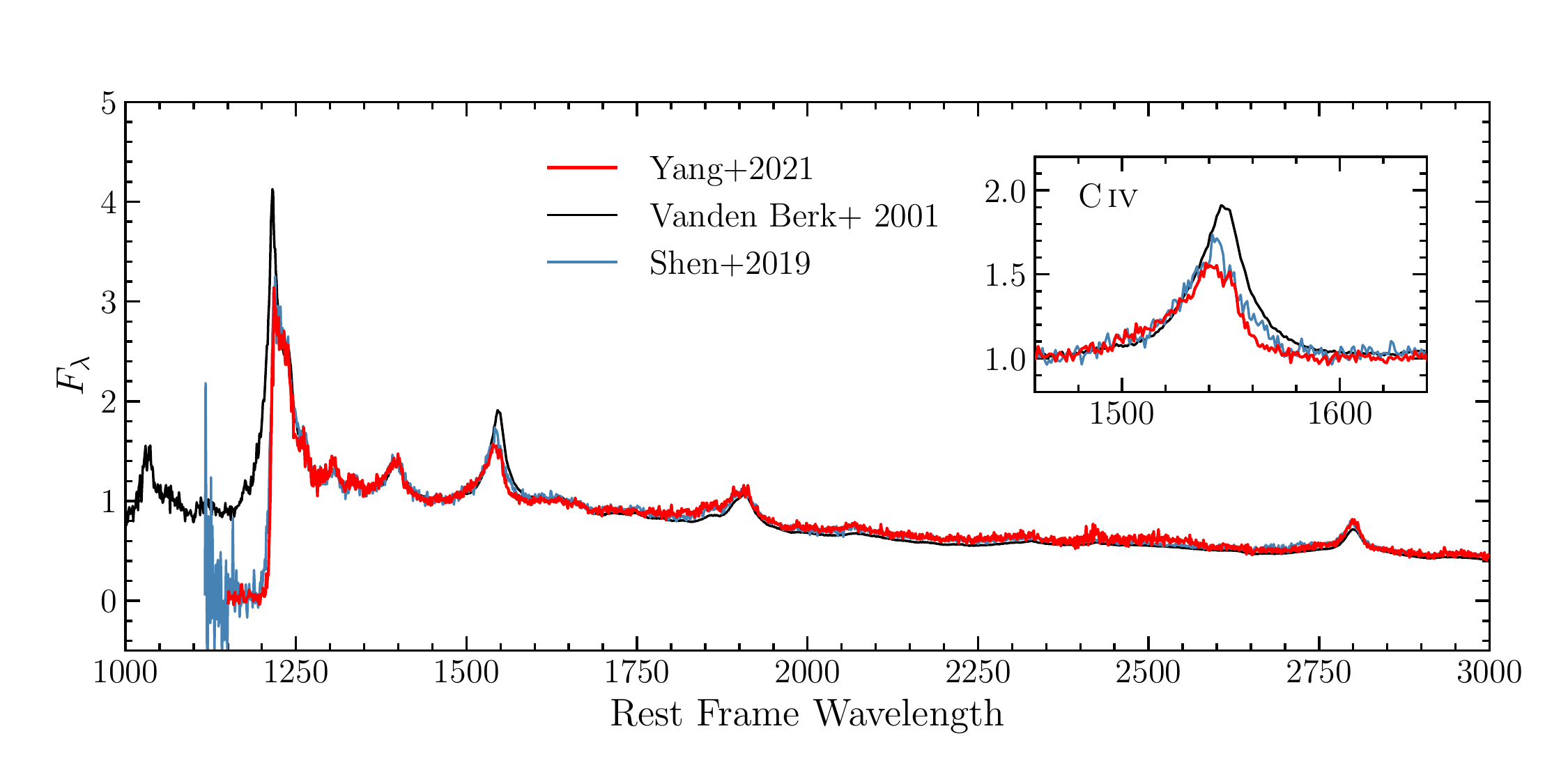}
\includegraphics[width=4in]{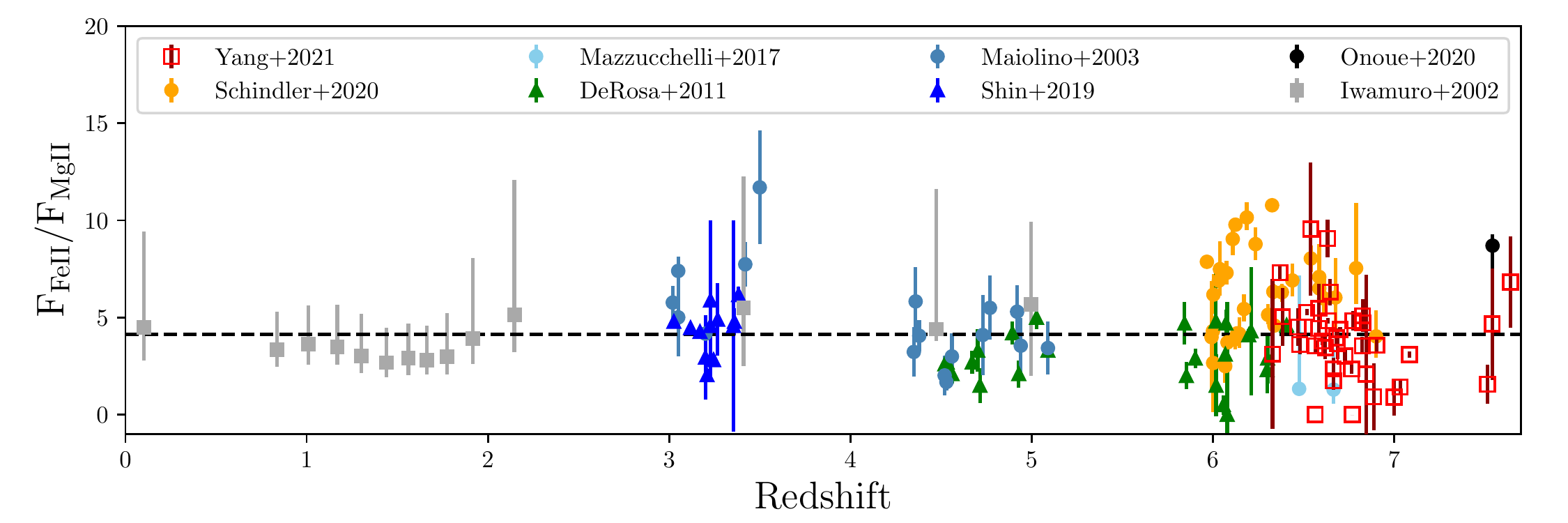}
\caption{Upper panel: Quasar composite spectrum (red solid line) from \cite{2021ApJ...923..262Y} compared with the low-redshift ($z\lesssim 3$) composite from  \cite{2001AJ....122..549V}; black line) and the $z\sim 6$  quasar composite from  \cite{2019ApJ...873...35S};  blue line). The average intrinsic spectrum of quasars does not exhibit significant redshift evolution. 
Lower panel: the evolution of the Fe/Mg emission line radio in quasars as a function of redshift. The quasar BLR is highly enriched even at the highest redshift. 
Adapted from \cite{2021ApJ...923..262Y}.
}
\label{fig:noevo}
\end{figure}

High-redshift galaxies have intrinsic blue SEDs that are dominated by young stellar populations. Metallicity, or chemical abundance, in their ISM has also been shown to be lower than in low-redshift galaxies \citep{2016ARA&A..54..761S}. However, the overall SEDs  and the chemical abundance in quasar broad line regions (BLRs) do not  evolve significantly with redshift, although quasar density decreases drastically at high-redshift. This apparent lack of spectral evolution in the UV properties of quasars were noticed as soon as the first $z\sim 6$ quasars were discovered (e.g.,  \citealt{2003ApJ...594L..95B, 2004AJ....128..515F, 2004ApJ...614...69I, 2007AJ....134.1150J}). 
%, and extends to the highest redshifts and larger samples \citep{2019MNRAS.487.1874R, 2019ApJ...873...35S, 2020ApJ...905...51S, 2021ApJ...923..262Y}.
\cite{2019ApJ...873...35S}  conducted an optical/NIR spectroscopic survey of 50 quasars at $z>5.7$. Their   observations covered the wavelength range from the Ly$\alpha$ to \mgii\ emission lines. \cite{2021ApJ...923..262Y} presented a spectroscopic survey of 34 quasars at $6.3 < z < 7.6$, covering a similar wavelength range. Both papers presented composite rest-frame UV spectra of their high-redshift samples. Fig~\ref{fig:noevo} compares the quasar composite spectra  at $z\sim 6 -7$ with the standard low-redshift SDSS quasar composite from \cite{2001AJ....122..549V}.  The average continuum slope and emission line strength/width do not evolve significantly with redshift, with the exception of a weaker and blueshifted \civ\ emission line (see discussion in Sec~\ref{subsec:feedback}). 

Broad emission line ratios in quasars have been used to constrain metallicities in the quasar BLR. \cite{2007AJ....134.1150J} and  \cite{2014ApJ...790..145D} studied emission line ratios in a series of UV lines, and found no evidence of evolution from lower redshift samples, with the gas metallicity a few times solar based on photoionization modeling. This is confirmed by studies using high S/N spectroscopy from the XQR30 survey (\url{https://xqr30.inaf.it/}) of quasars at $z=5.8 - 7.5$ \citep{2022MNRAS.513.1801L}. The line ratio of Fe\,\textsc{ii}/\mgii\ is of particular interest, because of the different enrichment histories of these two lines -- Mg is an $\alpha$-element predominately produced by core-collapse supernovae, while Fe is mainly from Type-1a supernovae with an evolutionary lifetime delay of a few hundred million years to 1 Gyr. The evolution of the Fe/Mg ratio in quasars has been studied extensively  \citep[e.g.,][]{2003ApJ...594L..95B, 2004ApJ...614...69I, 2014ApJ...790..145D, 2017ApJ...849...91M, 2019ApJ...873...35S,  2021ApJ...923..262Y}. \cite{2020ApJ...905...51S}  carried out detailed modeling of Fe\,\textsc{ii} and \mgii\ lines using a large sample of quasar spectra from VLT/Xshooter, and compared their measurements with those in the literature. There is no evidence of evolution in the Fe/Mg ratio up to the highest redshift, although scatter is large at any given redshift (Fig~\ref{fig:noevo}). The lack of spectral evolution in quasar rest-frame UV spectra, especially in the emission line properties, indicates that quasar BLRs are chemically enriched very rapidly, within the first few hundred million years after the initial star formation in the host galaxy. This is not surprising, as we will review in Sec~\ref{sec:host}: quasar host galaxies are sites of the most intense star formation in the early universe. 

However, there are a few noticeable evolutionary trends in quasars at cosmic dawn that could be related to the early phase of quasar growth. In Sec~\ref{subsec:feedback}, we discuss an increasing fraction of quasars with strongly blue-shifted high ionization UV emission lines and with strong BAL features as evidence for early quasar feedback. In Sec~\ref{subsec:proximity}, we present the discovery of $z\sim 6$ quasars with IGM absorption features indicative of very young ages. 

%\em Line Shifts.} High-redshift quasars show asymmetric shape and velocity offset of high ionization lines, in particular the CIV$\lambda 1450$ line. Low-redshift quasar population shows an overall blueshifted CIV line ($\sim 800$ km s$^{-1}$) 
%compared to the systemic redshift of the quasar \citep[e.g., traced by H$\beta$,][]{2011ApJS..194...45S}. This shift is generally understood in the context of strong accretion disk wine as an component of the high-ionization lines \citep[e.g.,][]{2011AJ....141..167R}. At $z>6$, the CIV line velocity shift is much stronger \citep[$\sim$ 1800 km s$^{-1}$,][] {2020ApJ...905...51S}. \cite{2019MNRAS.487.3305M} suggest that this redshift evolution can be understood in the context of the CIV winds being launched from the disk with an increased torus opacity at this redshift.

%{\em Broad-Absorption Line (BAL) Quasars.} A fraction of quasars show broad and highly blue-shifted absorption features in their rest-frame UV transitions. BAL features trace ionized winds in the broad line region and signatures of SMBH feedback. \cite{2022Natur.605..244B} used high quality spectra of quasars from the XQR30 survey that up to $\sim 50\%$ luminous quasars at $z\sim 6$ exhibit BAL features, compared to about 20\% observed in low-redshift samples \citep[c.f.,][]{2019ApJ...873...35S, 2020ApJ...905...51S}. This could be a result of the strong feedback associated with the rapid BH growth and galaxy assembly in the early Universe. 

A subset of high-redshift SDSS quasars have much weaker UV emission lines, in particular Ly$\alpha$ and \civ. \cite{2009ApJ...699..782D} identified them as  $>3\sigma$ outliers in the quasar emission line strength distributions.
%The unusual line width could be either a result of abnormal underlying UV continuum or special properties of BLRs related to quasar youth \citep{2015ApJ...805..123P}. 
They represent $\sim 1\%$ of quasar population at $z<4$, after correction for selection incompleteness \citep{2009ApJ...699..782D}. \cite{2016ApJS..227...11B} showed that this fraction increases to $\sim 14\%$ among $z>5.6$ PS1 quasars based on Ly$\alpha$ measurements. \cite{2019ApJ...873...35S} found a similar fraction using \civ\ measurements. It remains unclear whether weak line quasars are related to the young age or high accretion rate in the high-redshift quasar population \citep{2015ApJ...805..123P}.. 

%{\em Young quasars.} Quasars at the reionization epoch show strong Gunn-Peterson absorption in their spectra. However, the flux level doesn't immediately drop to undetectable levels blueward of Ly$\alpha$ emission lines. Instead, the strong UV radiation from quasars creates a large highly ionized region (quasar near-zone, or proximity zone, see Sec ...) with its sizes depending on quasar luminosity, age and IGM neutral fraction \citep{2006AJ....132..117F}. \cite{2017ApJ...840...24E} first discovered a subset of high-redshift quasars with very small proximity zone sizes, consistent with a quasar lifetime of $10^4 - 10^5$ years, significantly shorter than the typical Salpeter timescale of a few $10^7$ years. These young quasars might represent a unique phase of quasar evolution and provide special opportunities to study the triggering and feedback mechanisms of SMBHs at the highest redshift \citep{2020ApJ...900...37E} , while the identifications of such young quasars are difficult at lower redshift due to the fact that IGM is already highly ionized. 

X-ray emission originates from the accretion disk and surrounding hot corona close to the central SMBH, and provides crucial information about black hole accretion and AGN feedback \citep{2012ARA&A..50..455F}.  \cite{2001AJ....121..591B} detected the first $z\sim 6$ quasar in X-rays using the XMM/Newton telescope. More than 30 quasar at $z\gtrsim 6$ have now been detected in X-rays, using either the Chandra or XMM-Newton telescopes.  \cite{2017A&A...603A.128N} analyzed 29 quasars at $z>5.5$ with X-ray detections. They found a mean X-ray power-law photon index of $\Gamma \sim 1.9$, similar to that at low redshift. The optical-X-ray spectral slopes of the high-redshift also follow the relation established at low redshift. \cite{2019A&A...630A.118V} carried out a similar analysis, and found a slightly steeper X-ray power-law index, consistent with a generally higher Eddington ratio among SMBHs in these quasars at $z>6$. \cite{2021ApJ...908...53W} extended the X-ray analysis to quasars at $z\sim 7$  \citep[see also][]{2020MNRAS.491.3884P}. They also found a steepening of X-ray spectra with $\Gamma \sim 2.3$. 
The optical-X-ray power-law slope, $\alpha_{OX}$, traces the relative importance of the accretion disk and corono emission in quasars and AGN. At low redshift, there is a tight correlation between $\alpha_{OX}$ and the quasar UV luminosity \cite[e.g.][]{2007ApJ...665.1004J}. There is no evolution in this correlation up to $z\sim 7$ \cite[e.g.][]{2021ApJ...908...53W}. 
The X-ray observations show a consistent picture that the inner accretion-disk and hot-corona structure in quasars was established at the highest redshift with minimal evolution in their properties across cosmic time. 

At rest-frame NIR and MIR wavelengths, the radiation from quasars is dominated by reprocessed emission from the hot dust component beyond the accretion disk structure   \citep[see review by][]{2022Univ....8..304L}. At the highest redshift, dust observations (at rest-frame $>1.5\, \mu$m) have been largely limited to photometric observations of bright sources using  the Spitzer   \citep[e.g.,][]{2006AJ....132.2127J} and Herschel \citep{2014ApJ...785..154L} Telescopes, although this will change with the launch of JWST. These observations also showed a lack of evolution: the average IR-SEDs of quasars at $z\sim 5-6$ are consistent with low-redshift templates and the hot dust structure is already in place by $z\sim 6$. \cite{2010Natur.464..380J} reported the detection of two quasars at $z\sim 6$ with weak hot dust emissions  using Spitzer. Later studies suggested that these dust-deficient quasars are not unique to $z>6$, and examples with similar SEDs are found even in the low-redshift PG sample \citep[e.g.][]{2017ApJ...835..257L}. 

Quasars were initially discovered as point-like sources emitting strong radio emissions \citep{1963Natur.197.1040S}. However, the fraction of quasars that exhibit strong jet-powered radio emission (radio-loud quasars) is around 10\% with no significant dependence on redshift, at least up to $z\sim 6$ \citep{2015ApJ...804..118B,2021ApJ...908..124L,2021A&A...656A.137G}.  The highest redshift radio-loud quasar known is at $z=6.82$ \citep{2021ApJ...909...80B}, and the highest redshift blazar is at $z=6.10$ \citep{2020A&A...635L...7B}.  
It is still debated whether the seemingly constant radio-loud fraction is due to two intrinsically different populations of quasars or is a consequence of the duty cycle of jet emission during the life of a quasar. If the latter were the case, one should expect an increase in the radio-loud fraction at the
highest redshifts (where the available time for a quasar shortens), something we have not yet witnessed and depends on the lifetime of the radio-loud phase. 

The lack of evolution in the radio-loud fraction discussed above concerns the compact core radio emission. However, there is an important evolution with redshift if we focus on the extended radio emission:  the lack of giant ($\sim$100s\,kpc) radio lobes at $z>4$, which are common at lower redshifts \citep{2014MNRAS.442L..81F}. Indeed, the most extended radio jet known at $z\sim6$ is $<2\,$kpc \citep{2018ApJ...861...86M}. 
A plausible physical explanation is that at $z>4$, the CMB energy density [$\propto (1+z)^4$] exceeds the magnetic energy density in radio lobes. In that case, inverse Compton (IC) scattering losses will make any large lobes at high redshift radio weak and X-ray bright \citep{2014MNRAS.438.2694G}. The IC/CMB effect has been challenging to probe with current X-ray telescopes. Recent deep Chandra X-ray observations of the two $z\sim6$ quasars with the most extended radio jets resulted in tentative evidence of larger X-ray jets  \citep{2021ApJ...911..120C,2022A&A...659A..93I}. Obtaining this marginal evidence was expensive, even for our most powerful X-ray telescopes observing the best existing targets to test this effect. To robustly measure the IC/CMB effect in a sample of radio-loud quasars, we will likely need to wait for the next generation of X-ray telescopes.

%Studies based on both cross-matching wide-field radio survey with high-redshift quasar sample \citep{2015ApJ...804..118B} and dedicated deep radio observations \citep{2021ApJ...908..124L} yield a similar 10\% radio-loud fraction among $z\sim 6$ quasars. 

The discussions in this subsection present a picture that on average, the intrinsic SED of quasars at the highest redshift, with emission from the accretion disk (X-ray, UV/optical), broad emission line regions, hot dust, and radio jets, show little evolution from their low-redshift counterparts. The AGN structure is already in place and fully formed up to the highest redshift we have observed to date. 

JWST will enable detailed studies of the physical properties of high-redshift quasars in the rest-frame optical and IR wavelengths which are not possible from the ground.  It will be especially interesting to see if the dust component of the quasar population shows any evolution, since dust production mechanism  \citep[e.g.][]{2004Natur.431..533M}  and the ISM conditions in the host galaxies are likely different in early epochs. X-ray observations with Chandra and XMM of even the most luminous quasars at cosmic dawn can only yield a small number of photons. 
%ESA's Athena mission will allow high S/N spectroscopic observations of early quasars to characterize their accretion properties (including BH spin) and X-ray obscuration \citep[e.g.][]{2017A&A...608A..51M}. 
Future X-ray missions \citep[e.g.][]{2017A&A...608A..51M} could allow allow high S/N spectroscopic observations of early quasars to characterize their accretion properties (including BH spin) and X-ray obscuration. 

\section{GROWTH OF SUPERMASSIVE BLACK HOLES IN THE FIRST BILLION YEARS}
\label{sec:SMBH}

SMBHs are thought to be ubiquitous in the centers of massive galaxies, but their formation mechanism is still an outstanding question in astrophysics. Since the discovery of quasars at $z\gtrsim 6$ \citep{2001AJ....122.2833F}, the formation and growth of SMBHs have become even more intriguing and challenging to explain. In this section we will first review the current methods used to measure the BH masses at high redshift (Section~\ref{sec:bhm-measurements}), then give an overview of the properties of the observed population (Section~\ref{sec:demographics}), and the implications for early BH seeds and growth (Section~\ref{sec:seeds-constraints}). Finally, we will briefly describe additional methods that in the near future can be used to improve current BH mass estimates at high redshift (Section~\ref{sec:mass-future}). 

\subsection{Black Hole Mass Measurements in High-redshift quasars}
\label{sec:bhm-measurements}
%Farina2022 2022arXiv220705113F
%Yang2021 2021ApJ...923..262Y
%Shen2019 2019ApJ...873...35S
%Onoue2019 2019ApJ...880...77O

The spectra of $z\gtrsim6$ quasars show the typical signatures of broad (FWHM of thousands $\kms$) emission lines that are superimposed on the power-law continuum emission of the quasar's accretion disk. These broad lines are thought to emerge from a region very close to the accreting BH (so-called the ``broad-line region'') and thus provide a unique tool to constrain the properties of the central BH.
The size of the broad-line region scales with the quasar luminosity, and the width of the lines is a measure of the potential depth \citep{2004ApJ...613..682P}. Thus, a single-epoch spectrum gives enough information to estimate the mass of the central accreting BH \citep{2009ApJ...699..800V}. 

One of the most reliable estimators of BH mass and Eddington ratio is the 
H$\beta$ line (e.g., \citealt{2012ApJ...747...30P}). At $z > 6$, H$\beta$ is shifted to the MIR and is thus not 
accessible from the ground. As an alternative, the \civ\,$\lambda$1549 and \mgii \,$\lambda$2800 lines have been extensively used to measure BH masses of quasars at these redshifts (e.g., \citealt{2014ApJ...790..145D,2022arXiv220705113F}). Estimates based on the high ionization \civ\ line are uncertain, as this line shows large velocity offsets, implying significant non-virialized motions  (\citealt{2018MNRAS.478.1929M,2017ApJ...839...93P}). Moreover, there is mounting evidence that large \civ\ blueshifts ($>2000\,\kms$) are more common at $z>6$ than at lower redshifts \citep[e.g.,][see also Section~\ref{subsec:feedback}]{2019MNRAS.487.3305M,2020ApJ...905...51S}. Therefore, in this review, we only focus on results based on \mgii\ measurements, which are thought to be the most reliable until H$\beta$-derived masses are possible (\citealt{2019ApJ...875...50B}; see Section~\ref{sec:mass-future}).

One of the most widely used relations for BH mass estimation in high-redshift quasars is:
%-------------
   \begin{equation} \label{eq:V09bhmass}
      M\mathrm{_{BH}} = 10^{6.86} \, \left[ \frac{\mathrm{FWHM (\mgii)}}{1000\, \mathrm{km\,s^{-1}}} \right]^{2} \, \left[ \frac{L3000}{10^{44}\, \mathrm{erg\,s^{-1}}} \right]^{0.5} \, M_{\odot},
   \end{equation}

\noindent where FWHM(\mgii) is the Full Width at Half Maximum of the \mgii\ line, and L3000 is 
the luminosity at 3000\,\AA. \cite{2009ApJ...699..800V} presented the scaling relation from Equation~\ref{eq:V09bhmass}. The intrinsic scatter in the relation is 0.55 dex, which dominates the uncertainty for individual measurements.

The theoretical maximum luminosity that a source can achieve when the gravitational and radiation forces are in equilibrium is called the Eddington luminosity \citep{eddington1926}. The Eddington luminosity for pure ionized hydrogen in spherical symmetry is defined as 

\begin{eqnarray}\label{eq:Ledd}
    L_{\rm Edd} & = &\dfrac{4\pi c G m_p M_{\rm BH}}{\sigma_{T}}\\    % 1.26\times10^{38}\,\ergs{}\left(\frac{M_{\rm BH}}{M_\odot}\right) \\
         &= & 1.26\times10^{38}\,\ergs{}\left(\dfrac{M_{\rm BH}}{M_\odot}\right) \nonumber
\end{eqnarray}

\noindent where $c$ is the speed of light, $G$ is the gravitational constant, $m_p$ is the mass of a proton,  and $\sigma_T$ is the Thomson scattering cross-section. 
The ratio between the bolometric and Eddington luminosities (referred to as Eddington ratio; $\lambda_{\rm Edd} = L_{\rm Bol}/L_{\rm Edd}$) is useful to assess the quasar population's accretion properties. 
While several bolometric corrections are published in the literature (e.g., \citealt{2012MNRAS.422..478R,2012MNRAS.427.3081T,2020A&A...636A..73D}), in this review (e.g., in Fig.~\ref{fig:Lbol-BHmass}), we adopt the bolometric correction first presented by \cite{2006ApJS..166..470R} and used in several studies of high-redshift quasars \citep[e.g.,][]{2011ApJS..194...45S,2021ApJ...923..262Y,2022arXiv220705113F}:
%2012MNRAS.422..478R, Runnoe 2012
%2020A&A...636A..73D, Duras 2020
%2012MNRAS.427.3081T, Trakhtenbrot 2012
%2011ApJS..194...45S Shen 2011
%2006ApJS..166..470R Richards 2006

\begin{equation}\label{eq:lbol}
L_\textrm{bol} = 5.15 \times L3000,
\end{equation}

\noindent  \cite{2022arXiv220705113F} argued that bolometric luminosities estimated using equation \ref{eq:lbol} are on average in between the ones calculated using the bolometric corrections from \cite{2012MNRAS.422..478R} and \cite{2012MNRAS.427.3081T}. In the high-redshift quasar database presented in {\bf Supplementary Material} we report $L3000$ so the readers can apply their preferred bolometric correction.

\subsection{Current Demographics}
\label{sec:demographics}

\begin{figure}[h]
\includegraphics[width=0.8\textwidth]{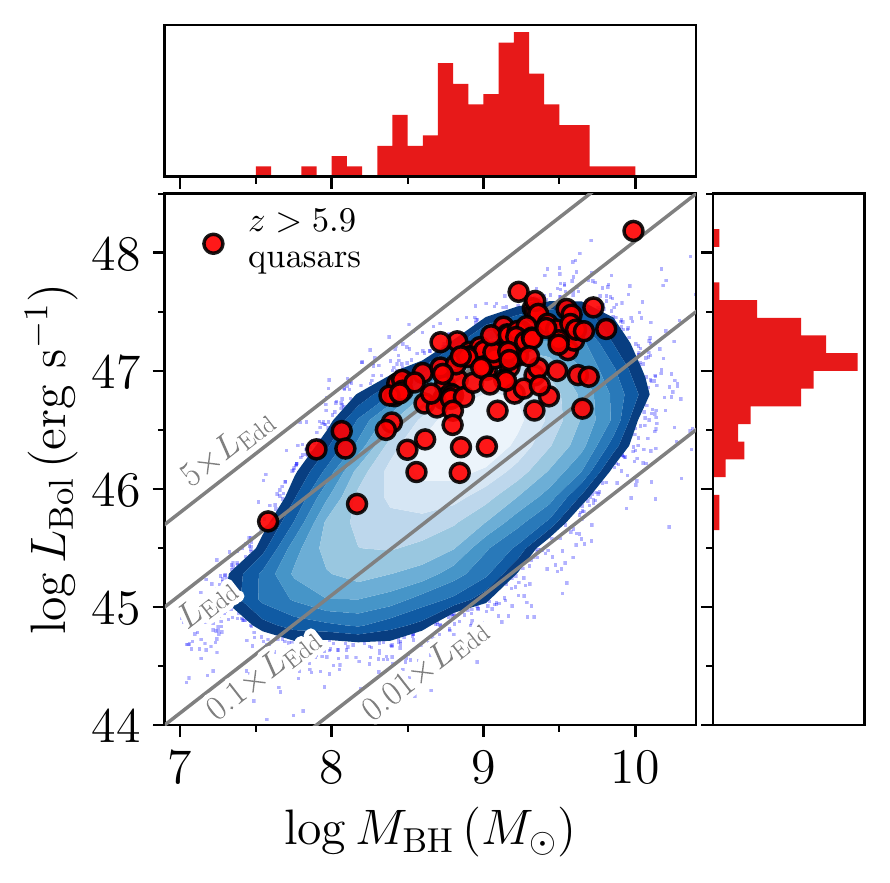}
\caption{Bolometric luminosity vs.\ BH mass. The blue contours and points show the distribution for low-redshift SDSS quasars while the red circles are measurements for $z>5.9$ quasars. Solid gray lines show the location of constant accretion rate at 0.01, 0.1, 1, and 5 times the Eddington luminosity. All BH masses are from \mgii\ measurements using equation~\ref{eq:V09bhmass} and the bolometric luminosities are derived using Equation~\ref{eq:lbol}.} 
\label{fig:Lbol-BHmass}
\end{figure}

At redshifts $6.0\lesssim z \lesssim 7.6$, the \mgii\ line falls in the NIR K band.  
The vast majority of the \mgii-based BH masses come from NIR spectra taken with  powerful ground-based telescopes such as Gemini, VLT, Magellan, and Keck \citep{2019ApJ...873...35S,2019ApJ...880...77O,2021ApJ...923..262Y,2022arXiv220705113F}. At the time of writing, there are \nqsomgii{} quasars at $z>5.9$ with reliable \mgii-based black hole mass estimates. Currently, there is a strong bias towards luminous ($\gtrsim 10^{46}\,$erg\,s$^{-1}$) quasars, but obtaining reliable measurements for fainter sources is challenging with current facilities (see \citealt{2019ApJ...880...77O}). We will need JWST and ELTs to enlarge the sample towards the faint end significantly.  Fig.~\ref{fig:Lbol-BHmass} shows the bolometric luminosity and BH mass distribution for all quasars with \mgii-based masses at $z>5.9$.  
The median BH mass is $1.3\times 10^{9}\,M_\odot$, while the least and most massive quasars are J0859+0022 at $z=6.39$ ($\sim$$4\times10^7\,M_\odot$;  \citealt{2019ApJ...880...77O}) and J0100+2802 at $z=6.33$ ($\sim$$10^{10}\,M_\odot$;  \citealt{2015Natur.518..512W}), respectively.

The first samples of NIR spectroscopic observations of $\gtrsim 6$ quasars showed that they accrete close to the Eddington limit, i.e., $\lambda_{\rm Edd}\approx 1$ \citep[e.g., ][]{2003ApJ...587L..15W,2007ApJ...669...32K,2011ApJ...739...56D}. Extending the measurements to larger samples \citep{2019ApJ...873...35S} or probing to fainter luminosities \citep{2019ApJ...880...77O} revealed a significant number of quasars accreting at sub-Eddington rates. The recent works by \cite{2021ApJ...923..262Y} and \cite{2022arXiv220705113F} analyzed some of the largest samples of NIR spectra of $z\gtrsim 6$ quasars. \cite{2021ApJ...923..262Y} studied 37 quasars at $z>6.3$ and found that these quasars have significantly higher Eddington ratios (mean of 1.08 and median 0.85) than a luminosity-matched quasar sample at low redshift. \cite{2022arXiv220705113F} studied the NIR spectra of 135 $z\gtrsim 5.7$ quasars (including BH properties derived from \mgii\ and \civ),  finding that at $z\sim 6$ the Eddington ratio distributions are consistent with a luminosity-matched low-redshift sample independent of the luminosity, while there is evidence for a mild increase in the median Eddington ratios for $z\gtrsim 6.5$ (in agreement with  \citealt{2021ApJ...923..262Y}).

In our current compilation (Fig.~\ref{fig:Lbol-BHmass}), the Eddington ratio ranges from $0.08$ to $2.70$, with a median of 0.79 and a mean of 0.92. Forty quasars exceed the nominal Eddington limit, and eight have  $\lambda_{\rm Edd}>2$. These values must be considered with caution as there is a substantial scatter in the BH mass estimations (see Section~\ref{sec:bhm-measurements}).

%Farina2022 2022arXiv220705113F
%Yang2021 2021ApJ...923..262Y

\begin{figure}[h]
\includegraphics[width=0.8\textwidth]{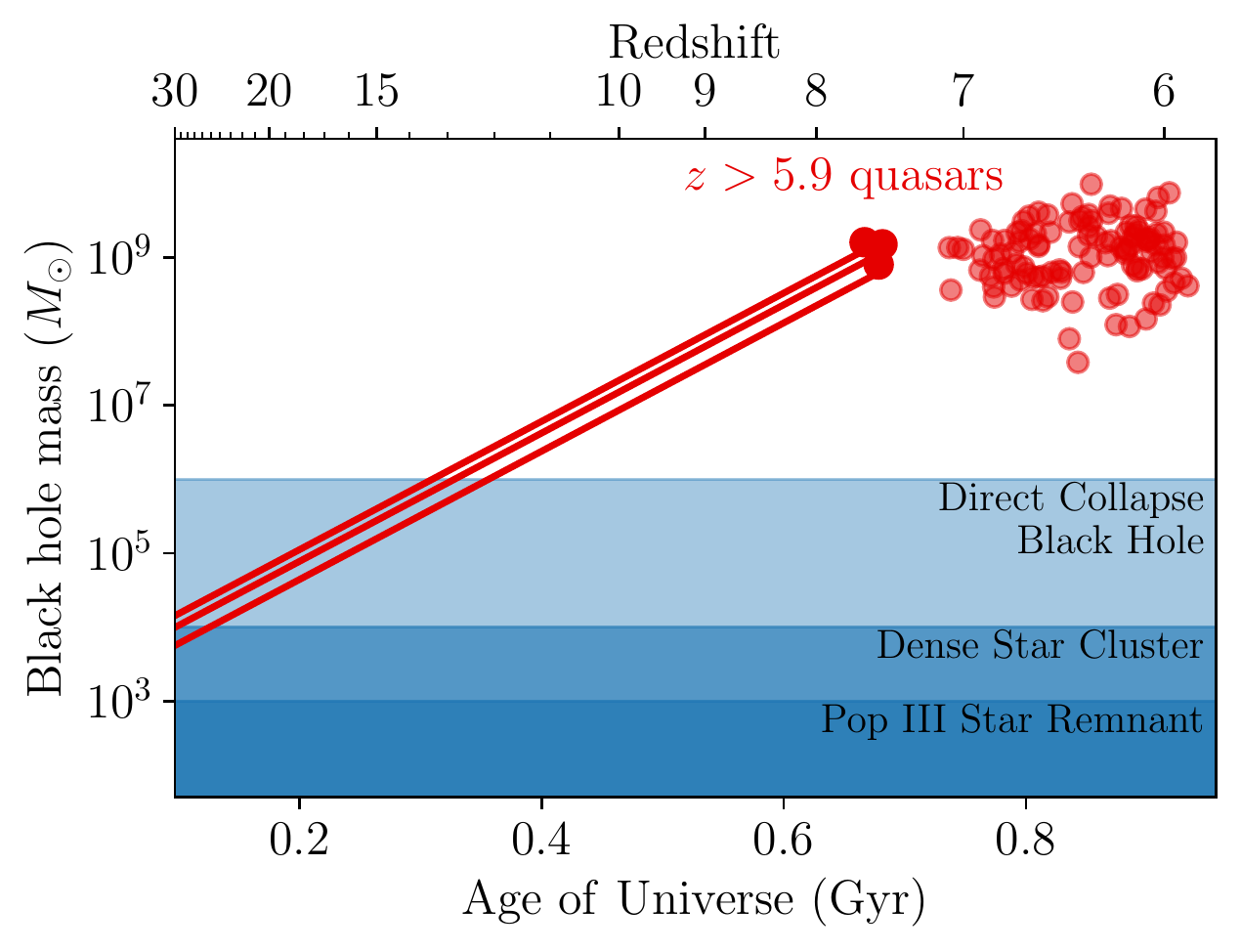}
\caption{BH growth history: BH mass vs.\ age of the Universe (redshift at the top). The red circles mark our compilation of robust \mgii\ BH masses for $z>5.9$ quasars. The red lines show the growth history (assuming constant, maximum Eddington-limited accretion) of the most distant quasars at $z>7.5$, which set the strongest challenges for BH formation theories. The shaded regions represent the mass ranges for popular BH seed formation scenarios (see \citealt{2020ARA&A..58...27I}).
}
\label{fig:bhgrowth}
\end{figure}

\subsection{Constraints on Early BH Seeds and Growth}
\label{sec:seeds-constraints}

As shown in Fig.~\ref{fig:bhgrowth}, the SMBHs powering the first quasars have quickly grown to enormous sizes, and explaining this mass growth remains challenging.  Assuming a 100\% duty cycle, the mass of the SMBH would grow exponentially as: 
% salpeter_time = (const.c * const.sigma_T/(const.G*4*np.pi *const.m_p)).to("Myr")
%     #t_salpeter = 0.45 * (eff/(1-eff)) / edd_ratio * u.Gyr
%     t_salpeter = salpeter_time * (eff/(1-eff)) / edd_ratio
    % M = Mi * np.exp((t/t_salpeter).decompose())
% \begin{equation}
%     M_{\rm BH}(t) = M_{\rm BH, seed} \times \exp\left[\frac{1-\epsilon_r}{\epsilon_r}\lambda_{\rm Edd} \frac{t}{t_{\rm Sal}}\right]
% \end{equation}

\begin{equation} \label{eq:bhgrowth}
    M_{\rm BH}(t) = M_{\rm BH, seed} \times \exp\left[(1-\epsilon_r)\lambda_{\rm Edd} \frac{t}{t_{\rm Sal}}\right]
\end{equation}
\noindent where $\epsilon_r$ is the radiative efficiency, which is the efficiency of converting mass to energy and the Salpeter time is defined as:
\begin{equation}
t_\mathrm{Sal}=\dfrac{\epsilon_r\sigma_\mathrm{T} c}{4 \pi G m_\mathrm{p}}\approx \epsilon_r \,450\,\mathrm{Myr}.
\end{equation}

Assuming $\lambda_{\rm Edd}=1$ and $\epsilon_r=0.1$, Equation~\ref{eq:bhgrowth} equals to

\begin{equation}
    M_{\rm BH}(t) = M_{\rm BH, seed} \times \exp\left[\dfrac{t}{50\,\mathrm{Myr}}\right]
\end{equation}
We used these assumptions for the growth tracks in Fig.~\ref{fig:bhgrowth}. We note that either a higher $\epsilon_r$ or a $<100\%$ duty cycle would make the constraints on BH growth even tighter. 
%Similarly, here we assume a 100\% duty cycle in BH growth; an episodic growth will further tighten the timescale constraint.   
Fig.~\ref{fig:bhgrowth} showcases how pushing the redshift frontier makes the formation and growth of SMBHs ever more challenging. Several potential theories have been developed to explain the existence of $\sim10^{9}\,M_{\odot}$ BHs at $z<6$. 
% For $\sim10^{9}\,M_{\rm BH}$ at $z<6$  several potential theories have been developed to explain their existence.
However, for the current redshift records at $z\sim 7.5$, at least under standard assumptions (i.e., Eddington-limited growth with $\epsilon_r=0.1$)  only the heaviest ``seeds'' seem to be big enough to form these sources in such a short time, although  they need to have sustained accretion their entire lifetime. For a thorough overview of the theoretical BH ``seeds'' we refer to the review by \cite{2020ARA&A..58...27I}. Other alternatives not represented in Fig.~\ref{fig:bhgrowth} involve radiative inefficient/highly obscured growth ($\epsilon_r \ll 0.1$; \citealt{2019ApJ...884L..19D}) or jet-enhanced accretion growth \citep{2008MNRAS.386..989J}. 
According to \cite{2022MNRAS.509.1885P},  the discovery of one $\sim 10^{10}\,M_{\odot}$ black hole at $z\gtrsim 9$ would exclude the entire parameter space available for $\sim 100\,M_\odot$ seeds. 
%I haven't read this paper yet but the last sentence of the abstract seems like a good quote
%https://ui.adsabs.harvard.edu/abs/2022MNRAS.516..138B/abstract
%"our results demonstrate that progenitors of  z  ~ 6 quasars have distinct BH merger histories for different seeding models, which will be distinguishable with Laser Interferometer Space Antenna observations."
%maybe add if convinced once I read it.

However, it is important to note that these stringent constraints mainly apply to the most massive BHs at the highest redshift. For objects at lower redshift ($z<5$), SMBH growth is no longer strongly limited by the available cosmic time. In other words, there could be multiple channels of SMBH growth, and the channel with massive seeds and high accretion rate is required only for the formation of the most extreme objects.

\subsection{The (Near) Future for Black Hole Mass Measurements at High Redshift}
\label{sec:mass-future}
\subsubsection{H$\beta$-derived masses} With JWST, robust H$\beta$-based mass estimates of the most distant quasars finally has become feasible (see Section~\ref{sec:bhm-measurements}). The JWST instruments NIRSpec and NIRCam can obtain spectroscopic observations of the H$\beta$ line in quasars at $5\lesssim z\lesssim 10$. 
H$\beta$ BH masses are expected to be one of the first results from the first observations of high-redshift quasars with JWST.

\subsubsection{Direct dynamical mass measurements}
The unprecedented resolution provided by ALMA can spatially resolve the sphere of influence of central black holes in nearby galaxies, enabling direct gas-dynamical BH mass measurements \citep[e.g., ][]{2021ApJ...919...77C}. The sphere of influence is the region where the SMBH dominates the gravitational potential; its radius is commonly defined as 

\begin{equation}
    r<\frac{GM_{\rm BH}}{\sigma^2}
\end{equation}

\noindent where $G$ is the gravitational constant, M$_{\rm BH}$ is the mass of the BH, and $\sigma$ is the stellar velocity dispersion. For reference, the radius of influence of a $10^9\,M_\odot$ SMBH in a galaxy with $\sigma=150\,\kms$ is $\sim$190\,pc. 

Obtaining a direct kinematic mass measurement of a SMBH at $z>6$ would be tremendously important to corroborate that the mass scaling relations used thus far (Eq.~\ref{eq:V09bhmass}) still hold at the highest accessible redshifts. Recent ALMA observations of the \cii\ emission line in $z>6$ quasar hosts are approaching resolutions comparable to the expected sphere of influence of SMBHs 
\citep[e.g., ][]{2022ApJ...927...21W}. We anticipate that in the upcoming years, such measurement will be possible, at least for some selected objects, and this will be an active area of research with the ngVLA \citep{2018ASPC..517..535C}.

\section{A MULTI WAVELENGTH VIEW OF QUASAR/GALAXY CO-EVOLUTION AT THE HIGHEST REDSHIFTS}
\label{sec:host}

In the local Universe there are tight correlations between the mass of the central SMBHs and the stellar bulge of the galaxy (see review by \citealt{2013ARA&A..51..511K}). When and how these tight correlations arose is an open question. Constraints at the highest redshifts can provide essential clues when the time to grow a galaxy and BH is limited.  Did the BH grow first, and the galaxy follow? Vice versa? Or did they grow together?  What are the main feedback mechanisms? How are the massive galaxies hosting quasars connected to their large-scale environment?

In this section, we will first review the current efforts to detect the gas reservoirs  required for the growth of the central SMBH and the formation of stars (Section~\ref{sec:qso-fueling}). We will then give an overview of the efforts to  measure the stellar UV light from the $z>6$ quasar hosts and the recent successes  in constraining their cold dust and gas via (sub)mm observations (Section~\ref{sec:FIR}).
We will then discuss how we can use this information to constrain BH/galaxy co-evolution (Section~\ref{sec:seeds-constraints}) and describe observational evidence (or lack of) quasar feedback (Section~\ref{subsec:feedback}). Finally, we will give an overview of the efforts for using $z\gtrsim 6$ quasars as signposts for overdensities and large-scale structure in the early Universe (Section~\ref{subsec:protoclusters}). 

%I'm following Alyssa's suggestion of calling one section QSO fueling -> Lya haloes and 
%Section 2 on "host galaxies" in general. 

\subsection{Quasar fueling - Ly$\alpha$ nebuale}
\label{sec:qso-fueling}

Enormous gas reservoirs are required to grow the SMBHs seen in the highest-redshift quasars continuously.   
% Other attempts to look for signs of UV light from $z\gtrsim 6$ host galaxies have focused on identifying extended \lya\ nebulae (also referred as \lya\ halos) around these quasars. 
Extended \lya\ nebulae (also referred to as \lya\ halos) trace the cold gas reservoirs likely feeding the central SMBH. In principle, such gas reservoirs are likely also star-forming regions  \citep{2001ApJ...556...87H,2017MNRAS.467.4243D}. The first efforts to look for these nebulae used narrow-band filters centered on the expected \lya\ emission \citep[e.g., ][]{2009MNRAS.400..843G,2012ApJ...756..150D,2019MNRAS.488..120M} and through deep long-slit spectroscopy \cite[e.g., ][]{2011AJ....142..186W,2014MNRAS.443.3795R}. The implications of these detections (as well as non-detections) were often difficult to interpret due to slit and narrow-band flux losses. Sometimes the same quasar observed with narrow-band imaging and long-slit spectroscopy yielded conflicting results. 

The search for these \lya\ nebulae had a significant step forward with the new Integral Field Spectrographs (IFS) mounted on 8-m class telescopes, particularly VLT/MUSE. IFS enabled 3D  morphology/kinematics of the \lya\ halos  \citep{2019ApJ...881..131D}. The most comprehensive search for \lya\ halos to date is the REQUIEM survey
\citep{2019ApJ...887..196F}, targeting 31 $z>5.7$ quasars with VLT/MUSE. REQUIEM revealed 12 \lya\ nebulae with a range of luminosities and morphologies extending up to 30\,kpc. Recently, \cite{2022ApJ...929...86D} 
reported that the \lya\ morphology and kinematics seem decoupled from the gas in the host galaxies as traced by \cii\ emission. This result might imply that the observed \lya\ halos are being powered by the central SMBH instead of tracing star formation, in line with recent simulations \citep{2022arXiv220311232C}.   Nevertheless, the resonant nature of the \lya\ line makes it difficult to disentangle the AGN/star-formation contribution completely. IFS observations of additional non-resonant lines (e.g., H$\alpha$ with JWST) will be needed to confirm these results.

% \subsection{Quasar Host Galaxy Observations in UV}
% \label{sec:stellar-light}

\subsection{Quasar Host Galaxy Observations - Sites of Massive Galaxy Evolution}
\label{sec:FIR}

Given the extreme luminosities and inferred BH masses of these quasars, we would expect that they reside in hosts with significant stellar mass already in place (\citealt{2013ARA&A..51..511K}). However, the brightness of their central accreting BHs has prevented direct detection of the underlying starlight at (observed) optical and near-infrared
wavelengths even with HST \citep[e.g.,][]{2012ApJ...756L..38M,2020ApJ...900...21M}. 
 JWST's unprecedented resolution, sensitivity, and wavelength coverage should overcome most of the shortcomings of previous attempts, as \cite{2021MNRAS.506.1209M} have demonstrated with simulations. Several cycle 1 JWST programs attempt to reveal the stellar light of $z\gtrsim 6$ quasar hosts for the first time. %commenting it out as we won't be discussing JWST results here.  %and it has recently been demonstrated that this is really feasible, at least at lower redshifts \citep{2022arXiv220903359D}.

\begin{figure}[h]
\includegraphics[width=4in]{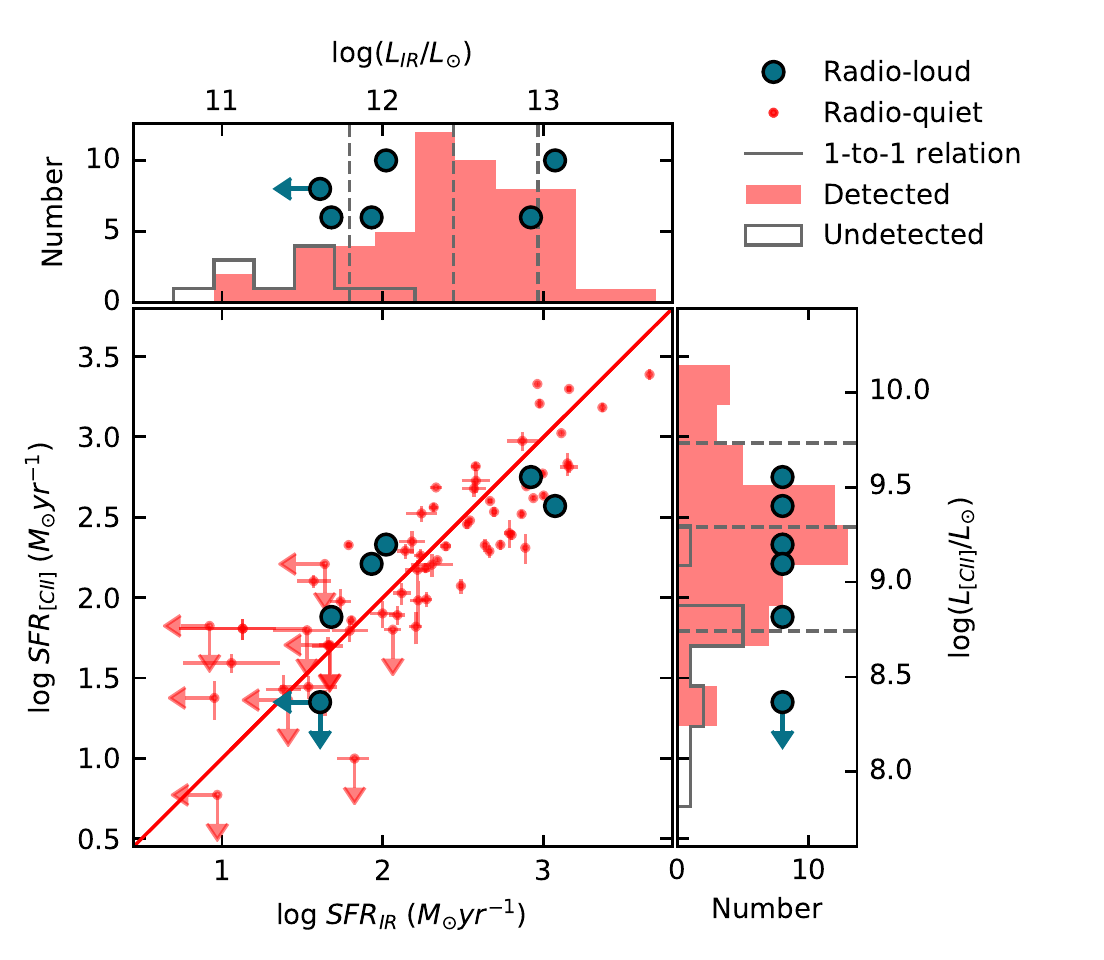}
\caption{Star-formation distributions for radio-quiet (small points) and radio-loud (circles) quasars at $z\gtrsim 6$, derived from \cii\ and IR luminosities (Figure adapted from \citealt{2022A&A...664A..39K}).
}
\label{fig:sfr-distribution}
\end{figure}
%Decarli 2018 2018ApJ...854...97D
  
% \subsection{Quasar Host Galaxy Observations in FIR - Sites of Massive Galaxy Evolution}
% \label{sec:FIR}

The radiation at observed far-IR  (FIR) to mm wavelengths of $z\gtrsim 6$ quasars is dominated by the reprocessed emission from cool/warm dust in the host galaxy. These wavelengths provide the best window to study host galaxy properties, minimizing contamination from the quasar's accretion disk emission. Pioneering works pushed the limits of mm and radio facilities to characterize the dust and the molecular and atomic gas in the first  bright $z\gtrsim 6$ quasars discovered (e.g., \citealt{2003A&A...406L..55B,2003Natur.424..406W,2005A&A...440L..51M,2006ApJ...642..694B,2007AJ....134..617W,2008ApJ...687..848W}). These studies were only sensitive to the most luminous systems, and found that about 1/3 of high redshift quasar host galaxies have luminosity comparable to those of  hyper-luminous IR galaxies ($L_{\rm FIR} \sim 10^{13}$ L$_\odot$). Assuming the dust heating came from starburst activity, this
suggested enormous star formation rates of $100 - 1000 \,M_{\odot}\, \rm yr^{-1}$. 
 Therefore, these early studies showed that early SMBH growth can be accompanied by extended, intense star formation and large reservoirs of dense and enriched molecular gas.
 
 Rapid progress has been made in FIR observations of high-redshift quasar host galaxies with the advent of ALMA and major upgrades on the NOEMA interferometers (conveniently located in different hemispheres). The greatest advances can be divided into three major areas: \textit{(i)} sensitivity, \textit{(ii)} resolution, \textit{(iii)} multi-line tracers of the ISM. 
 
 \textit{Sensitivity:} The \ciilong\  fine-structure line, which is the primary coolant of the cold neutral atomic medium, has become the workhorse for studies of $z\gtrsim 6$ quasar hosts. This was the natural choice as \cii\ is one of the brightest emission lines in star-forming galaxies and its frequency at $z>6$ is conveniently located in a high-transmission atmospheric window visible with ALMA and NOEMA. 
 \cii\ detections using the early generation of interferometers were challenging. By the time of the review of \cite{2013ARA&A..51..105C}, there were only two $z\gtrsim6$ quasars with \cii\ detections, while now that number is  $\sim 80$. \cii\ line and IR continuum luminosities measurements provide independent estimates of SFR.  Fig~\ref{fig:sfr-distribution} shows a recent compilation of the \cii\ and IR luminosities and SFRs for $z\gtrsim 6$ quasar hosts. 

Early ALMA results demonstrated its ability to study \cii\ emission from quasar hosts of both UV-bright and UV-faint $z\gtrsim 6$ quasars  \citep{2013ApJ...773...44W,2013ApJ...770...13W}. These studies motivated the push to larger samples, marking the transition from studies of individual interesting sources to the first statistical samples. \cite{2018ApJ...854...97D} and \cite{2018ApJ...866..159V} presented results from an ALMA snapshot survey of a large sample of  luminous $z\sim 6$ quasars to study their FIR continuum and \cii\ properties. Remarkably, even with integration time of $\lesssim 15$ min on ALMA, the detection rate is $\gtrsim90$\% for these quasars. They found typical \cii\ luminosities of $10^9 - 10^{10}\, L_{\odot}$, FIR luminosities of $0.3-13\times10^{12}\,L_\odot$, and estimated dust masses of $10^7-10^9\,M_\odot$ with star formation rates ranging from 50 to 2700\,$M_\odot$\,yr$^{-1}$. ALMA also enabled detections of FIR continuum and \cii\ emission in a significant number of UV-faint quasars at high redshift (e.g., \citealt{2018PASJ...70...36I,2019PASJ...71..111I}). Overall, there are only weak correlations between the bolometric luminosity of quasars (dominated by emission from accretion disk in the rest-frame UV/optical) and the FIR luminosity of the quasar hosts (dominated by star formation).

 \begin{figure}[h]
\includegraphics[width=3.5in]{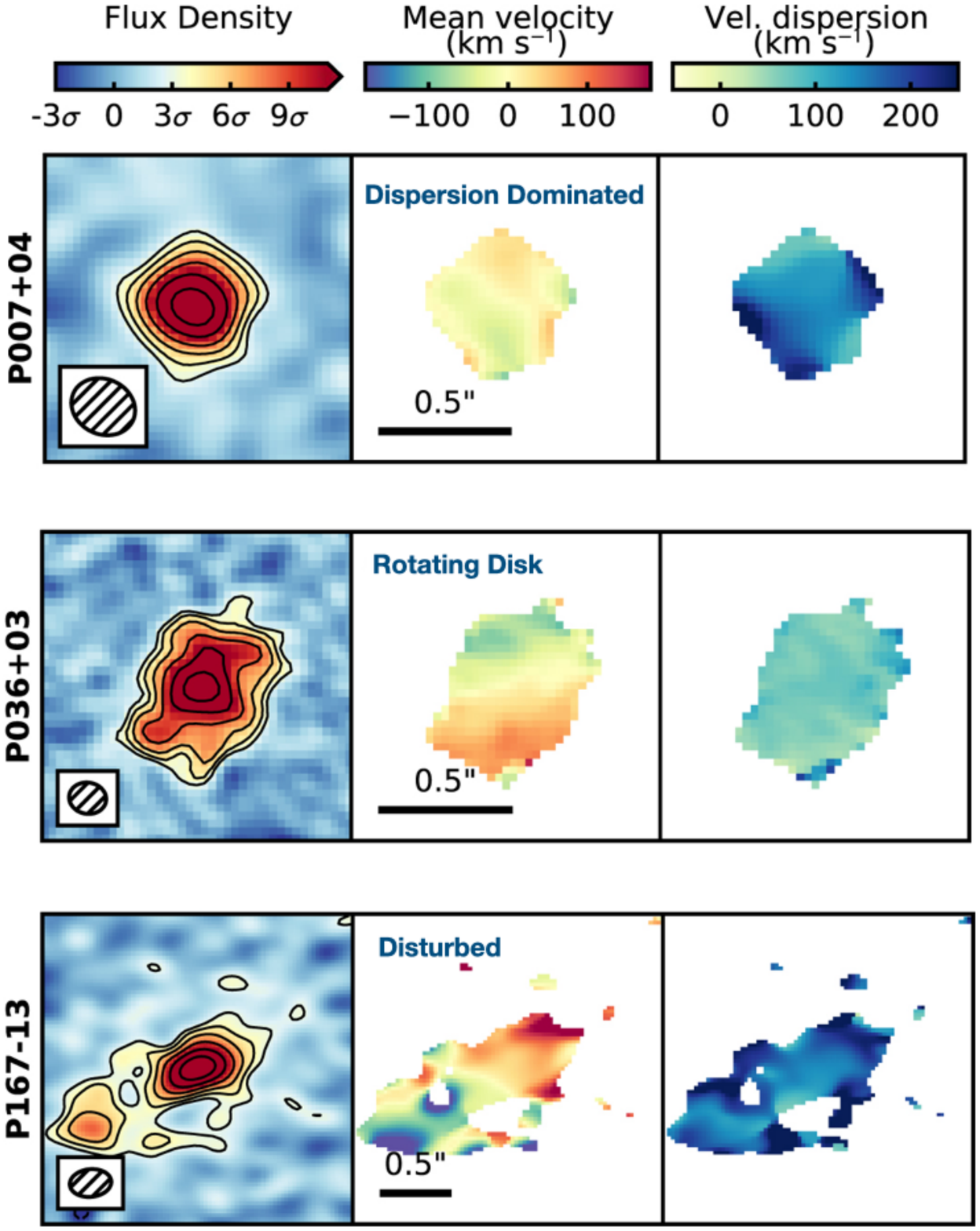}
\caption{ALMA \cii\ observations of quasar hosts reveal diverse morphologies and kinematics (figure adapted from \citealt{2021ApJ...911..141N}). The top panel shows a dispersion dominated galaxy, the middle panel  a galaxy consistent with a rotating disk, and the bottom panel shows a disturbed galaxy with an on-going merger.}
\label{fig:almahosts}
\end{figure}

 \textit{Resolution:} The unprecedented spatial resolution of ALMA allowed imaging of the ISM of $z>6$ quasar host galaxies at sub-kpc resolution for the first time. The first studies at kpc scales ($\sim0.2-0.3^{\prime\prime}$) revealed interesting morphological characteristics. For example, \cite{2017ApJ...845..138S} found ordered motion in a $z=6.13$ host that can be modeled with a rotating disk, \cite{2017ApJ...837..146V} report a remarkably compact ($\lesssim 1\,$kpc) host galaxy at $z=7.08$ that does not exhibit ordered motion on kpc scales, and \cite{2019ApJ...881L..23B} showed that a galaxy merger hosts a $z=7.54$ quasar.

 \cite{2020ApJ...904..130V} presented a survey of 27 host galaxies at $z\sim 6$ imaged at kpc scale with ALMA. They showed that the \cii\ emission in the bright, central regions of the quasars have sizes of 1.0–-4.8\,kpc and are typically more extended than the dust continuum emission.  The implied star formation rate densities at the center of these quasar hosts are a few hundred $M_{\odot}$\,yr$^{-1}$  kpc$^{-2}$,   below the Eddington limit for star formation (although there are examples of host galaxies forming stars at near the maximum possible rate, e.g., \citealt{2020ApJ...903...34A,2021ApJ...917...99Y}).  \cite{2021ApJ...911..141N} modeled the kinematics of these 27 host galaxies. They found a large diversity in the quasar host properties. About 1/3 of the galaxies show smooth velocity gradients consistent with emission from a gaseous disk. About 1/3 have no evident velocity gradients, with their kinematics dominated by random motion. The final 1/3 exhibit signatures of close companion interaction or galaxy merger activities (see Fig.~\ref{fig:almahosts} for examples). 
 %Although in principle ALMA could provide spatial resolutions as high as $\sim$50\,pc at these redshifts, 
 The highest resolution observations that currently exist for $z>6$ quasars are 400\,pc for a quasar host at $z=6.6$ \citep{2019ApJ...874L..30V} and 200\,pc for a host galaxy at $z=6.9$ \citep{2022ApJ...927...21W}. These two studies highlight the power of ALMA by revealing complex morphologies, \cii\ cavities in the gas distribution, and in one case very compact dust continuum and \cii\ emission, reaching extreme densities in the central 200\,pc from the SMBH. We expect that observations at comparable resolution or higher will be an active area of investigation in the near future.

A key result of these morphological studies is that the population of host galaxies of luminous early quasars is diverse. While, on average, they are sites of massive galaxy assembly, they also appear to be in different evolutionary phases. Apparently, the peak of quasar activity is not tied to a particular stage of early galaxy formation in those systems. 

\begin{figure}[h]
\includegraphics[width=0.5\textwidth]{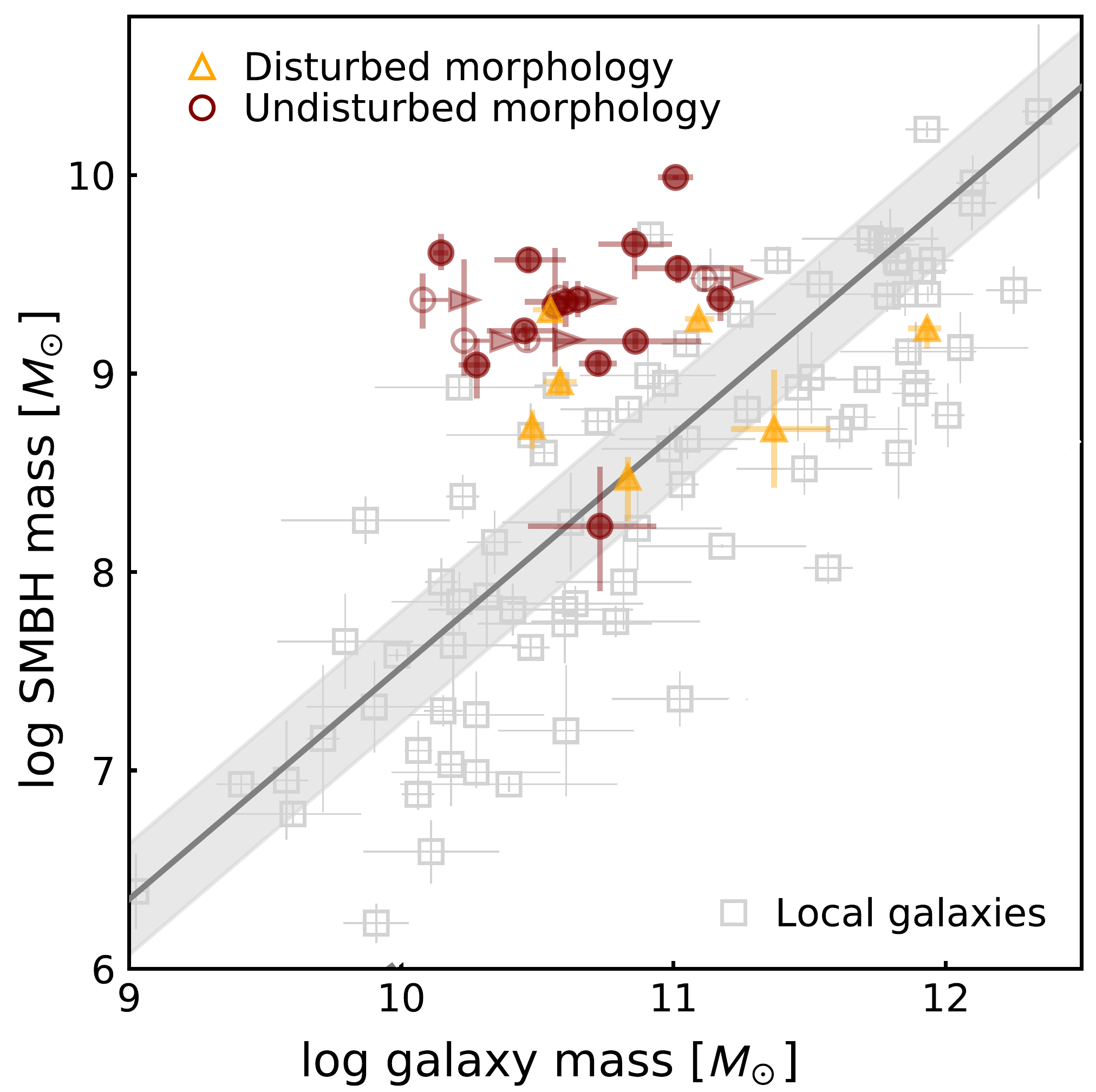}
\caption{Black hole mass vs.\ host dynamical mass for $z\gtrsim 6$ quasars (figure from \citealt{2021ApJ...911..141N}). The gray points are local galaxies and the shaded region is the best-fit local relation \citep{2013ARA&A..51..511K}. The $z\gtrsim 6$ sample is divided by their \cii\ morphology (see legend) and it is generally above the local relation. 
}
\label{fig:MdynMBH}
\end{figure}

 \textit{Multi-line diagnostics:} Even though the \cii\ emission line is important for redshift determination and gas dynamics (as discussed above), it cannot constrain the physics of the ISM by itself. A suite of emission lines tracing various phases of the ISM is required to characterize the physical conditions such as temperature, ionization, density, and metallicity. At the time of writing, this is an area of rapid development, and about a dozen quasar hosts have been detected in more than one tracer. Some of the main targeted lines (in addition to \cii) are:
 
 \begin{itemize}
     \item \textit{The neutral carbon emission [CI]\,369\,$\mu$m} line, which traces the atomic medium in the external shells of molecular clouds \citep[e.g., ][]{2019ApJ...880..153Y,2021A&A...652A..66P}.
     \item \textit{Two transitions of the [NII] line (205\,$\mu$m and 122\,$\mu$m)}, which is a pure tracer of the ionized medium and can be used to measure the fraction of \cii\ associated with the neutral medium (e.g., \citealt{2019ApJ...881...63N,2020ApJ...900..131L}).
     \item \textit{The [OI] 146 $\mu$m line}, which is sensitive to the temperature of the atomic gas in the galaxy  (e.g., \citealt{2020ApJ...900..131L,2022ApJ...927..152M}).
     %and in combination with \cii constrains the density and intensity of the radiation field in photon-dominated region models.
\item \textit{The [OIII] 88\,$\mu$m line}, which provides critical diagnostics for the conditions of star formation and the properties of massive stars (e.g., \citealt{2018ApJ...869L..22W,2019PASJ...71..109H,2019ApJ...881...63N}).
\item \textit{Several rotational CO transitions}: considered as the main probe of the molecular gas in dense star-forming regions \citep[e.g.,][]{2014MNRAS.445.2848G,2017ApJ...845..154V,2019ApJ...881...63N,2022A&A...662A..60D,2022ApJ...930...27L,2021A&A...652A..66P,2019ApJ...880..153Y}
     \item \textit{Several water lines} that arise in the warm, dense phase of the interstellar medium and \textit{a number of OH molecular lines} that are usually associated with outflows of dense material (e.g., \citealt{2019ApJ...880..153Y,2021A&A...652A..66P,2022arXiv220804335P}).
 \end{itemize}

 A combination of these different tracers yields molecular gas masses in the range $M_{\rm H2}\sim 10^{10}-10^{11}\,M_\odot$, and ISM metallicities comparable to the solar value. Generally, the observed luminosities are better modeled by photo-dissociation regions instead of X-ray-dominated regions (see the review by \citealt{2022ARA&A..60..247W}). The current small sample of quasars with multiple ISM tracers is highly biased towards objects bright in \cii. We expect that the characterization of physical conditions of the ISM in early massive galaxies will be expanded to much larger samples.

%  %Novak 2019 Pisco
%  \citealt{2019ApJ...881...63N}
% %Pensabene Water
% \cite{2022arXiv220804335P}
% %Decarli CO 76 65 with [CII] 
% \cite{2022A&A...662A..60D}
%entire sample in the range MH2 ∼ 1010-1011 M⊙.
%We compared the observed luminosities of dust, [C II], [C I], and CO(7-6) with predictions from photo-dissociation and X-ray dominated regions. We find that the former provide better model fits to our data, assuming that the bulk of the emission arises from dense (nH > 104 cm−3) clouds with a column density NH ∼ 1023 cm−2, exposed to a radiation field with an intensity of G0 ∼ 103 (in Habing units).
% %Similar Venemas CO 76 65 in z~7 
% \cite{2017ApJ...845..154V}
% %same CO 76 65 for z~6.6
% \cite{2022ApJ...930...27L}
% %Today, we count >50 quasars at z>6 detected in at least one ISM line, and some detected in more than 10 different tracers, including several fine structure lines ([CII], [CI], [NII], [OIII], [OI]) and multiple molecular line transitions (CO, H2O, OH). More than half of these quasars has been observed at sub-kpc angular resolution. 
% %
% %Meyer J1148 [CII] N[II] [Oi]
% \cite{2022ApJ...927..152M}
% %Walter [OIII]
% \citealt{2018ApJ...869L..22W}
% %Hashimoto [OIII]
% \cite{2019PASJ...71..109H}
% %Pensabene multi-line [NII] [CI] CO OH, H20
% \cite{2021A&A...652A..66P}
% %Yang CI OI CO  H20 in lensed quasar
% \cite{2019ApJ...880..153Y}
% %Li [NII] [OI] [Oiii]
% \cite{2020ApJ...900..131L}

\subsection{Constraints on Galaxy Mass and Black Hole/Galaxy Co-Evolution}

Dynamical masses of the quasar hosts can be measured using spatially and kinematically resolved line observations (see Section~\ref{sec:FIR}). The dynamical mass enclosed within a radius $R$ is typically expressed as:

\begin{equation}
\label{eq:dynmass}
    M_{\rm dyn}=\frac{v^2_{\rm circ}}{G}R,%=2.33 \times   10^5\,v^2_{\rm circ}\,R \label
\end{equation}

\noindent where $G$ is the gravitational constant and $v_{\rm circ}$ is the maximum circular velocity of the gas disk. Obtaining $v_{\rm circ}$ is not trivial, as it requires knowing the inclination angle of the galaxy.  Estimates of dynamical masses combined  with robust BH masses (Section~\ref{sec:bhm-measurements}) make it feasible to push BH/galaxy co-evolution studies to the highest accessible redshifts.

\cite{2010ApJ...714..699W} measured dynamical masses using CO emission from 8 quasar hosts at $z\sim 6$. Assuming an average inclination angle of $40^\circ$ and $R=2.5\,$kpc,  they found that the $z \sim 6$ quasars have, on average, SMBHs a factor 15 more massive than expected from the local BH -- bulge mass relation. This result suggests  that BHs in high redshift quasars either got a major head-start or grew faster than their host galaxies; if the M-$\sigma$ relation  existed at $z\sim 6$, would show a strong cosmic evolution. Numerous subsequent works have focused on the correlation of SMBH mass and galaxy dynamical mass at these redshifts (e.g., \citealt{2016ApJ...830...53W,2018ApJ...854...97D}). Most of these studies had to assume an inclination angle and/or an average size for the emitting gas.  \cite{2021ApJ...911..141N} carried out careful dynamic modeling of their \cii\ observations (including inclination and size). They found a mean dynamical mass of  $\sim 5\times 10^{10}$ M$_\odot$ for a sample of luminous $z\sim 6$ quasars with $\sim 10^{9}$ M$_\odot$  BHs. As shown in Fig.~\ref{fig:MdynMBH}, this places them about one order of magnitude above the local relation. It is important to note that the relationship could be strongly affected by potential biases from selection and by using gas tracers (\citealt{2018MNRAS.478.5063H,2011MNRAS.417.2085V}). Indeed, observations of low luminosity quasars show a narrower \cii\ line width (e.g., \citealt{2017ApJ...850..108W,2018PASJ...70...36I}), placing them close to the local relation.

\cite{2022MNRAS.511.3751H} analyzed the co-evolution of SMBHs and their host galaxies in six cosmological simulations with different models for SMBH growth. Although the simulations are all consistent at $z\sim 0$, they diverge at $z>5$, highlighting the importance of obtaining robust observational constraints at these redshifts. 
\cite{2022ApJ...931L..11L} argued that to robustly confirm whether the highest-redshift quasar population resides above the local BH - bulge mass relation, an improvement in the accuracy of mass measurements and an expansion of the current sample to lower black hole masses is required. We expect that JWST will enable significant advances for BH and galaxy host masses.

%Li 2022 -> Need better BH mass estimates and larger sample of lower MBH

%Habouzit -> discrepancy of simulations and the need to measure BH/host at low luminosity quasars
% \cite{2022MNRAS.511.3751H}

\subsection{Evidence of Quasar Feedback }
\label{subsec:feedback} 
Central accreting SMBHs play an important role in shaping galaxy evolution \citep[see review by ][]{2012ARA&A..50..455F}.
Indeed, to reproduce the observed distribution of galaxy masses at $z=0$, simulations require that strong AGN feedback was already in place at $z\sim 6$  (e.g., \citealt{2017MNRAS.467.4739K}). Below we list some of the  observational evidence (with caveats) that these feedback mechanisms are taking place in the $z\gtrsim 6$ quasar population:

{\em UV Line Shifts.} High-redshift quasars show asymmetric shape and velocity offset of high ionization lines, in particular the \civ\ line. The low-redshift quasar population shows an overall blueshifted \civ\ line ($\sim 800$ km s$^{-1}$) 
compared to the systemic redshift of the quasar \citep[e.g., traced by H$\beta$,][]{2011ApJS..194...45S}. This shift is generally understood in the context of a strong accretion disk wind that contributes to the high-ionization lines \citep[e.g.,][]{2011AJ....141..167R}. At $z>6$, the \civ\ line velocity shift is much stronger \citep[$\sim$ 1800 km s$^{-1}$,][]{2020ApJ...905...51S}. \cite{2019MNRAS.487.3305M} suggested that this redshift evolution can be explained by the \civ\ winds being launched from the disk with an increased torus opacity at this redshift.

{\em BAL Quasars.} A fraction of quasars show broad and highly blue-shifted absorption features in their rest-frame UV transitions. BAL features trace ionized winds in the broad line region and are recognized signatures of SMBH feedback. \cite{2022Natur.605..244B} used high-quality spectra of quasars from the XQR30 survey to show that up to $\sim 50\%$ of luminous quasars at $z\sim 6$ exhibit BAL features (with outflow velocities up to 17\% of the speed of light), compared to about 20\% observed in low-redshift samples. 
\cite{2021ApJ...923..262Y} studied 37 quasars at $6.3<z<7.64$ and reported a BAL fraction of $\sim 24\%$, smaller than the \cite{2022Natur.605..244B} work but still slightly larger than what is observed at lower redshifts. 
This potential evolution of the BAL fraction could be a result of the strong feedback associated with the rapid BH growth and galaxy assembly in the early Universe. 

{\em Ly$\alpha$ halos.} The existence of extended \lya\ nebulae around $z\gtrsim 6$ quasars was discussed in Section~\ref{sec:qso-fueling}. Recently, \cite{2022arXiv220311232C} performed a suite of cosmological, radiation-hydrodynamic simulations to understand the origin and properties of the observed Ly$\alpha$ halos. 
The simulations unambiguously require quasar-powered outflows to match the observed properties at $z\gtrsim 6$, providing indirect evidence for AGN feedback. 

{\em Radio jets}. Only six galaxies hosting radio-loud  quasars at $z\gtrsim 6$ have been studied with ALMA and NOEMA (\citealt{2021ApJ...920..150R,2022A&A...664A..39K}).  
 %These six host galaxies show a diverse range of properties, and---
Assuming no AGN contribution, their, their \cii- and FIR-derived star-formation properties are consistent with those reported for the much more studied radio-quiet quasars (c.f., Decarli et al.\ 2018, Venemans et al.\ 2020; see Fig.~\ref{fig:sfr-distribution}). However,  \cite{2021ApJ...920..150R} and \cite{2022A&A...664A..39K} show indirect evidence that the FIR emission of radio-loud quasars can be strongly affected by synchrotron emission. In that case, their IR-derived star formation rates can be overestimated, implying that  we might be witnessing negative AGN feedback at $z>6$. However, more measurements of the FIR continuum and/or an enlarged sample are required to quantify the potential impact on the population.

{\em Broad components of [CII] emission.} Broad wings ($\gtrsim 1000\,\kms$) in the \cii\ emission lines are thought to be caused by AGN outflows. The observational evidence for these \cii\ broad wings remain tentative, as different analyses of similar datasets provide inconsistent results. For example, \cite{2012MNRAS.425L..66M} and \cite{2015A&A...574A..14C} reported a strong \cii\ outflow in the host galaxy of the quasar J1148+5251 at $z=6.42$.  \cite{2022ApJ...927..152M}, on the other hand, found no evidence of a broad velocity component but reported that J1148+5251 has the most spatially extended \cii\ emission ($\sim 10\,$kpc) among $z\sim 6$ quasar hosts. \cite{2021ApJ...914...36I} report a broad ($\sim1000\,\kms$) \cii\ wing in the spectrum of a quasar at $z=7.07$, and \cite{2022A&A...664A..39K} reported a tentative \cii\ broad component that could be as wide as $\sim1400\kms$ in a BAL radio-loud quasar at $z=6.12$.  \cite{2019A&A...630A..59B} stacked the \cii\ spectra of 48 quasars at $4.5<z<7.1$ and reported a broad \cii\ component, while \cite{2018ApJ...854...97D} stacked the \cii\ spectra of 23 $z\sim 6$ quasar hosts and found no evidence for a \cii\ broad component. \cite{2020ApJ...904..131N} stacked the \cii\ spectra using different techniques (spectral stacking and $uv$-plane stacking) and found no evidence  for \cii\ broad-line emission. \cite{2020ApJ...904..131N} argued that the results can depend on the stacking techniques and resolution (e.g., if the resolution is low nearby companions can be confused with \cii\ wings).  

{\em OH absorption.} The hydroxyl molecule (OH) in absorption traces high-velocity molecular inflow or outflows (see review by \citealt{2020A&ARv..28....2V}). 
\cite{2020A&A...633L...4H} report the first tentative detection ($3\sigma$) of the OH 119\,$\mu$m doublet in absorption towards a $z\gtrsim 6$ quasar, suggesting the presence of a molecular outflow. Outflow signatures from both atomic and molecular lines will be a focus of future high-quality ALMA observations. 

\begin{figure}[h]
\includegraphics[width=0.9\textwidth]{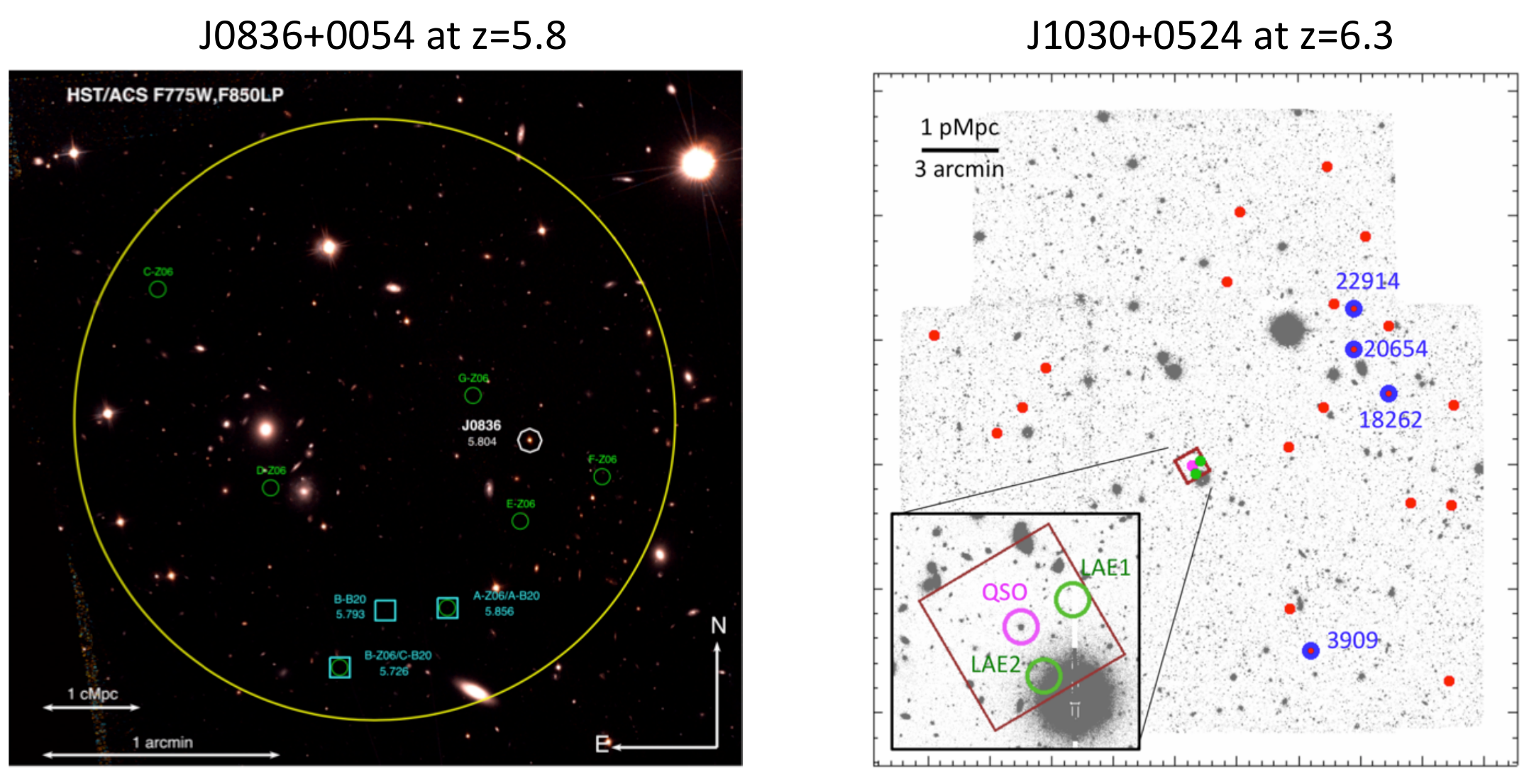}
\caption{The two $z\gtrsim 6$ quasars with robust overdensities on $\gtrsim 1$\,Mpc scales.  \textit{Left:} Field around J0836+0054 at $z=5.8$ (figure from \citealt{2022ApJ...926..114O}). Cyan squares (green circles) show spectroscopically-confirmed (photometrically-selected) galaxies at a redshift consistent with the quasar. 
\textit{Right:} Field around J1030+0524 at $z=6.3$ (figure from \citealt{2020A&A...642L...1M}). Red  dots show photometrically selected galaxies at a redshift consistent with the quasar. Blue and green circles are spectroscopically confirmed galaxies at a redshift consistent with the quasar. 
}
\label{fig:qsoenv}
\end{figure}

\subsection{Protoclusters and Large Scale Structure } \label{subsec:protoclusters}
%costa 2014 2014MNRAS.439.2146C
%mignoli 2020A&A...642L...1M

Theoretical models predict that $z\gtrsim 6$ quasars should be highly biased tracers of  the underlying dark matter distribution, signposting the first overdensities of galaxies on Mpc scales, i.e., protoclusters (e.g., \citealt{2009MNRAS.394..577O,2014MNRAS.439.2146C}; but some scatter is expected, see e.g., \citealt{2021ApJ...917...89R}). Observationally, this has been challenging to demonstrate. Photometric selection around high-redshift quasars has found evidence for overdensities \citep[e.g.,][]{2010ApJ...721.1680U,2014A&A...568A...1M}, densities comparable to random fields \citep[e.g.,][]{2013ApJ...773..178B,2014MNRAS.442.3454S,2017ApJ...834...83M}, and even underdensities \citep[e.g.,][]{2009ApJ...695..809K}. 

There are currently two $z\gtrsim 6$ quasar fields with robust overdensities (i.e., including spectroscopic confirmation) on Mpc scales (\citealt{2022ApJ...926..114O,2020A&A...642L...1M}; see Fig.~\ref{fig:qsoenv}). 
These are SDSS J1030+0524  at $z=6.3$ and SDSS J0836+0054 at $z=5.8$. 
%In both cases, it took several years to obtain the necessary datasets to confirm the overdensities. 
The works by \cite{2005ApJ...622L...1S,2009ApJ...695..809K,2014MNRAS.442.3454S,2017A&A...606A..23B,2019A&A...631L..10D} were crucial for confirming the overdense environment of J1030+0524. Similarly, \cite{2006ApJ...640..574Z,2006PASJ...58..499A,2020ApJ...896...49B} were fundamental for confirming the large-scale structure around J0836+0054. Recently, \cite{2021ApJ...921L..27Y} report the first quasar pair known at $z\sim 6$ (with a projected separation $<10\,$kpc), implying a rich environment that still awaits confirmation at larger scales.

ALMA observations of $z\sim 6$ quasars hosts (see Section~\ref{sec:FIR}) have serendipitously detected a population of \cii-bright companion galaxies in the immediate environment of $\sim$20-50\% of the targeted quasars (\citealt{2017Natur.545..457D,2019ApJ...882...10N,2020ApJ...904..130V}; see also bottom panel of Fig.~\ref{fig:almahosts}). The large  fraction of such quasar-galaxy pairs exceeds by orders of magnitude the expectations based on the current constraints of the number density of \cii-bright galaxies at these redshifts.  Three out of five quasar companions with deep HST observations remain undetected \citep{2019ApJ...881..163M,2019ApJ...880..157D}, implying a significant part of a potential large-scale structure might be obscured.   The two HST-detected companions show tentative evidence of AGN activity (\citealt{2019ApJ...887..171C,2019A&A...628L...6V}; but see also \citealt{2021A&A...649A.133V}).

% What remains unclear is whether these close companions are a signature of a protocluster or the mechanisms to grow  these supermassive black holes. 
The field of view of ALMA is too small to study whether these gas-rich companions exist in large numbers over Mpc scales as predicted by simulations. MUSE observations of one of these quasar/companion fields reveal two additional LAEs in close proximity, strengthening the case for an overdensity \citep{2022ApJ...927..141M}. 
It is likely that the next leap on our understanding of the environment of the highest-redshift quasars will come with JWST. JWST deep imaging, multi-object, and slitless spectroscopy capabilities will provide a new opportunity to probe the galaxy population in $z\gtrsim 6$ quasar environments to unprecedented depth over the  Mpc scales.

\section{QUASARS AS PROBES OF COSMIC REIONIZATION}
\label{sec:reionization}
%%% 11 Pages Target %%%%%
Quasars are both agents, and powerful probes of the cosmic reionization history. As rare but luminous sources with hard ionizing spectra, they contribute to the overall photon budget driving reionization, together with stellar photons from galaxies. 
%\cite{2021arXiv211013160R} reviewed in detail the contributions of star forming galaxies and quasars/AGN to the reionization photon budget, and concluded that radiation from quasar/AGN is a minor contributor based on the current QLF measurements (see discussions in Section~\ref{subsec:QLF}.  \cite{2022NatAs...6..850J} further confirmed this point by placing upper limit on the contributions of low luminosity AGN to reionzation using deep HST survey data. On the other hand quasar/AGN will dominate the UV background in the IGM at lower redshift ($z<4$) in a post-reionization universe with their rapidly rising density towards lower redshift. 
%Fundamentally however, reionization concerns the response of %diffuse matter to emergent radiation from galaxies and AGN. 
Absorption measured in quasar spectra from foreground gas has yielded many of the most sensitive  constraints on the density, ionization, and chemical enrichment of this tenuous intergalactic material. While quasars' utility as reionization probes has been understood since foundational work by \citet{1965ApJ...142.1633G}, it has taken 40-50 years for quasar surveys to uncover objects deep into the reionization epoch and fully exploit the information that their absorption spectra encode \citep{2006AJ....132..117F}. 

While absorption studies have yielded many of the most precise constraints on the IGM opacity and neutral hydrogen fraction, they are subject to limitations on both systematic accuracy, and physical interpretation. These limitations arise from multiple sources, but the most significant factors discussed below are (a) uncertainty in the quasar's intrinsic/unabsorbed spectral continuum, (b) the large on-resonance oscillator strength of \hi\,\lya, which leads to a wide gap in sensitivity between neutral fractions of $\xhi \sim 10^{-4}$ (where resonance absorption saturates) and $\sim 10^{-1}$ (where damping wings appear), and (c) computational challenges of simulating representative volumes with inhomogeneous and rapidly evolving radiation fields. 
%For a recent review of modeling and simulation of reionziation, see \cite{2022arXiv220802260G}. ]
%\vspace{-2cm}
\subsection{AGN and the Ionizing Photon Budget}

Quasar accretion produces a hard ionizing spectrum, making the AGN population a contributing source to the budget of photons needed for reionization. However as detailed in the recent review by \citet{2021arXiv211013160R} and references therein, there is an emerging consensus that AGN play a subdominant role compared to galaxies as an overall driver of the phase transition, even when one extrapolates observed luminosity functions to thresholds fainter than current detection limits.  
\cite{2022NatAs...6..850J} further reinforced this conclusion by placing an upper limit on the contributions of low luminosity AGN to reionization using deep HST survey data. \cite{2022arXiv220702233F} have synthesized measurements across the lierature of galaxy+AGN UV luminosity functions, to estimate their relative contributions to the ionizing photon budget, using strong assumptions about the ionizing radiation escape fractions of each population. They estimate that AGN supply $\sim 10\%$ of the total ionizing photon budget at $z\sim 5$, roughly the lower limit of redshifts considered in our review. They found that this fraction drops rapidly toward higher redshift because of differential downward evolution in the quasar luminosity function relative to galaxies.   By $z\sim 7$ they estimate that AGN contribute only $0.3-1.0\%$ of all ionizing photons, declining by yet another order of magnitude by $z\sim 9$.  In contrast, \cite{2019ApJ...884...19G} suggested that AGN contributions to the UV background can still be significant at $z\sim 5.6$ based on X-ray selected objects.

The quantitative uncertainties in these analyses are still considerable; \cite{2019ApJ...879...36F} found a $10\times$ higher contribution from AGN at $z\sim 7$ using different assumptions.  Much of this may be attributed to uncertainties in the faint end slope of the quasar luminosity function. Still, the qualitative picture that reionization is heavily dominated by star-forming galaxies appears to remain fairly robust with respect to these uncertainties.  The fractional contribution of quasars and AGN increases toward lower redshift because of their rapidly rising number densities, such that they are more likely to dominate the meta-galactic UV background in the at $z<4$ post-reionization universe. 
New surveys with the Nancy Grace Roman Space Telescope and JWST should revolutionize this field, providing both deep, near-IR color selected objects deep into the EoR, and also sensitive spectra to confirm the AGN nature of faint candidates and separate them from star-forming galaxies.   

\subsection{Neutral Hydrogen Absorption in the Diffuse IGM}

\begin{figure}[t]
\includegraphics[width=5.0in]{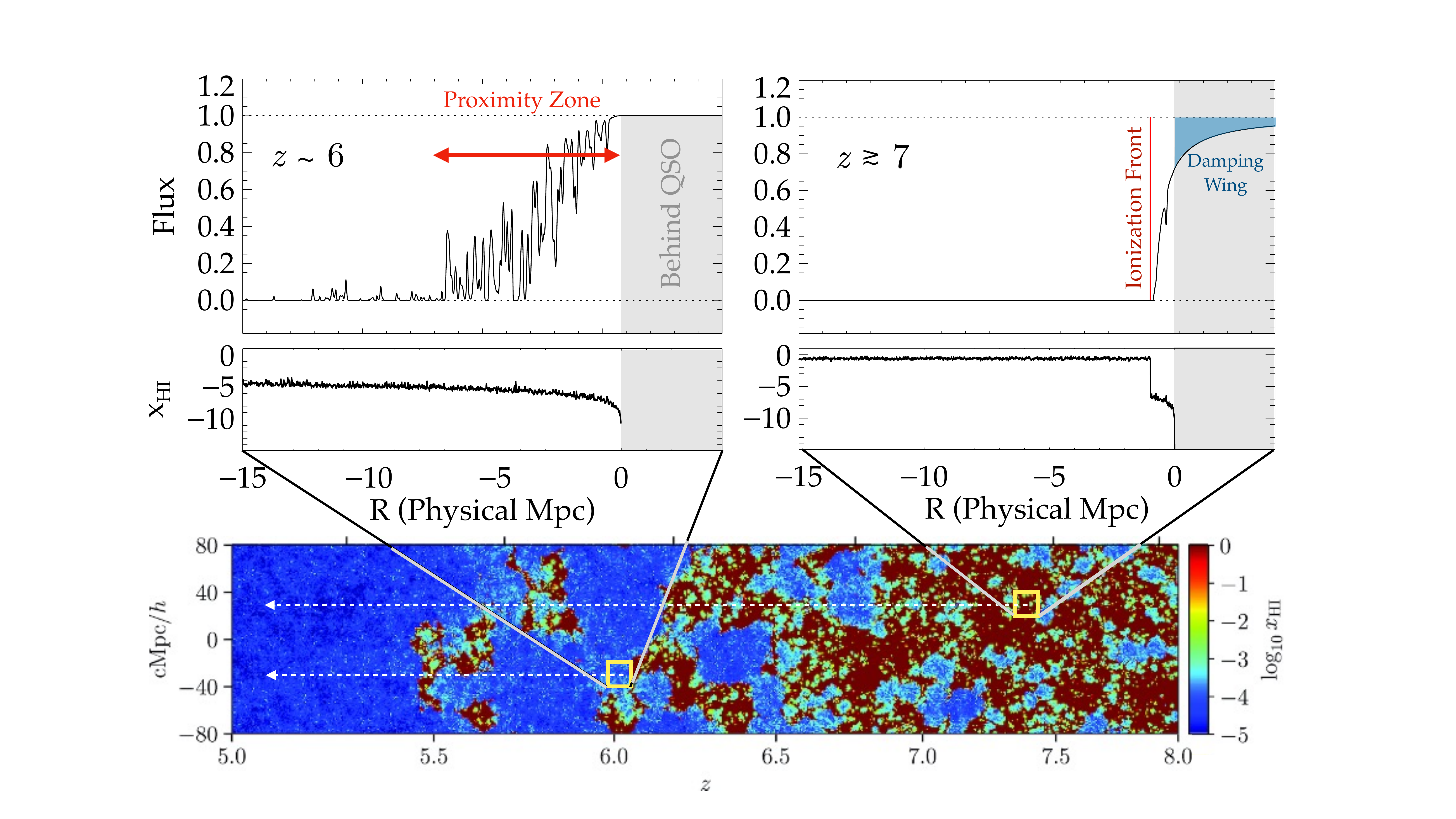}
\caption{Observable properties and interpretation of Proximity Zones (upper left) and Damping Wings (upper right).  In the simulation panel (bottom), each box represents the location of a background quasar whose sightline extends to the left.  Proximity zones are studied at $z\sim 5.8-6.5$ and represent regions of enhanced ionization from the quasar in an already-ionized IGM.  There is no absorption redward of the quasar's Ly$\alpha$ emission line (marked with a gray shaded region), and a region of low optical depth extends $R_p\sim 5-10$ proper Mpc.  Damping wings have only been detected at $z\gtrsim 7$, and represent ionization fronts penetrating into an IGM with a $O(\sim 0.1)$ neutral fraction, shown with red color scale in the simulation. On-resonance absorption is saturated up to and very near the quasar's emission redshift, and the off-resonance Damping Wing (shaded blue) extends redward of Ly$\alpha$ in the quasar's rest frame. Constraints on the IGM neutral fraction come from this blue shaded zone, as well as the very short run of unsaturated pixels between the QSO's systemic redshift and the end of the proximity zone.}
\label{fig:proximity}
\end{figure}

The optical depth from neutral Hydrogen in a matter-dominated universe (appropriate for the redshifts considered here) is:
\begin{equation}
\label{eqn:tau}
   \tau(\lambda_{obs}) = {{\pi e^2  f_{\rm 12}}\over{m_e c}} \left( {{{c}\over{H_0\Omega_M^{1/2}}} \int_{0}^{z} {{n_{\rm HI}(z)\over{(1+z)^{5/2}}} \phi(c/[\lambda_0(1+z)])dz}} \right)
\end{equation}
The leading fraction has units of cross section per unit frequency in terms of the dimensionless oscillator strength $f_{12}$, which is an atomic constant determined for each ion ($f_{12}=0.416$ for Ly$\alpha$).  The expression in brackets reflects a column density, i.e. an integral of the radial line element, weighted by the \hi\  density field. The function $\phi(\nu)$ is a unity-normalized line shape. It is typically approximated as a Voigt profile, which consists of a deep Gaussian line core whose width is determined by internal velocities, convolved with a much wider but also much weaker Lorentzian wing.  This so-called ``damping wing" arises from an energy-time uncertainty in the \lya\ transition, leading to a small but non-zero absorption cross section in the rest frame for photons far from the resonance wavelength $\lambda_{\rm rest}=1215.67$\,\AA.  Equation \ref{eqn:tau} indicates that the optical depth at each observed wavelength $\lambda_{\rm obs}$ contains contributions from neutral hydrogen at every intervening foreground redshift. 

However because $\phi(\nu)$ is so sharply peaked around line resonance, the integral is almost always dominated by matter at the redshift where $\lambda_{\rm obs}/(1+z)=\lambda_{\rm rest}$. Directly at this resonance wavelength, the \lya\ cross section is very large---roughly $4\times 10^{-14}$ cm$^2$. Moreover the mean baryonic density increases toward high redshift as
\begin{equation}
    \bar n_H(z)\sim {{\Omega_b \rho_c}\over{\mu m_H}} (1+z)^3 = 8.6\times 10^{-5} \left({{1+z}\over{7}}\right)^3 {\rm cm}^{-3},
\end{equation} 
which increases both the total gas column density, and also the \hi\  recombination rate, favoring a higher neutral gas fraction $\xhi$.  Collectively these lead to a very large \lya\ optical depth of $\tau\sim 10^5$ if the IGM is 100\% neutral. 

However, observations show $\tau=O(\sim 1)$ at $2<z<4$; the presence of transmitted flux implies that the \lya\ forest is only neutral at the $\xhi \sim 10^{-5}$ level. The optical depth increases toward high redshift in tandem with the neutral fraction, but by the time the neutral fraction reaches $\xhi \sim 10^{-4}$, the optical depth already exceeds $\tau>5$, at which point $>99\%$ of the background quasar's light is absorbed.  On-resonance line absorption is simply too strong to provide useful measurements of the neutral fraction $\xhi$ above this value.

In contrast, the atomic cross section for a photon with wavelength in the Lorentzian damping wing is suppressed by 5--6 orders of magnitude relative to the peak at line center---even at small velocity offsets (100--200$\,\kms$).  The eponymous Damped Lyman Alpha (DLA) phenomenon is therefore only seen in small, rare high-overdensity environments with large column densities ($N_{\rm HI}>10^{20}$) at $z<5$. 

Reionization is characterized by an increasing neutral fraction in the diffuse IGM, which eventually becomes strong enough to produce damping wings \citep{1998ApJ...501...15M}. However these are convolved with on-resonance absorption from \hi\ at other redshifts, with the unique exception of the quasar's immediate foreground, where the damping wing can extend redward of the \lya\ emission line.

Bridging this gap between strong on-resonance absorption, whose measurements probe $10^{-6}< \xhi < 10^{-4}$, and weak off-resonance damping wings which only emerge at $\xhi > 10^{-1}$ is a recurring theme for IGM measurements near the reionization epoch. On-resonance measurements are most useful in regions of locally enhanced ionization or at lower redshifts ($z<6.3$) toward the late stages of reionization.  Damping wing measurements are most powerful at early times and higher neutral fractions seen for $z>7$. Observational constraints are much less precise for the epochs in the range $6.3 < z < 6.8$, when the neutral fraction evolved from $0.01\%$ to $10\%$--- the \lya\ line absorption is saturated, but damping wings have not yet emerged.

\subsubsection{Numerical Simulations}

Cosmological simulations play an essential role in interpreting the spectra of reionization-epoch quasars, and placing their associated transmission measurements into a broader physical context. We refer the reader to an excellent recent review \citep{2022arXiv220802260G} on theoretical and numerical reionization models for a comprehensive overview of key methods and simulation projects.

Briefly, simulations must confront several intersecting challenges in the attempt to reproduce the quasar spectral observables summarized below, together with luminosity functions of reionization-era galaxy populations.  An idealized simulation should have:
\begin{itemize}
    \item A large box ($L>100$ cMpc) to cover many scale lengths of the galaxy-galaxy auto-correlation function, and to sample the rare high-density peaks that could plausibly host $M>10^9 M_\odot$ SMBHs.
    \item Sufficient spatial and mass resolution to model disk galaxies realistically and resolve low-mass halos that produce light and heavy elements, as well as small-scale features in the \lya forest and Lyman Limit systems that act as ionizing radiation sinks.
    \item Ray tracing of emergent radiation from individual galaxies, to account for spatial variations in the density and ionizing background, shadowing from optically thick absorbers, or the timing of reionization.
    \item Full and self-consistent coupling of radiation to hydrodynamics, to capture the effect of ionization heating, pressure smoothing and adiabatic cooling on the evolving density field and ionization fraction. These will also impact the formation of subsequent stars and galaxies.
\end{itemize}
Computational limits prevent practical simulations from meeting all four these requirements simultaneously, though rapid progress in recent years has allowed some groups to progress from analytic or semi-numerical efforts to meet two or even three of the above bullets in single simulations. The Gnedin \& Madau review contains a helpful classification of leading numerical projects according to their box sizes, hydro prescriptions, treatment of the UV background, and radiation-hydro coupling.

While most large-scale simulations focus on reproducing global reionization signals such as the volume-averaged neutral fraction, a separate numerical toolkit has evolved specifically to interpret observations of proximity zones (Sec~\ref{subsec:proximity}) and damping wings (Sec~\ref{subsec:dampingwings}) produced in the immediate foreground of luminous quasars.

These investigations begin with global simulations of the reionization-era IGM, and introduce the effect of a high-luminosity quasar by hand in the simulation volume.  
%A number of these simulations make the simplifying assumptions of a spatially-uniform ambient UV background from field galaxies (prior to the quasar's active phase), and decoupling of radiative transfer from the hydrodynamic physics.  The latter assumption is motivated by the rapid time scale on which ionization fronts or enhancements are driven into the IGM---much faster than the time scale for hydrodynamic response.  Model fidelity is improved when spatial variations in the UV background from clustered sources is treated in detail; however this places additional demands on the baseline simulation and is also often omitted.
A large parent simulation is searched for high-mass halos ($M\ge 1-2\times 10^{11}M_\odot$) representative of hosts where SMBHs are first thought to form. A bright quasar is then inserted at the center of the halo, emitting ionizing photons with a luminosity scaled to the absolute 1450\AA\ magnitude of known quasars, and extrapolated to the Lyman edge using carefully constructed template SEDs. The simulation is then evolved forward to follow expansion of the ionization front or proximity zone into the IGM.  Multiple sightlines are drawn through the simulation box, terminating at the quasar's host halo, and simulated transmission spectra are extracted using the density, temperature, ionization and velocity cubes.

This methodology has been adapted by several groups to place joint constraints on quasar lifetimes and the IGM neutral fraction at $z>7$ \citep{2011MNRAS.416L..70B,2015MNRAS.454..681K,2018ApJ...864..142D,2000ApJ...542L..75C}, and to estimate the ages of $z\sim 6$ quasars from their proximity zone sizes \citep{2017ApJ...840...24E,2007MNRAS.374..493B,2021ApJ...911...60C}.  In ionized proximity zones, the simulations can also be used to calibrate methods to reconstruct the fluctuating density field near the central host \citep{2022ApJ...931...29C}.   

Essentially the simulations are used to generate statistical realizations of the matter distribution around massive quasar host halos, capturing non-linear physics and scale-dependent correlations in a way that analytic PDFs of the optical depth and/or excursion set formalism  cannot accomplish. 

A completely separate class of simulations is beginning to address challenging questions about heavy element enrichment of the early IGM and CGM \citep{2016MNRAS.461..606K}, for comparison with observation of heavy element absorption in $z>6$ QSO spectra. These simulations rely on subgrid feedback prescriptions for metal production, which are then transported with standard hydro treatments as at lower redshift. Translation from metal abundances into observable column densities is more complicated than for HI, because the ionization balance of multi-level ions depends on the UV background at other, often harder energies than the Lyman edge. A notable early effort in multi-frequency treatment of an inhomogeneous background is the Technicolor Dawn simulation \citep{2018MNRAS.480.2628F}, which tracks 24 separate frequencies and measures ionization balances of C IV and Si IV in the $z>5.5$ universe, though with a much smaller box size ($L\sim 12$ cMpc) than the aforementioned simulations designed to capture quasar host halos. 

\subsubsection{Transmission measurements}

%\begin{figure}[t]
%\includegraphics[width=4in]{Yang2020_tau.jpeg}
%\includegraphics[width=3in,angle=-90]{fig/fig_tauevo.pdf}
%\caption{(Left): Evolution of the Ly$\alpha$ optical depth with redshift, adapted from %\citet{2020ApJ...904...26Y}. (Right) Evolution of the mean free path for ionizing photons, illustrating the rapid evolution approaching $z\sim 6$. Adapted from %\citet{2021MNRAS.508.1853B}.}
%\vspace{-1cm}
%\label{fig:tauevo}
%\end{figure}

\begin{figure}[ht]
 \begin{minipage}{0.45\linewidth}
 \vspace{0cm}
\includegraphics[width=2in,angle=-90]{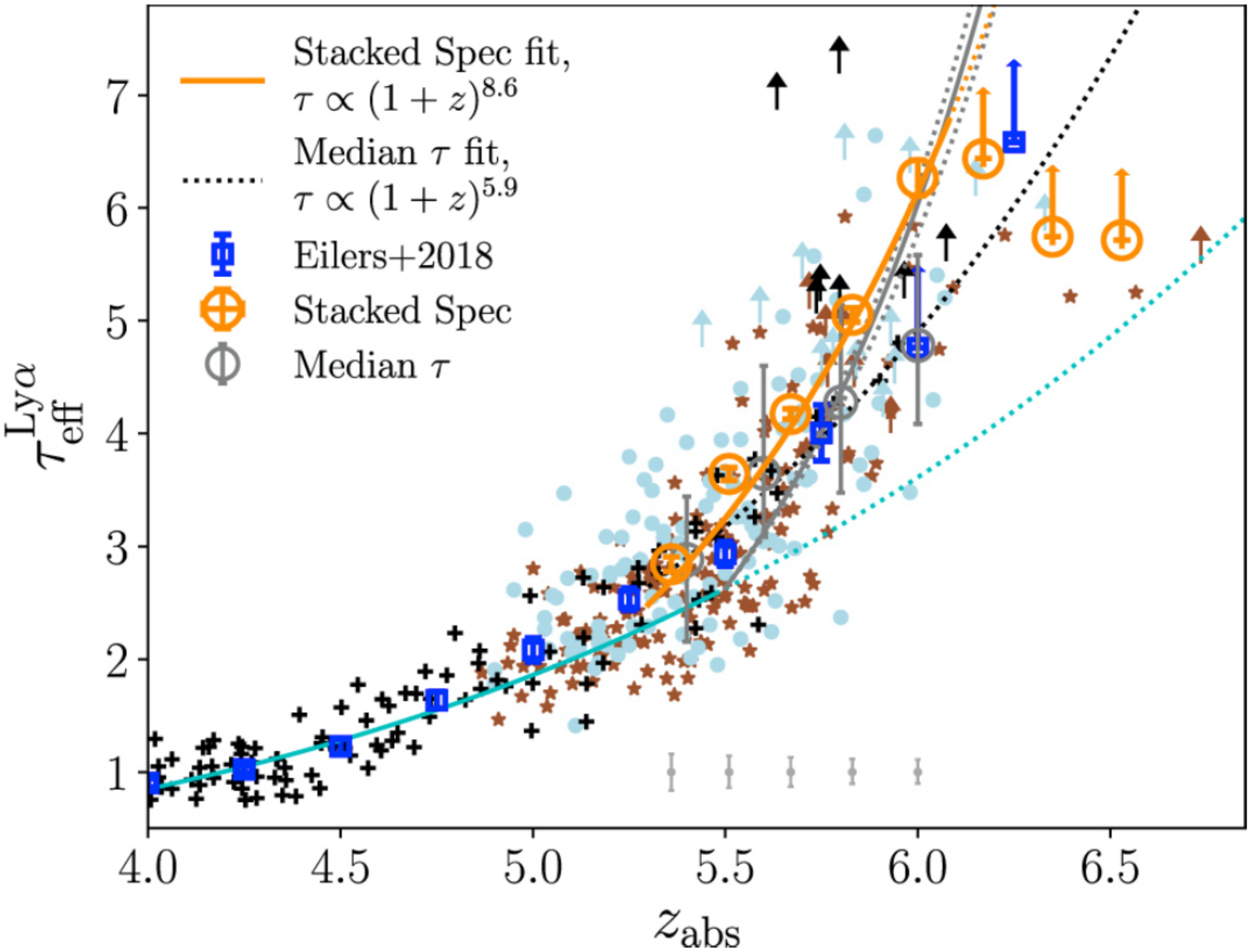}
 \end{minipage} 
\begin{minipage}{0.45\linewidth}
\hspace{-4cm}
\includegraphics[width=2.4in]{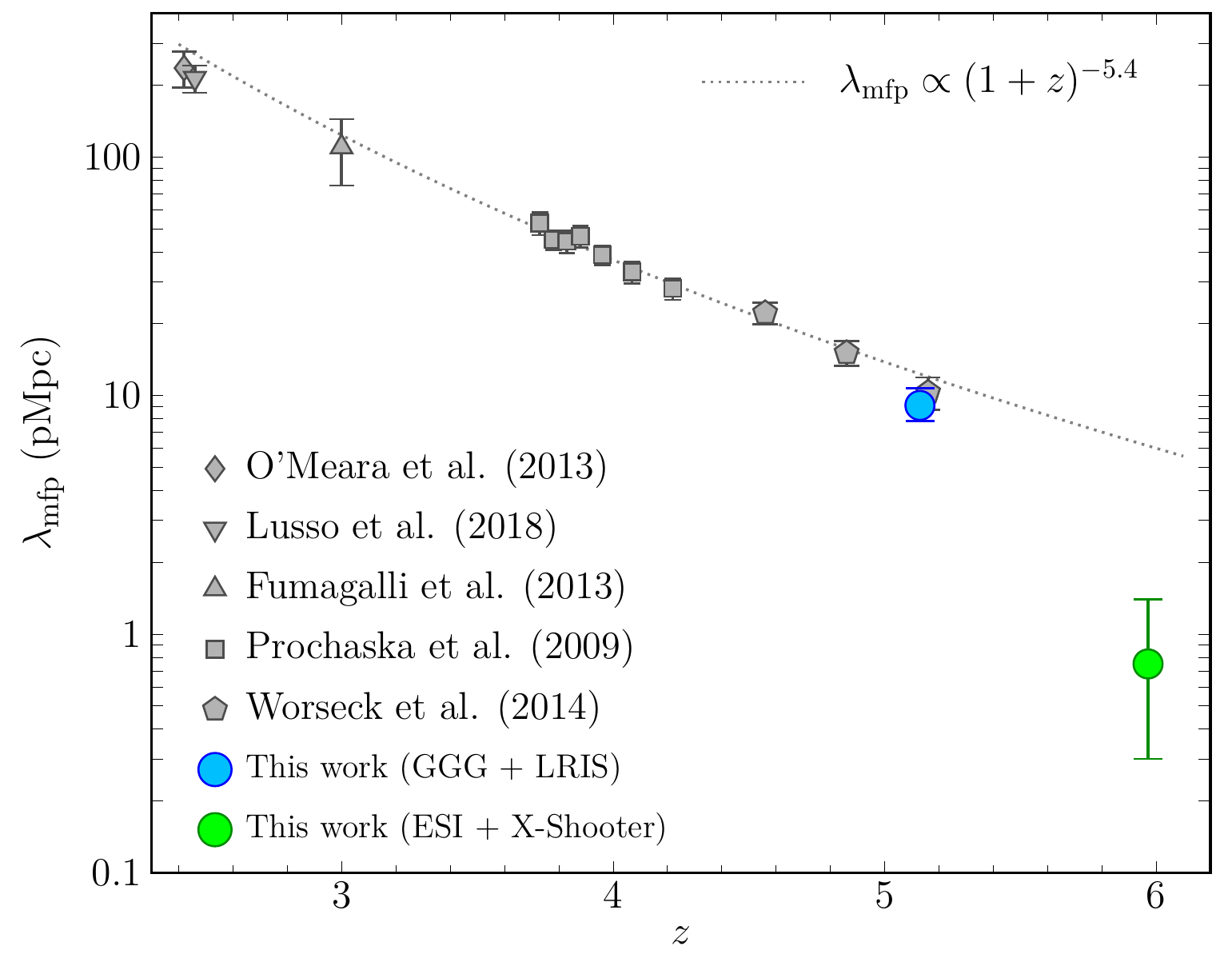}
\end{minipage}
%\caption{Evolution of the mean free path for ionizing photons, illustrating the rapid evolution approaching $z\sim 6$. Adapted from \citet{2021MNRAS.508.1853B}.}
\caption{(Left): Evolution of the Ly$\alpha$ optical depth with redshift based on spectroscopy of large quasar samples, adapted from \citet{2020ApJ...904...26Y}.
The optical depth increases smoothly towards high redshift at $z<5.5$, then appears to evolve more strongly, with the emergence of complete Gunn-Peterson absorption troughs (upper limits). 
(Right) Evolution of the mean free path for ionizing photons, illustrating the rapid evolution approaching $z\sim 6$. Adapted from \citet{2021MNRAS.508.1853B}.}
%\vspace{-1cm}
\label{fig:tauevo}
\end{figure}

Before the discovery of $z\gtrsim 6$ quasars, the evolving \lya\ forest optical depth, which gradually increases with $z$, had been measured extensively at lower redshifts \citep{1997ApJ...489....7R,2007ApJ...662...72B} and interpreted as a product of the $\bar{n}\propto(1+z)^3$ density field together with a smoothly evolving UV ionizing radiation background \citep{2013MNRAS.436.1023B}. 
%Early observations with the Sloan Digital Sky Survey uncovered a population of $z>6$ quasars using $(i-z)$ dropout selection that led to a reexamination of underlying assumptions about smooth evolution in $\xhi$ approaching reionization \citep{2001AJ....122.2833F}. 
Early discoveries of $z>6$ SDSS quasars \citep{2001AJ....122.2833F} led to a reexamination of underlying assumptions about smooth evolution in $\xhi$ approaching reionization.
In particular, deep medium-resolution spectroscopy of one of very first such objects --- SDSS J1030+0524 at $z=6.3$ --- revealed a region of Lya forest absorption consistent with zero flux at wavelengths corresponding to $5.95<z<6.16$ \citep{2001AJ....122.2850B}.  This work has been interpreted as the first detection of a robust Gunn-Peterson absorption trough, and has been followed by many similar analyses to improve how the IGM opacity is quantified \citep{2006AJ....132..117F,2020ApJ...904...26Y,2018MNRAS.479.1055B,2018ApJ...864...53E,2007ApJ...662...72B,2002AJ....123.2151P}.

The simplest absorption indicator from a conceptual standpoint is the \lya\ opacity, $\tau = -\ln(I/I_0)$, where $I_0$ represents an estimate of the unabsorbed continuum flux level. Deeper spectra of SDSS J1030+0524 and many subsequently discovered objects consistently uncover similar spans of total absorption in \lya\ resonance line at $z>6$, implying a large lower bound on $\tau$ limited only by the signal-to-noise ratio (SNR) of the spectral data. This highlights an important challenge in such measurements: the physical information is contained in measuring flux that is {\em not} seen. The logarithmic nature of optical depth means that even in spectra with SNR$\sim 100$ one can measure limits of $\tau \gtrsim 4$ (per pixel) at best, and the zero points can be affected by systematic errors in data reduction, including accurate subtraction night sky emission and correction for telluric absorption. Comparison of observations of the same quasars observed with different instruments and reduced with different pipelines indicate discrepancies at the 1--2\% flux level, for a mean transmitted flux of 0.5--1.0\% at $z\sim 6$ \citep{2011MNRAS.415.3237M,2018ApJ...864...53E}. 

Additional dynamic range can often be achieved by observing Ly$\beta$ or higher order transitions from the Lyman series. These have progressively decreasing oscillator strengths and require correspondingly larger \hi column to saturate, allowing access to larger values of $\tau$. This advantage is offset by complications arising from unknown blending with \lya\ forest absorption at lower redshift, and uncertainty in extrapolating the intrinsic quasar continuum from \lya\ to Ly$\beta$.

Measurements of the \lya\ opacity are always smoothed to suppress noise and minimize the effect of differing spectral dispersion and resolution among instruments used in the observations. Following \citet{2006AJ....132..117F}, most early studies have used observationally-motivated redshift bins of $\Delta z=0.15$ for ease of comparison, definining the "effective optical depth" $\tau_{\rm eff}=-\ln \left< F \right>$ over the smoothing interval. More recent studies advocate for a physically-motivated bin size of 50 cMpc, because a constant redshift interval evolves by $\sim 40$\% in comoving length over $5<z<7$ where such measurements are made \citep{2018ApJ...864...53E,2015MNRAS.447.3402B}. The two values ($\Delta z = 0.15$ and 50 cMpc) are comparable at $z\sim 7$.  Fig~\ref{fig:tauevo} (left panel) presents a recent summary of \lya\ optical depth measurements. 

Average $\tau$ measurements are simply the first moment of an underlying optical depth distribution, which contains richer information and has therefore been measured extensively and presented as a cumulative distribution function (CDF) of optical depth. Evolution of the CDF is characterized by three phases, again reflecting the logarithmic nature of the measurements.  At $z\sim 5.0-5.3$ the distribution is fairly narrow and centered around $\tau\sim 2$, corresponding to a \lya\ forest with significant ($\sim 90\%$) absorption at line resonance, but not yet fully saturated.  At $5.5<z<6.0$ the CDF widens to span $2<\tau<6$, as an increasing number of smoothed windows exhibit saturated absorption (i.e. zero transmitted flux or high $\tau$), but many significant transmission/low neutral fraction windows remain.  Above $z>6$ it becomes increasingly rare to find unsaturated windows with $\tau < 4$, and the distribution once again narrows because spectral SNR renders the $\tau > 6$ regime inaccessible.  

Several groups have explored alternative ionization metrics that use counts of transmission spikes in the Lya forest, motivated by the many challenges of measuring optical depths in realistic, heavily absorbed data.  \citet{2019ApJ...876...31G} constructed distribution functions $d^2N/dLdz$, capturing the density of spikes per cMpc and unit redshift, above a threshold line flux.  These can be counted in spectra of varying resolution and SNR using similar techniques as are used for blind searches for metal absorption lines in quasar spectra.  Their analysis indicates that, after accounting for instrumental broadening and noise which suppress the spike count in realistic observed $5<z<6$ spectra by an order of magnitude, the observed spike counts are roughly consistent with their self-consistent simulations of the post-reionization IGM transmission. In the simulation boxes, transmission spikes arise from regions that are under-dense in hydrogen and galaxies, but over-ionized relative to expectations from equilibrium with the average radiation field, likely because of proximity to local ionizing sources.

At $z>6$ one expects transmission spikes to be exceedingly rare, so the detection even of single spikes implies the existence of early ionized cavities.  Searches for spikes in high-redshift spectra have yielded several detections at $6<z<6.3$ that provide upper bounds on the local optical depth, complementing the lower bounds derived from measurements of saturated spectral regions \citep[e.g.,][]{2020ApJ...904...26Y}.

%\begin{figure}[h]
%includegraphics[width=4in]{Becker21_mfp.pdf}
%\includegraphics[width=3.5in]{fig/fig_mfp.pdf}
%\caption{Evolution of the mean free path for ionizing photons, illustrating the rapid evolution approaching $z\sim 6$. Adapted from \citet{2021MNRAS.508.1853B}.}
%\label{fig:mfp}
%\end{figure}

A very different way to conceptualize reionization is as a rapid increase in the mean-free-path (mfp) of ionizing photons. Although this is physically equivalent to an increase in the volume-averaged ionization fraction, this alternative framing allows a different analysis approach that does not require high-SNR measurements of localized pixels or spikes.  Instead, samples of quasar spectra are stacked after shifting into the rest frame. Then a low-dimensional parametric model is fit to the shape of the stacked spectrum below the Lyman limit at $912$\AA, corresponding to a probabilistic flight length of ionizing photons traveling from the quasar toward Earth. This model was originally developed at lower redshift to study the evolution of the ionizing background far from the background quasar.  At high redshift, additional complexity must be added to the model to account for local ionization effects near the quasar, since the mean free path can approach the length scale of proximity zones of enhanced ionization around the quasar.  

Fig~\ref{fig:tauevo} (right panel) presents the measurements of IGM mpfp using the most recent stacking analyses from \cite{2021MNRAS.508.1853B}. 
%The most recent stacking analyses, including sightlines at $z\sim 6$, 
It suggests a rapid change in the mean free path from $8-10$ pMpc at $z=5.1$ to $0.75-1.5$ pMpc at $z=6.0$ (the range reflects  potential fluctuations from cosmic variance/sample size, and also uncertainty in the correlation between the \hi\ opacity and ionization rate).  This evolution is much stronger than would be inferred from extrapolation of the evolving mfp at lower redshifts, as one would expect in the late stages of a phase transition like reionization. 

\subsubsection{Dark Gaps and the late stages of reionization}

The inverse phenomenon of \lya\  transmission spikes---which trace the first environments to complete reionization at $z>6$---several quasar sightlines exhibit long, dark gaps of zero flux at $z<6$. These gaps are now understood to represent residual islands of neutral matter during late-stage reionization. As first explored by \citet{2015MNRAS.447.3402B} after the discovery of a $>110$ cMpc dark span towards the quasar ULAS J0148+0600, measurements of gap length have the advantage of being relatively insensitive to the details of estimating the quasar continuum, or absolute flux levels.  Long gaps are not seen at $z\sim 5$, and indeed recent work by \cite{2021ApJ...923..223Z} finds that 90\% of quasar spectra at $z\sim 6$ have a dark gap of 30 cMpc or longer, but only 15\% do at $z=5.6$.

Gap statistics have generated significant theoretical interest, as they appear to be inconsistent with numerical models approximating the ionizing UV background field as spatially uniform at $z=5.5-6$ \citep{2021ApJ...923..223Z}. 
Various mechanisms have been proposed to explain the phenomenon, including (a) upward fluctuations in gas density (yielding a higher recombination rate and neutral fraction), (b) downward fluctuations in the ionizing radiation background, or (c) upward fluctuations in galaxy number density, leading to lower-than-average temperatures \citep{2015ApJ...813L..38D}. The latter scenario is less intuitive, arising from a subtle effect where regions of high galaxy density reionize early, and then have more time to cool off from the associated photoionization heating. These first areas to reionize have a lower temperature than surrounding regions that reionize later and have not yet cooled. Because the \hi\ recombination rate scales as $T^{-0.7}$, cooler regions that reionized earliest would have higher neutral fraction and opacity characteristic of dark gaps. 

This yields a testable prediction to distinguish between dark gap models arising from UV background versus temperature-driven fluctuations \citep{2018ApJ...860..155D}. Galaxy surveys along sightlines with gaps should see an enhancement in galaxy counts relative to the field if the gaps arise in cool regions that reionized early; conversely there should be a galaxy deficit if the gaps arise in islands of enhanced neutral fraction and/or low ionizing radiation background (two effects which are not straightforward to distinguish).  \citet{2021ApJ...923...87C}, \citet{2018ApJ...863...92B}, and \citet{2022MNRAS.515.5914I}  conducted narrowband Lya surveys in fields centered on well-known dark gaps to test this prediction, and found an underdensity of Lyman-alpha emitters within $20h^{-1}$ Mpc of the quasar sightline, favoring a scenario where the dark gaps represent residual neutral islands persisting in the late-reionization universe.  Spectroscopic galaxy surveys have been difficult and expensive at these redshifts from the ground, but early observations with JWST indicate that prospects are very favorable for correlating galaxies observed in rest-frame optical emission lines with the IGM Lya opacity.

Overall, a consistent picture emerges from a suite of distinct IGM observations using quasar spectra, including (a) the average \lya\ optical depth and its CDF, (b) dark gaps, (c) transmission spikes, and (d) the mean free path. The first three of these four measurements result from on-resonance \lya\ absorption and the fourth measures the Lyman continuum, making them all most useful toward the tail end of reionization at $5.5 < z < 6.3$, when the volume-averaged neutral fraction is $\sim 10^{-4}$, but with significant spatial fluctuations. Taken together there is evidence that large neutral islands persisted well after $z\sim 6$, possibly in regions of low galaxy density and ambient ionizing radiation. During this time there was a tenfold increase in the mean free path for ionizing photons, eventually reaching cosmological length scales of 10 pMpc or more. The earliest regions to fully reionize can be traced by transmission spikes at $z\sim 6.3$, and are likely to reside in regions of low gas density that happen by chance to fall near a source of ionizing radiation.

\subsubsection{Thermal signatures}

Ionizing sources emit a spectrum of radiation including photons with energy above the $E>13.6$ eV Hydrogen ionization potential.  The excess photon energy heats the post-reionization IGM, which subsequently cools over time due to adiabatic cosmic expansion.  Such heating and cooling is clearly seen during the reionization of \heii\ at lower redshift \citep{2019ApJ...872...13W}; an even larger effect should be present for Hydrogen reionization.  Unfortunately unlike \heii, it is not possible to measure heating from Hydrogen reionization in ``real time'', because Gunn-Peterson absorption fully blankets the signal at the redshifts of interest.  Nevertheless the temperature of the IGM at lower redshift---after reionization is complete---still encodes information about the past thermal history. 

High-resolution quasar spectra are excellent thermometers of the IGM, because thermal gas motions projected along the line of sight smooth or broaden absorption features. At lower redshifts where individual unblended lines can be resolved in the \lya\ forest, temperature maps directly to line profile width through the Voigt profile parameter $b=\sqrt{2kT/m_{\rm ion}}$, though this signature is convolved with similar broadening from turbulence and bulk gas motions.  

At $z>4.5$ it becomes impossible to distinguish individual lines, so the IGM's temperature signal must be measured statistically. Effectively, thermal effects impose a low-pass filter on the power spectrum, smoothing out small scale structure in the pattern of \lya\ transmission.  The band limit of this filter is a partly a proxy for the instantaneous temperature, but it also retains a memory of past heating, which is partly manifested through a pressure (i.e. Jeans) broadening beyond the normal Hubble flow. These effects are calibrated by statistical comparison to cosmological simulations. 

The simulations are generally run on a grid which varies the mean temperature of the IGM at specified redshifts $T_0(z)$, the slope $\gamma$ of the IGM's temperature-density relation $T(\Delta)=T_0\Delta^{\gamma-1}$, and the thermal energy injected into the IGM during reionization $u_0$, in units of eV per baryon. Synthetic spectra are drawn directly from  simulation volumes and run through identical analysis software as the true quasar spectra, to ascertain which combination of $T_0, \gamma$ and $u_0$ produce 1D flux power spectra most closely matching observed values.

Using these methods, \citet{2019ApJ...872..101B} find IGM temperatures of $T_0=7000-8000$ K at $4.2<z<5.0$, which is where the effective optical depth of the forest is $1<\tau_{\rm eff}<2$, making the method most sensitive.  Toward higher redshifts where Gunn-Peterson saturation begins to manifest, thermal measurements become increasingly challenging.  However recent constraints have been extracted by fitting the widths of emission line spikes, indicating slightly higher $T_0=10,000-12,000$K at earlier times $z=5.3-5.9$ as expected according to the paradigm of heating and adiabatic cooling \citep{2020MNRAS.494.5091G}.  

The simulations can also explore a range of different reionization histories to see which parameters best reproduce the post-reionization power spectrum. Toy models of instantaneous reionization provide joint constraints on the redshift of thermal injection and its amplitude, favoring $z\sim 8-10$ as the reionization redshift, and peak post-reionization temperatures of $T=10,000-20,000$ K depending on the choice of ionizing background spectrum.  Temperature constraints on the reionization history are limited because the IGM tends to rapidly cool back to its ionization equilibrium temperature, washing out the signal from the initial thermal impulse.  As mentioned above, the rapid thermal broadening is accompanied by a pressure broadening that also manifests in the power spectrum, and is somewhat longer lived. It is not possible to fully disentangle pressure and temperature smoothing effects from line-of-sight measurements. However measurements made using pairs of adjacent quasar sightlines can break this degeneracy, since they only measure smoothing on a transverse spatial scale by construction \citep{2017Sci...356..418R}.

Summarizing, while the power spectrum and thermal history cannot provide precision constraints on the timing of hydrogen reionization, it supports a general result that by $z\sim 5$, the initial thermal energy injected into the IGM has already cooled substantially, by perhaps a factor of $2-3$.  This sets a lower bound on the elapsed time between the period of thermal energy injection and $z\sim 4-5$ where the measurements are made.  This in turn implies that the peak heating associated with reionization occurred at $7<z<10$, broadly consistent with CMB results reported by Planck \citep{2020A&A...641A...6P}.

\subsection{Neutral Hydrogen Absorption in the Quasar Environment}

Each quasar spectrum offers the opportunity to study radiative feedback in the neighborhood of a luminous source of ionizing photons: the background object itself. At $2<z<3$, a clear deficit is seen in the density of Lyman alpha forest lines as a quasar's emission redshift is approached \citep{1986ApJ...309...19M,1988ApJ...327..570B}. For these lower redshifts where the ambient IGM is clearly optically thin, the so-called ``proximity effect'' reflects a local enhancement of ionization in an already highly-ionized medium, and can be used to derive estimates of the quasar's ionizing flux over the metagalactic background.

This measurement becomes increasingly valuable at high redshift, because the diffuse IGM is optically thick to \lya\  absorption.  However the observational manifestation of interactions between quasar radiation and the IGM differs according to whether the surrounding medium has already undergone reionization --- in which case one observes a ``proximity zone'' --- or the surrounding medium still has a high neutral fraction --- in which case one observes a damping wing. Both phenomena are now well established; in the sections below we aim to clarify the distinction between the two, reconcile some differences in definitions, and clarify at which redshifts each one is most powerful.

\subsubsection{Proximity zones and the effects of quasar radiation on recently reionized matter}
\label{subsec:proximity}

\begin{figure}[t]
\includegraphics[width=4in]{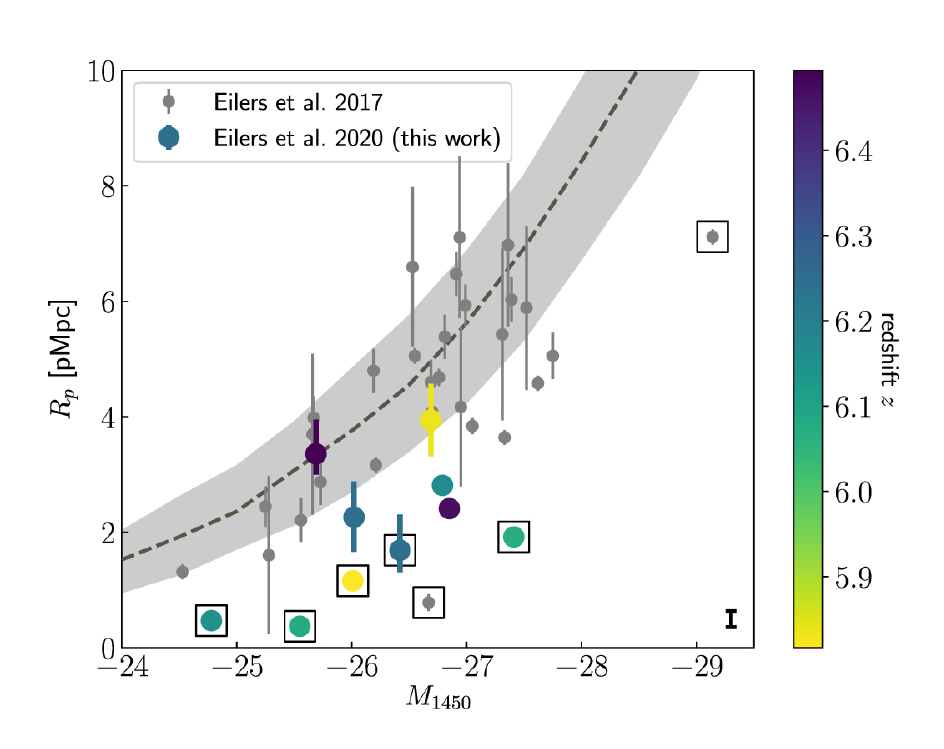}
\caption{The size distribution of proximity zones at $5.9<z<6.4$ \citep{2021ApJ...917...38E}, versus their host quasar's absolute magnitude or luminosity.  Shaded region shows expectations from a numerical simulation assuming the quasar has reached ionization equilibrium with its surroundings.  This matches a representative sample of quasars at these redshifts (gray points), but separate samples have been identified with small proximity zones that have been interpreted as evidence of recent UV accretion episodes.}
\label{fig:proxzonesize}
\end{figure}

At $z_{\rm em}\sim 5.9-6.5$, on-resonance (Gunn-Peterson) \lya\  has sufficient column density to fully absorb a quasar's continuum. However the volume-averaged neutral fraction is only $\sim 10^{-4}$ during late-stage reionization, so a quasar's local ionization boost can suppress the local \lya\ opacity to optically thin values $\tau < 1$, allowing some flux to transmit at wavelengths just blueward of the \lya\ emission line.  In exact analogy to the classic proximity effect at lower redshift, these ``proximity zones" at $z\sim 6$ trace a local ionization enhancement in an already ionized medium. They should not be confused with Str{\"o}mgren spheres, because the outside medium has a neutral fraction $\xhi \ll 1$.

The size of a quasar's proximity zone depends on its ionizing photon luminosity, its lifetime (i.e. how long it has been shining in a UV-bright phase), and the density and ionization of the surrounding medium. By convention, observers have defined the proximity zone as the total path from $z_{em}$ to where a quasar's spectral flux density remains $>10\%$ of the continuum level, after smoothing by a boxcar window of width 20\AA\  \citep{2006AJ....132..117F}.  While observationally convenient and consistent, this definition is not motivated by any physical transition or process.  It is loosely related to a radius of enhanced ionizing background, but at a 10\% flux threshold the enhancement is still very strong --- the quasar contribution  exceeds the ambient field by an order of magnitude \citep{2017ApJ...840...24E}. In select cases one may compare \lya\ and Ly$\beta$ measurements to improve dynamic range in analogous fashion to IGM measurements \citep{2004ApJ...611L..69M}. In this case, one is probing the sharpness of the boundary of enhanced ionization fraction.

Uniformly selected samples have yielded typical proximity zone sizes of 2--6 pMpc, with a strong (expected) dependence on quasar luminosity, and comparatively weak dependence on redshift between $6.0<z<6.5$ \citep{2017ApJ...840...24E,2020ApJ...900...37E}.  Intriguingly, after adjusting  proximity zone sizes to normalize out the effect of quasar luminosity, \citet{2017ApJ...840...24E} found that $\sim 10\%$ of their sample had exceptionally small proximity zones. They argued that the most likely explanation is that we are observing these quasars before their radiation field has reached ionization equilibrium with the surrounding IGM, i.e. their current UV-luminous period is $t<100,000$ years old, with extreme objects even at $t<10,000$ years \citep{2021ApJ...917...38E}.  Stacking analysis of multiple quasars' near zones indicates that the median quasar lifetime is larger by approximately an order of magnitude \citep{2021ApJ...921...88M}. Fig~\ref{fig:proxzonesize} presents a summary of proximity zone size measurements, along with ``young quasars'' identified through this type of measurement. 

If this interpretation is correct, then proximity zone size distributions can be used to constrain the UV duty cycle of early quasars. The size is sensitive only to the age of the most recent or current UV-bright episode, which is not necessarily the first in its history.  However the growth timescale problem for building SMBHs at $z>6$ is well known \citep[see Sec~\ref{sec:SMBH}; and ][]{2010A&ARv..18..279V}, and this challenge is only exacerbated for any accretion duty cycle below unity.  It may be that the usual assumed correspondence between robust accretion and UV-bright episodes may need revision, and significant black hole growth can occur during a dusty, obscured phase.

\subsubsection{IGM Damping Wings}
\label{subsec:dampingwings}

\begin{figure}[t]
\includegraphics[width=3.5in]{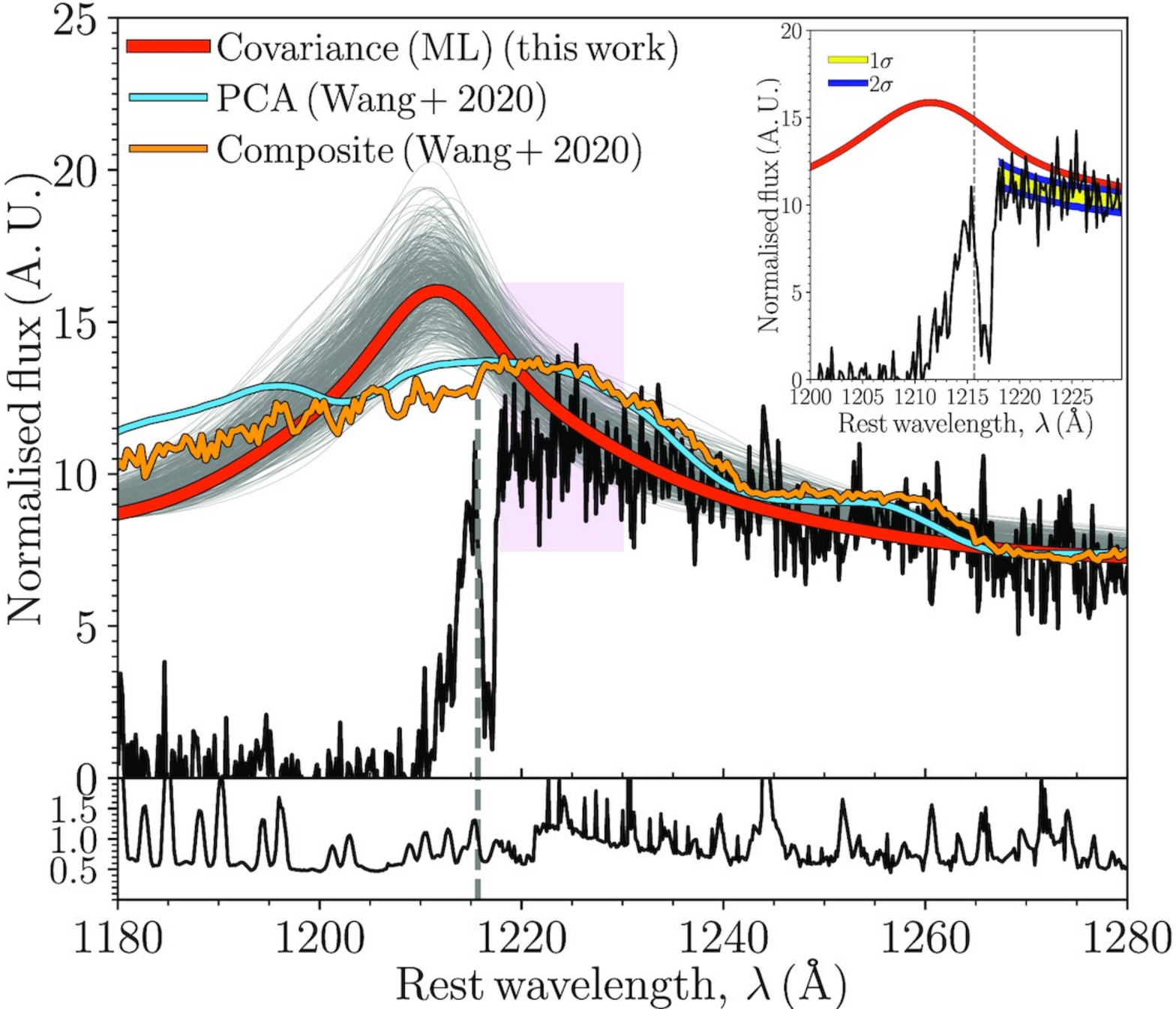}
\caption{Damping wing measurements for J0252-0503 ($z=7.00$).  Quasar spectrum is shows with black curve, and three continuum models are shown for the damping wing analysis.  The quasars systemic redshift is shown with a vertical dashed line. Adapted from \cite{2022MNRAS.512.5390G}.}
\label{fig:damping}
\end{figure}

As one penetrates deeper into the reionization epoch near $z\sim 7$, the ambient neutral fraction increases and the corresponding length scale of proximity zones becomes smaller.  When the neutral fraction approaches 1--10\%, the effects of off-resonance \lya\ damping wings become significant \citep{1998ApJ...501...15M}. Details of this process have been reviewed in \citet{2016ASSL..423..187M}. Briefly, high column densities of neutral hydrogen in the nearby foreground of a quasar lead to a non-negligible probability that a photon emitted redward of \lya\ in the quasar's rest frame will still be absorbed by neutral intergalactic matter en route to Earth.  This last fact is critical---in the far-foreground \lya\ forest region there is an unresolvable degeneracy at any wavelength between off-resonance damping wing absorption from neutral gas, and on-resonance absorption from ionized gas at a different redshift.  In contrast, absorption redward of the \lya\ emission line can only come from off-resonance interactions.
%, since the on-resonance line would be at a redshift behind the quasar and therefore not geometrically configured for absorption.  

The IGM damping signature is expected around quasars meeting two conditions: they have recently turned on, and are embedded in an IGM with high global neutral fraction. In contrast to $z\sim 6$ proximity zones that have already been pre-ionized by surrounding galaxies, the $z\sim 7$ quasars with damping wings are driving true ionization fronts into the surrounding IGM with a size that increases with the quasar's active phase lifetime and ionizing photon flux, and decreases with larger overdensity. 

The strength of a damping wing is most sensitive to the neutral fraction of matter just outside the quasar's ionization front.  Gas closer to the quasar is inside the ionization front, so its neutral fraction is too low to produce off-resonance \lya\ absorption wings. Gas that is far outside the ionization front can produce off-resonance absorption, but its line-of-sight velocity offset is too large for the off-resonance wing to produce appreciable optical depth redward of Lya emission.  

Damping wing measurements are significantly affected by systematic uncertainties in modeling the detailed shape of quasars' intrinsic \lya\ profile. Different groups have tackled this challenge using training sets of low-redshift SDSS or HST/COS spectra where the \lya\ emission profile and unabsorbed continuum can be estimated more reliably. The objective is to predict the unabsorbed \lya\ emission line shape and forest continuum (and associated uncertainties) using pixels to the red of the quasar's \lya\ emission line.  

Multiple groups have used Principal Component Analysis \citep{2018ApJ...864..143D,2011Natur.474..616M}, Machine/Deep Learning models  \citep{2021MNRAS.502.3510L}, Neural Networks \citep{2020MNRAS.493.4256D}, covariance matrices \citep{2017MNRAS.466.4239G,2022MNRAS.512.5390G}, stacking of nearest-neighbor observed spectra \citep{2012Natur.492...79S,2015MNRAS.452.1105B,2021MNRAS.503.2077B}, or other hybrid approaches to train and validate continuum+\lya\ software on lower redshift quasars.  Most methods also calculate some envelope of formal uncertainty on the continuum estimate before applying them to $z\sim 7$ quasar spectra. When comparing different reconstruction techniques in the Ly$\alpha$/Ly$\beta$ forest region and around the \lya\ emission line, the difference between continuum models for the same object using different codes often exceeds expectations from their formal error estimates, indicating the presence of residual systematics. It has therefore been important to test observations of the same objects taken using different instruments and analysis software to build confidence in the fundamental conclusion about damping wing strengths and the neutral fraction. Fig~\ref{fig:damping} illustrates some of the different techniques in predicting the quasar intrinsic spectrum in the damping wing region of a quasar at $z=7.0$. 

While it is impossible to know if any particular continuum realization is ``correct'', a consensus is indeed emerging that certain specific objects --- most notably ULASJ1342+0928 \citep[$z=7.54$; ][]{2018Natur.553..473B}, DESJ0252$-$0503 \citep[$z=7.00$; ][]{2020ApJ...896...23W}, J1007+2115 \citep[$z = 7.51$; ][]{2020ApJ...897L..14Y} and ULASJ1120+0641 \citep[$z = 7.09$; ][]{2011Natur.474..616M} --- show strong evidence for IGM damping wings for all continuum models explored in the literature. Various analyses derive slightly different IGM neutral fractions for the same objects on account of modeling differences \citep{2022MNRAS.512.5390G,2019MNRAS.484.5094G,2018ApJ...864..142D,2011MNRAS.416L..70B}. But a robust result remains that $67\%$ of known quasars at $z>7$ have (a) negligible proximity zone sizes, and (b) flux measurements at wavelengths slightly redder than the Lya emission line that fall below all continuum reconstructions based on low-redshift quasars.  This combination of small proximity zone and damping wing has not been measured in a single quasar at $z<7$, despite the much larger sample sizes available for testing.

Because most of the damping wing is generated just outside the ionized bubble, there is a risk that the same signature could arise from chance alignment of a classical damped \lya\ absorber (i.e. from the ISM of a bound proto-galaxy) near the quasar in an otherwise ionized IGM (Section \ref{sec:metals_near_hosts}). Indeed there are examples of such ``proximate DLAs" in the literature, including several at $z>6$. The a priori odds of such a chance alignment are small, but were considered carefully in studies of the first damping wing \citep{2012Natur.492...79S}. There are two arguments now suggesting that this phenomenon is not affecting IGM measurements at $z>7$.  First, none of the four sightlines listed above exhibit heavy-element absorption at the redshift of the putative DLA, to very sensitive limits.  Second, the presence of damping wings in multiple new $z>7$ discoveries points to a global phenomenon, as the joint probability of seeing proximate DLAs in all of these becomes lower still. With very high-SNR data it is in principle possible to distinguish between DLA absorption from the condensed CGM/ISM of a proto-galaxy and an extended patch of neutral but gravitationally unbound IGM using the damping profile alone. However no studies at present have claimed such a detection.

A final unknowable factor in damping wing analysis is the spatial distribution of gas density and accompanying fluctuations in ionization inside and outside the proximity zone. Early papers used analytic models based on distribution functions of overdensity, and assumed that the gas was in ionization equilibrium with a uniform background radiation field \citep{2011Natur.474..616M,2012Natur.492...79S}. Later and more sophisticated treatments have drawn distributions of sightlines from cosmological simulations to capture the line-of-sight variations in density and ionization \citep{2019MNRAS.484.5094G}. Large simulation boxes are required for this method, to identify the rare, high-mass halos where early quasars are thought to reside. Many sightlines are constructed which terminate by design at the high-mass halo, to generate many realizations of the damping wing for different viewing orientations.  

Summarizing, among all quasar absorption methods, IGM damping wings yield the only two-sided constraint on $\xhi$ at $z>5.7$, and are therefore one of our most important probes of reionization history.  They are nevertheless subject to substantial, and probably unresolvable systematic uncertainties concerning quasar continuum shape and details of the density and ionization fields around each quasar.  Nevertheless a tentative consensus has emerged that for the majority of objects yet discovered at $7.0<z<7.5$, the near-zone size and damped absorption rule out models with global ionization fractions $\xhi<1\%$ or $\xhi\approx 100\%$, instead favoring values from $20-80\%$. Adding more objects will potentially increase confidence in the result, but may not reduce its fractional uncertainty to better than $\pm 40\%$, since the uncertainty envelope is dominated by model assumptions and is therefore not random. 

\begin{figure}[t]
\includegraphics[width=5in]{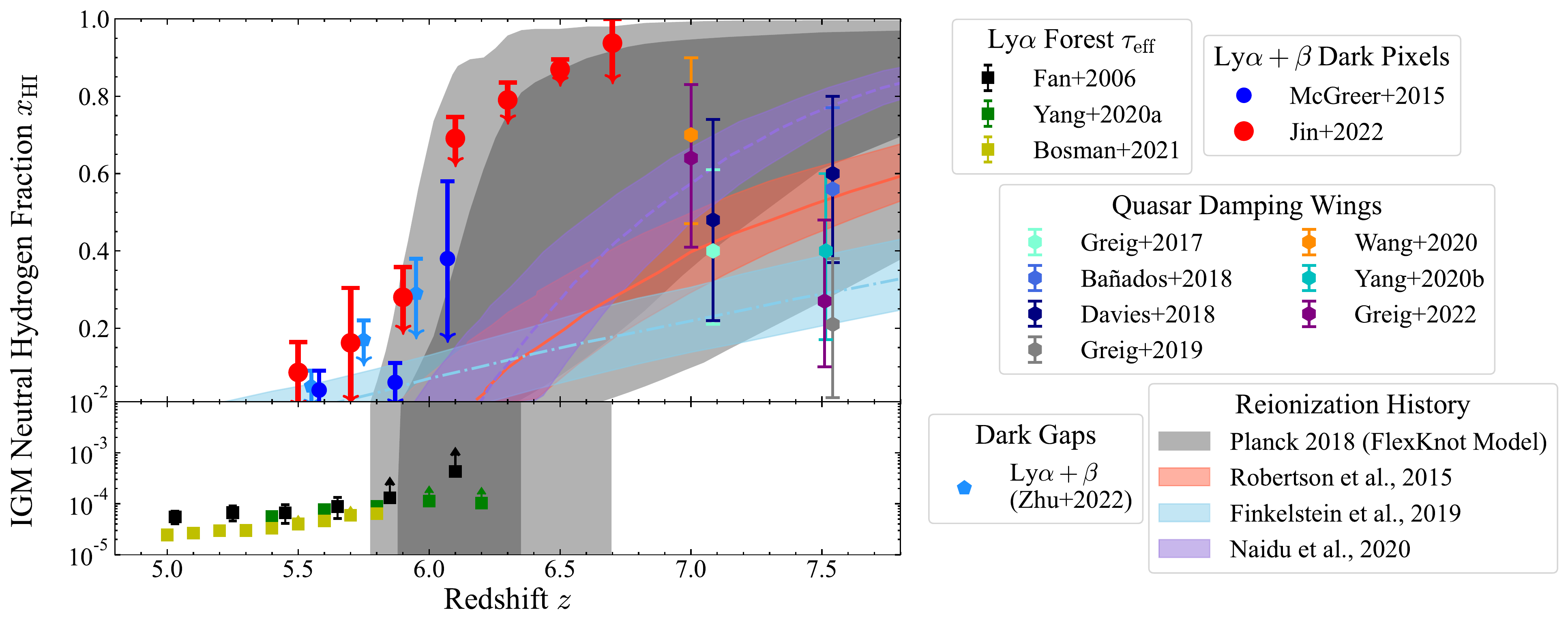}
% \caption{Summary of quasar absorption constraints on the reionization history. All data points are Adapted from \cite{2022arXiv221112613J}}
%EB version
\caption{
Summary of quasar absorption constraints on the reionization history (all data points; Figure adapted from \citealt{2022arXiv221112613J}). The dark (light) gray-shaded region is the 1$\sigma$ (2$\sigma$) constraint from the Cosmic Microwave Background \citep{2020A&A...641A...6P}. The red, blue, and purple shaded regions display the 1$\sigma$ reionization histories from the \cite{2015ApJ...802L..19R}, \cite{2019ApJ...879...36F}, and \cite{2020ApJ...892..109N} models, respectively. See \cite{2022arXiv221112613J} and references therein, for more details.
}
\label{fig:nhsummary}
\end{figure}

Still this result is extremely significant because any measurement placing $\xhi\sim 50\%$ (or at a similar order of magnitude) represents a working definition of the reionization midpoint. Fig~\ref{fig:nhsummary} summarizes the various reionization constraints from different quasar absorption observations, compared with both constraints from the measuremnts of CMB polarization  \citep{2015ApJ...802L..19R} and a number of empirical reionization history models. Optical depth measurements provide one-sided lower limits on $\xhi$ at $z<6.3$, transmission spikes provide one-sided upper limits at $5.5<z<6.5$, and damping wings provided two-sided constraints with large errors at $7<z<7.5$. This combination of limits and measurements permits models with a range of reionization histories with midpoints centered anywhere from $6.5<z<8.0$.  Future 21-cm experiments will potentially measure these values with more precision and accuracy; these quasar constraints will help such experiments optimize their redshift window. Still, when expressed in terms of lookback time rather than redshift, quasar absorption constraints correspond to a 1--2\% measurement of the timing of reionization, to $94.5^{+0.9}_{-0.5}$\% lookback, or $760^{+80}_{-120}$ Myr after the Big Bang.  

\subsection{Heavy Element Enrichment of Intervening Quasar Absorbers}

The radiative feedback that drove reionization was accompanied by chemical feedback that can also be observed in the spectra of high redshift quasars.  At $z>6$, large samples of high-resolution quasar spectra taken with VLT/XShooter, Magellan/FIRE, Keck/NIRES and Gemini/GNIRS have revealed foreground heavy element absorption lines redward of \lya\, where the quasar's continuum can be easily modeled. Because \hi\ is saturated in the \lya\ forest at $z>6$, metal absorption lines provide the only accessible information about diffuse gas at these epochs.

Absorption strength in these papers is sometimes reported as column density $N_{\rm ion}$ and other times as a rest-frame equivalent width $W_r$.  The IR spectrographs listed above have intrinsic resolutions of $\Delta v\sim 50$ km/s or coarser, which is insufficient to resolve the fine $5-10$ km/s velocity structure seen in metal-line absorbers at lower redshift.  Equivalent width measurements are therefore conservative and will be consistently measured across all spectrographs.  Column densities are easier to interpret physically, but must be approached with care, accounting for the possibility of unresolved saturation in the line core, or blending of unresolved components. The detection sensitivity of current spectrographs limits metal line analyses to column densities characteristic of circumgalactic gas at lower redshift. Deeper spectra or statistical stacks will be needed to search for metals in the diffuse IGM.

Early studies of heavy element enrichment in the reionization epoch \citep{2011ApJ...743...21S,2013MNRAS.435.1198D,2006MNRAS.371L..78R,2009ApJ...698.1010B,2006ApJ...653..977S} focused on the \civ\ % 1549\AA
doublet, which is the most ubiquitous IGM metal line at lower redshifts and a tracer of highly-ionized circumgalactic gas. At $z>5.5$ \civ\ shifts into the $J$ band.  Most surveys focus on a similar set of observables, including the comoving absorption density $dN/dX$\footnote{The quantity $dX=dz\sqrt{\Omega_M(1+z)^3+\Omega_\Lambda}$ is equivalent to a redshift path at $z=0$, but accounts for pure cosmological evolution in the comoving radial line element. A population of absorbers with constant product $n(z)\sigma$ of comoving space density $n(z)$ and absorption cross section $\sigma$ will have a constant $dN/dX$ at all redshifts.}, and the equivalent width distribution $d^2N/dXdW$.  

These surveys established that the frequency of \civ\ absorbers decreases toward higher redshift, such that individual \civ\ systems become rare at $z>6$ --- most sightlines have no \civ\ doublets in this range. For example, \citet{2018MNRAS.481.4940C} surveyed deep spectra of four quasar sightlines, finding a four-fold decrease in comoving line density $dN/dX$ between $z=4.8$ and $z=5.7$, and zero detections above $z>6$, consistent with earlier but smaller studies in the literature \citep{2011ApJ...743...21S}.  There is very tentative evidence of a steepening in the column density distribution function (CDDF) $f(N)=d^2N/dNdX$, because of differential evolution of the strongest \civ\ absorbers at high redshift. However these analyses  still utilize modest numbers of sightlines ($<10$), so shot noise and saturation become significant limitations at the high-$W_r$ end.  There are now relatively large samples of high quality spectra for $z>6.5$ quasars, which should lead to improved estimates of the CDDF and \civ\ evolution in the near future.

Evolving enrichment levels are sometimes summarized in terms of the \civ\ contribution to closure density $\Omega_{\rm CIV}$, which is a mass-weighted integral of the CDDF. Like $dN/dX$ or the normalization of $f(N)$, $\Omega_{\rm CIV}$ declines toward higher redshift.  While it is convenient to express global metal enrichment in terms of this single number, the power-law slope $f(N)\propto N^{-1.5}$ of the CDDF is shallower than $\alpha=-2$, implying that the integral $\int Nf(N)dN$ diverges toward high column densities.  The single strongest absorber in any sample therefore dominates the integral, meaning that by construction $\Omega_{\rm CIV}$ is affected by shot noise and pathlength searched. 
%It is often assumed that at some large column density, the power-law CDDF will have an exponential cutoff allowing $\Omega_{\rm CIV}$ to converge.  However current high-redshift surveys have not searched enough pathlength to establish where this cutoff occurs.

The disappearance of \civ\ at high redshift could be driven either by a decline in heavy element abundances, or by a change in ionization. If abundance evolution predominates one might also expect to see a decline in the frequency of low-ionization species such as \mgii, OI, SiII, CII, or FeII.  However there have been robust detections of all these low-ionization lines at $z>6$, suggesting an important change in the ionization state of metal absorption line systems coincident with the reionization of intergalactic hydrogen.

The first detections of low-ionization metal absorption at $z>6$ focused on the OI 1302, CII 1334, and SiII 1260 lines, which have rest wavelengths only slightly to the red of \lya\ and can therefore be detected in high-resolution optical (rather than infrared) spectra \citep{2011ApJ...735...93B}. However the small offset between these and \lya\ also leaves a very short search pathlength per sightline; if they are too far in the quasar's foreground these lines fall within the \hi\ Gunn-Peterson trough and become impossible to detect.

\begin{figure}[t]
\includegraphics[width=5in]{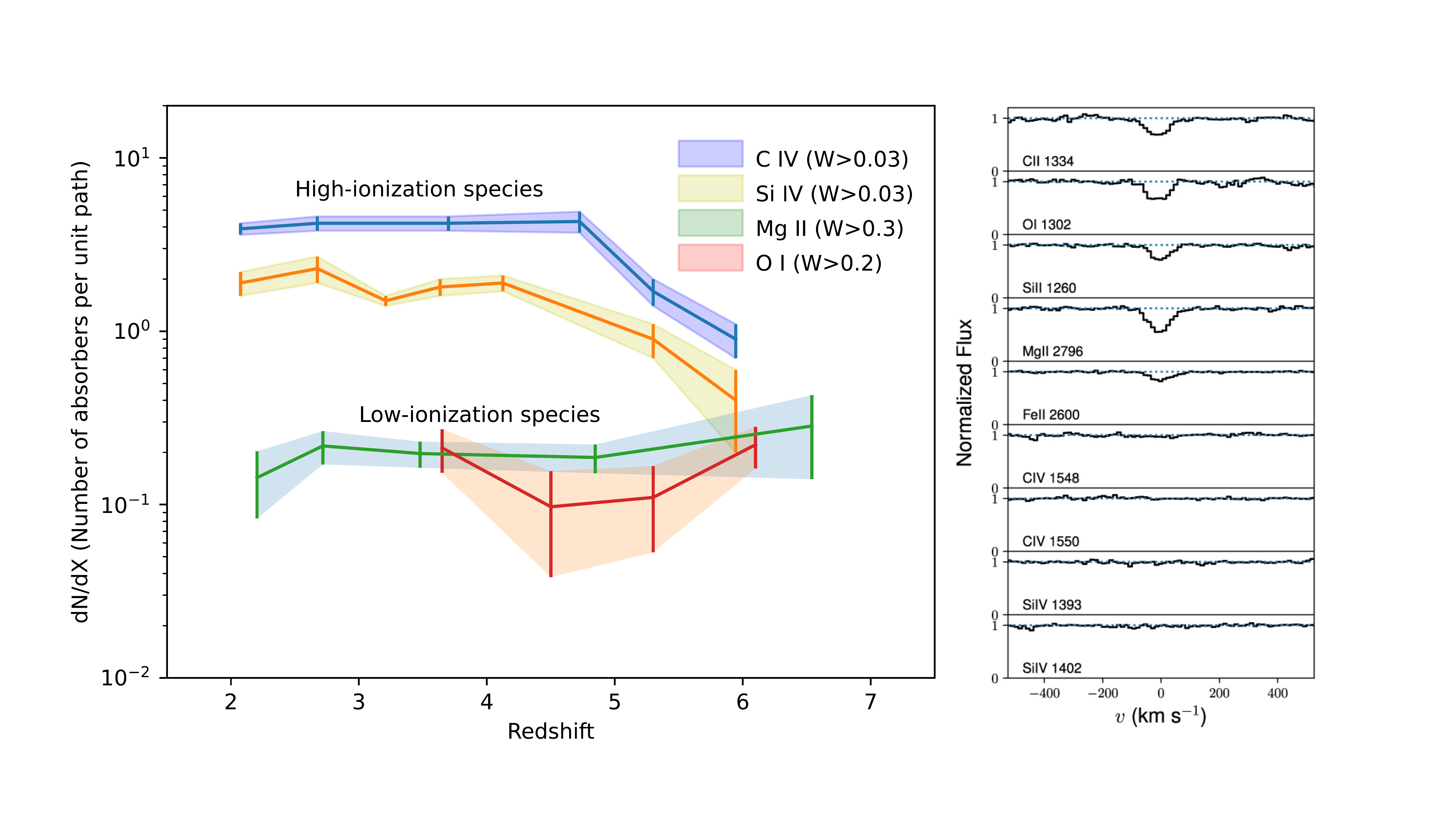}
\caption{(Left) Evolution in number density $dN/dX$ of heavy element absorption systems, illustrating the different behavior between highly ionized species \civ\ and  SiIV, which decline with increasing redshift, and low-ionization species \mgii\ and OI, which do not decline.  (Right) A stack of $z>5.7$ absorbers selected individually by MgII detection, which shows other low-ionization lines, but no evidence of \civ\ or SiIV.}
\label{fig:metal}
\end{figure}

With improved IR spectra it became possible to search systematically for low-ionization \mgii\ doublets at $z>6$, albeit with complex selection windows caused by broadband atmospheric absorption between the $J/H$ and $H/K$ bands, discrete telluric absorption, and bright hydroxyl emission lines.  After accounting for these effects, \mgii\ surveys find trends consistent with the other neutral ions---there is no clear evolution in the comoving line density $dN/dX$ for systems with $W_r>0.3$\AA\ \citep[although if one restricts to very strong absorbers, there is a slight excess at $z\sim3$;][]{2017ApJ...850..188C,2012ApJ...761..112M}. \mgii\ absorbers are now detected out to the highest known redshift of an absorption system, at $z=6.84$ \citep{2020arXiv201110582S}.

The statistical decrease in \civ\ frequency together with statistical non-evolution of low-ionization species requires that as reionization is approached, one should eventually find individual absorbers with CII, OI, SiII, and \mgii\ but no detection of a highly ionized \civ\ phase. This is indeed the case; in fact such low-ionization-only systems predominate at $z>6$, composing over 2/3 of all metal line absorbers \citep{2019ApJ...882...77C}. 

In sharp contrast, ``low-ionization only" absorbers are exceedingly rare at all redshifts $z<5.5$, indicating a qualitative change in the CGM's observable properties during the reionization epoch.  At $2<z<4$ OI, CII, and SiII are much more rare than \civ;  nearly every known low-ionization absorber is accompanied by stronger \civ\ and SiIV lines offset by $<100$ km/s, and there are many examples of \civ\ with no low-ionization lines \citep{2019ApJ...882...77C}.  Because absorption surveys are cross-section selected, the canonical interpretation is that ionized \civ\ traces a widespread low-density CGM, while OI, CII, and SiII trace cooled and condensed clumps in the ISM of a nearby galaxy, or small, cold precipitates embedded in the warm ionized CGM \citep{2016ApJ...830...87S}.  
The evolution of the number density of heavy element absorption systems is summarized in Fig~\ref{fig:metal}.

The observational disappearance of the ionized CGM implies that during this epoch where galaxies are in early assembly, their circumgalactic gas has not yet been enriched to observable levels, and/or the ambient radiation field is insufficient in intensity or hardness to maintain a high degree of ionization.  Unfortunately it is impossible to calculate abundances directly at $z>6$, because \hi\ column density measurement is prevented by the Gunn-Peterson trough. However the number density of these low-ionization metal lines is similar to extrapolations of $dN/dX$ for DLAs and sub-DLAs \citep{2019ApJ...882...77C,2019ApJ...883..163B} and is either flat or slightly increasing at earlier times.  If one assumes \hi\ column densities in this range ($N_{\rm HI} > 10^{19.8}$), the detected low-ionization lines are consistent with a metallicity of [C/H]$\sim -2.3$, in a warm neutral medium. If there exists a warm ionized medium of similar metallicity in pressure equilibrium with the warm neutral phase, it would be clearly detected in \civ.  To evade detection, the ionized medium must have $\sim 100\times$ lower abundance, or [C/H]$<-4$ \citep{2020arXiv201110582S}.  If the heavy elements seen in low ionization lines are produced locally to that environment, this implies that those metals have not yet been mixed into the ionized CGM.  Evidently the galactic environments traced by $z>6$ metal absorbers bear some resemblance to metal-poor DLAs at lower redshifts, but they are at an early and immature stage of chemical evolution and mixing.

\subsection{Metal Absorption in the Vicinity of QSO Hosts}
\label{sec:metals_near_hosts}

At all redshifts one can find a rare but important class of metal absorbers in the immediate foreground of their respective quasar hosts \citep{2008ApJ...675.1002P}.  These ``proximate DLAs" have a special importance, and four have been discovered at $z>6$. They are the only discrete absorbers in the reionization epoch for which it is possible to constrain $N_{\rm HI}$ directly using the off-resonance wing. Because DLAs with $N_{\rm HI}>10^{20.3}$ cm$^{-3}$ are self-shielded and largely neutral, the pDLAs therefore also permit direct measurement of heavy element abundances, with no ionization correction needed.

The pDLAs are distinguished from IGM damping wings on the basis of coincident metal absorption.  They are also found at lower redshift than the four $z>7$ IGM damping wing candidates---two pDLAs are at $z\sim 6.3-6.4$ \citep{2019ApJ...885...59B,2022AJ....163..251A}, one at $z\sim6$ \citep{2020MNRAS.494.2937D}, and one is at $z=5.93$, 40\,kpc from a galaxy detected with ALMA at coincident redshift \citep{2018ApJ...863L..29D}.  At these redshifts no other quasars have exhibited metal-free damping wings, and many observations of transmission spikes indicate that reionization is well underway \citep{2020ApJ...904...26Y}.

The pDLAs studied by \cite{2019ApJ...885...59B} and \cite{2018ApJ...863L..29D}   have abundances near $0.001Z_\odot$, which would place them amongst the most metal poor DLAs known at lower redshift \citep{2011MNRAS.417.1534C}. Neither shows evidence for anomalies in relative abundances, as might be expected from a medium enriched with Population III supernova debris.  High ionization \civ\ and SiIV are not reported in either system, consistent with the discussion in the previous section about the disappearance of the high-ionization CGM phase. This has been cited as evidence that the pDLA originates outside of the quasar host, where the UV radiation field should be enhanced and an ionized medium would be expected.

The evidence so far suggests that pDLAs in the reionization epoch are representative of randomly intervening metal absorbers at similar redshift. 

%- Figure 11: evolution of optical depth
%
%- Figure 12: some figure highlighting late end of reionization
%
%- Figure 13: damping wing measurement
%
%- Figure 14: proximity zone sizes

\section{Summary}
\label{sec:summary}

Since the first discoveries of reionization-era quasars more than twenty years ago, tremendous progress has been made in extending the frontiers of quasar research deeper into the cosmic dawn, in characterizing and understanding these objects in the context of  early SMBH growth and galaxy formation, and in using them as beacons to shed light on the evolution of the IGM and the history of cosmic reionization.
 % With the new facilities  planned for the next decade, we expect that the quasar redshift frontier will reach to redshift 9--10, when the first generation of luminous quasars are forming.  At the same time, deep X-ray and JWST surveys will detect the earlier phase of SMBH growth at $z>10$. Synergetic observations of JWST and ALMA will probe the connection of early SMBH growth and galaxy formation from within the BH sphere of influence to the scale large scale structure. Meanwhile, detailed spectroscopic observations of early quasars will map the entire history of reionization from $z\sim 10$ to 5.

\begin{summary}[SUMMARY POINTS]
\begin{enumerate}
\item The current frontier of quasar research is at $z\sim 7.6$, with a large statistical sample assembled at redshifts up to 7.  High-redshift quasar selection is a challenging data mining problem because of the rarity of the targets, the limited training sets, and the overwhelmingly large number of contaminant sources, both astrophysical and instrumental. We provide a database of the properties of all published quasars at $z>5.3$ as {\bf Supplementary Material} that accompanies this review. 
\item  The rapid evolution of the quasar LF has been well established. However, the exact shape of the LF is still uncertain, both at the high luminosity end due to small number statistics, and at the low luminosity end due to limitations of deep surveys.  In sharp contrast, early quasars ``look'' almost identical to their low-redshift counterparts in spectral properties.  Even the earliest  known quasars have a well-established AGN structure.
\item  There are more than 100 $z\gtrsim 6$ quasars with robust BH mass measurements.
%(from the \mgii\ emission line, currently the most reliable tracer at these redshifts). 
Their masses range from $\sim$$4\times10^7\,M_\odot$ to $\sim$$10^{10}\,M_\odot$. The BHs are growing at high Eddington ratios ($\lambda_{\rm Eff,median}=0.79$ and $\lambda_{\rm Eff,mean}=0.92$), with mild evidence of an increase of the accretion rate for $z\gtrsim 6.5$ quasars in comparison to luminosity-matched quasar samples at lower redshifts.
\item   The UV stellar light in $z\sim6$ quasar hosts remains undetected, while on the other hand, (sub)mm observations have revealed the existence of copious amounts of gas and cold dust in many of these galaxies. Quasar hosts are massive, star-forming galaxies showing diverse morphology and kinematics; their average dynamical mass  appears lower than that based on the local scaling relation between BH and galaxy masses. 
\item Ly$\alpha$ absorption in quasars' foreground IGM provides many of the most stringent constraints on the history of reionization, though the details of the observations vary with redshift.  Toward the end of reionization at $z<6$, counts of transmission spikes, dark gaps, and pixel optical depths indicate that universe is mostly ionized, but may harbor substantial islands of neutral matter as late as $z\sim 5.3$.
At $z>7$, four quasars have been discovered with prominent, metal-free damping wings that are thought to represent off-resonance absorption from an IGM with neutral fraction $>10\%$, and possibly as high as $80\%$. 
\item {Heavy element absorption is seen in quasar spectra well into the epoch of reionization, tracing chemical feedback from stars whose radiation is also transforming the global ionization balance. It appears that these absorbers are high-redshift, low-metallicity analogs of damped Ly$\alpha$ absorbers observed at lower redshift. However the highly ionized phase of the circumgalactic medium, seen universally in CIV at $z<6$, abruptly vanishes in the EoR. This either indicates that the metals in the CGM are undergoing their own ionization transition, or that the ionized phase of the CGM has not yet been enriched with heavy elements at $z>6$.}
\end{enumerate}
\end{summary}

% Future Issues
\begin{issues}[FUTURE ISSUES]
\begin{enumerate}
\item  With the new facilities planned for the next decade, we expect the quasar redshift frontier to reach redshift 9--10, when the first generation of luminous quasars is forming. A key challenge will be the spectroscopic confirmation of these faint high-redshift sources. 
\item  SMBH mass measurements using the H$\beta$ line with JWST and gas-dynamical estimates with ALMA (and eventually the ngVLA) will provide more accurate measurements than what is currently possible. These results will be crucial to test whether the current methods calibrated at low redshift (e.g., \mgii\ single-epoch virial mass)  still hold at the highest accessible redshifts. To measure the BH masses of the progenitors of the currently observed $z=6-7$ quasar population, NIR spectroscopic follow-up with the ELTs will be key.
%\item Current observations are only sensitive to luminous, type-1 quasars with SMBH masses $>10^{7-8} M_{\odot}$, likely the end product of BH growth in the early Universe. Future deep X-ray and JWST surveys will detect the earlier, and more obscured phase of SMBH growth.  
\item JWST will likely be able to finally reveal stellar light from $z\gtrsim 6$ quasar hosts. ALMA will continue playing a critical role in understanding and characterizing the interstellar medium of quasar host galaxies via multi-line diagnostics. 
\item {New quasar samples uncovered by Euclid and the Roman Space Telescope and observed with JWST and ELTs should substantially improve confidence in $z\ge 7$ measurements of IGM damping wings. They will allow studies even deeper into the past.} 
\item{JWST already shows promise of being transformative for surveys of galaxies in fields where high-redshift quasars have been used to constrain the IGM density field.  This may open the possibility of studying cross-correlations between \lya, metals, and galaxies.}
\end{enumerate}
\end{issues}

%Disclosure
\section*{DISCLOSURE STATEMENT}
The authors are not aware of any affiliations, memberships, funding, or financial holdings that
might be perceived as affecting the objectivity of this review. 

% Acknowledgements
\section*{ACKNOWLEDGMENTS}
We dedicate this review to the memory of Professor Maarten Schmidt, who passed away in September 2022 at the age of 92. Maarten identified the first quasar 3C273 in 1963. His ground-break work on quasars transformed our understanding of the distant universe and inspired generations of quasar astronomers.  

The authors thank Thomas Connor, Xiangyu Jin, Yana Khusanova, Chiara Mazzucchelli, Marcel Neeleman, Jan-Torge Schindler, Feige Wang, and Jinyi Yang for permissions to use or adapt their figures, and Irham Andika, Silvia Belladitta, Alessandro Caccianiga, Kuenley Chiu, Emanuele Faarina, Anniek Gloudemans, Linhua Jiang, Nobunari Kashikawa, George Khorunzhev, Yongjung Kim, Yoshiki Matsuoka, Israel Matute, Ian McGreer, Christopher Onken, Sophie Reed,  Suhyun Shin, Daniel Stern,  Bram Venemans, Jonah Wagenweld,  Feige Wang, Lukas Wenzl, Chris Willott, Zhang-Liang Xie and Jinyi Yang, for providing their quasar spectra in digital form, Xiangyu Jin, Jan-Torge Schindler, Feige Wang and Jinyi Yang for re-formatting their published figures for this review, and George Becker, Roberto Decarli, Alyssa Drake, Zoltan Haiman, Yoshiki Matsuoka, Daniel Mortlock, Michael Strauss, Feige Wang and Jinyi Yang, and the editor Joss Bland-Hawthorn for their thoughtful comments on an earlier draft of the review.

% \textbf{EB needs to add additional names}
XF acknowledges supports from US NSF Grant AST 19-08284.
% References
%
% Margin notes within bibliography
%\section*{LITERATURE\ CITED}

%To download the appropriate bibliography style file, please see \url{https://www.annualreviews.org/page/authors/general-information}. \\

%\noindent
%Please see the Style Guide document for instructions on preparing your Literature Cited.

%The citations should be listed in alphabetical order, with no titles. For example:

\bibliography{qso_2022_ref}
\bibliographystyle{ar-style2}

\end{document}